\documentclass[10pt,reprint,aip,pop,amsmath,amssymb]{revtex4-1}
\pdfoutput=1 
\usepackage{graphicx}
\usepackage[dvipsnames]{xcolor}
\usepackage{epstopdf}

\newcommand{\Order}{\mbox{$\mathcal{O}$}} 
\newcommand\Alfven{Alfv\'en }

\newcommand{\V}[1]{\mathbf{#1}} 
\newcommand{\T}[1]{\texttt{#1}} 
 
\newcommand{\zhat}{\mbox{$\hat{\mathbf{z}}$}} 
 
\newcommand{\yhat}{\mbox{$\hat{\mathbf{y}}$}} 
\newcommand{\eqnref}[1]{Eqn.~\ref{#1}}
\newcommand{\figref}[1]{Figure~\ref{#1}}
\newcommand{\secref}[1]{\S\ref{#1}}

\begin{document}


\title{Characterizing Velocity-Space Signatures of Electron Energization in Large-Guide-Field Collisionless Magnetic Reconnection} 



\author{Andrew J. McCubbin}
\email[]{andrew-mccubbin@uiowa.edu}
\affiliation{Department of Physics and Astronomy, University of Iowa, Iowa City, IA 52240, USA}
                            
\author{Gregory G. Howes}
\affiliation{Department of Physics and Astronomy, University of Iowa, Iowa City, IA 52240, USA}

\author{Jason M. TenBarge}
\affiliation{Department of Astrophysical Sciences, Princeton University, Princeton, NJ 08544, USA}


\date{\today}

\begin{abstract}
Magnetic reconnection plays an important role in the release of magnetic energy and consequent energization of particles in collisionless plasmas. Energy transfer in collisionless magnetic reconnection is inherently a two-step process: reversible, collisionless energization of particles by the electric field, followed by collisional thermalization of that energy, leading to irreversible plasma heating. Gyrokinetic numerical simulations are used to explore the first step of electron energization, and we generate the first examples of field-particle correlation (FPC) signatures of electron energization in 2D strong-guide-field collisionless magnetic reconnection.  We determine these velocity space signatures at the x-point and in the exhaust, the regions of the reconnection geometry in which the electron energization primarily occurs. Modeling of these velocity-space signatures shows that, in the strong-guide-field limit, the energization of electrons occurs through bulk acceleration of the out-of-plane electron flow by parallel electric field that drives the reconnection, a non-resonant mechanism of energization. We explore the variation of these velocity-space signatures over the plasma beta range $0.01 \le \beta_i \le 1$. Our analysis goes beyond the fluid picture of the plasma dynamics and exploits the kinetic features of electron energization in the exhaust region to propose a single-point diagnostic which can potentially identify a reconnection exhaust region using spacecraft observations.
\end{abstract}

\pacs{}

\maketitle 


%
\section{Introduction}
Vast amounts of energy can be stored in the magnetic field of space and astrophysical plasmas. Upon reconfiguration, this embedded field may undergo reconnection that releases substantial energy, energizing particles, sometimes explosively. Magnetic reconnection occurs in a host of plasma regimes, from fusion device disruptions to the birth of the solar wind and solar flares. Additionally, magnetic reconnection occurs often in the dynamic solar wind, especially at interfaces with planetary magnetic fields. Identification and quantification of particle energization may help describe the physics needed to answer such questions as the coronal heating problem. \cite{Klimchuk:2006} In diffuse plasmas that are nearly collisionless, as those composing the solar wind and present throughout most of the heliosphere, kinetic descriptions of energy transfer are necessary to understand reconnection at particle kinetic length scales. \cite{Hesse:2011, Hesse:2020}

Significant work through theoretical, numerical and observational studies have developed a more complete picture of the various mechanisms potentially responsible for particle energization during magnetic reconnection in collisionless plasmas. \cite{Drake:2005, Egedal:2008, Egedal:2012, Drake:2012, Loureiro:2013b, Jiansen:2018, Pucci:2018, Munoz:2018, Dahlin:2020} Fermi acceleration and direct $E_{\parallel}$ particle acceleration have been identified to account for electron energization in collisionless reconnection. \cite{Egedal:2012, Dahlin:2014, Dahlin:2015} \citet{Dahlin:2016} demonstrated that there is a clear transition between Fermi acceleration and direct electric field acceleration with a guide field $B_g / B_R = 0.6 $, where $B_R$ is the in-plane reconnecting magnetic field magnitude. This first order Fermi acceleration is nearly completely suppressed at values $B_g / B_R > 1$. \citet{Pucci:2018} also found electron energization switches from perpendicular ($j_\perp E_\perp$) to parallel ($j_\parallel E_\parallel$) energization at $B_g / B_R = 0.6$  for driven reconnection. \citet{Guo:2017} identified evidence of perpendicular electric field energization of electrons in strong-guide-field steady reconnection, $B_g / B_R = 3$. Using 2D PIC simulations, they found perpendicular energization is due to polarization drifts, sustained by charge-separation generated electrostatic fields near the x-point and along the separatrices, and global curvature drifts. Additionally, they proposed charge-separation is sustained long enough to break the frozen-in condition in the presence of a large guide field, allowing perpendicular energization mechanisms due to additional sustained non-ideal effects which are not present in anti-parallel reconnection. 

Numerous studies on anti-parallel reconnection with no guide-field have been carried out. However, large-guide-field investigations have received less attention in the literature. This neglect is likely due to historically few observations of very large guide fields in the heliosphere, (i.e $B_g / B_R > 5$). Larger guide fields ($B_g / B_R > 1$) are expected to be found within the corona, which may be confirmed observationally as Parker Solar Probe completes its mission. In the last decade, larger guide fields have been found to exist during reconnection at the magnetosheath and magnetosphere \cite{Burch:2016, Eriksson:2016, Oieroset:2016, Oieroset:2017, Zhou:2018a}, and magnetotail \cite{Nakamura:2008, Rong:2012}. Additionally, many previous studies have focused on categorizing and identifying reconnection events through their fields and local macroscopic plasma parameters, both in numerical and spacecraft investigations. \cite{Eastwood:2010, Eastwood:2018, Zhou:2018a, Phan:2020} Some recent kinetic investigations have been enabled by advances in particle detection on recent spacecraft missions, which have identified evidence of crescent distributions that are likely common in asymmetric reconnection as found by the Magnetospheric Multiscale (MMS) mission. \cite{Price:2016,  Hesse:2017, Genestreti:2018} Until recently, few investigations have focused directly on kinetic particle energization in the non-relativistic large-guide-field limit using a formal kinetic analysis framework.

Most \textit{in situ} measurements in collisionless plasmas are performed by single spacecraft. Even in cases with multiple spacecraft, such as in the MMS mission, the few points of spatial information are insufficient to describe the larger scale spatial distribution of particle energization. Thus, analysis techniques must be structured to use localized (usually single-point) measurements to make observations and determinations of plasma behavior. Understanding the kinetic behavior of large-guide-field reconnection with single-point measurement techniques may help as a diagnostic tool for spacecraft identification of such events. We present an analysis of electron energization through the characterization of kinetic velocity-space signatures at specific spatial locations in a 2D gyrokinetic reconnection simulation. This work is built from a geometry proposed by \citet{Porcelli:2002} implemented in the Astrophysical Gyrokinetics Code $\T{AstroGK}$, as previously used by \citet{Numata:2015}. Using the field-particle correlation (FPC) framework developed by \citet{Klein:2016a}, we identify and analyze the velocity-space signatures of particle energization that arise from strong-guide-field collisionless reconnection. 

\section{Kinetic Mechanisms of Electron Energization}
In analyzing energy transfer in a weakly collisional plasma, it is important to point out that the energy transfer is inherently a two-step process \citep{Howes:2018}. First, collisionless interactions between the electromagnetic fields and the plasma particles serve to transfer energy between the fields and the particles.  Energy transferred from the fields to the particles will generate fluctuations in the particle velocity distribution function (VDF).  The collisionless energy transfer in this first step is inherently reversible, so the energy associated with those fluctuations is non-thermal. Subsequently, these fluctuations in the VDF can undergo linear \cite{Loureiro:2013b, Howes:2017a} or nonlinear phase mixing \cite{Tatsuno:2009} to sufficiently small scales in velocity space that arbitrarily weak collisions can serve to smooth out those fluctuations. This second collisional step is irreversible, effectively thermalizing the energy that was transferred to the particles, heating that plasma species and increasing the entropy.  

In this investigation, we will focus on the first step in this process, whereby the electromagnetic fields do reversible work on the plasma particles.  In magnetic reconnection, this process effectively releases magnetic energy and converts it into other forms (bulk plasma flows or non-thermal energization). Note that the collisionless energization can be facilitated through a resonant process, as in the case of Landau damping of kinetic \Alfven waves, or through non-resonant processes, \emph{e.g.}, direct particle acceleration by electric fields.  The second collisional step of the particle energization in collisionless magnetic reconnection was the focus of the analysis by \citet{Numata:2015}.  

In the analysis of the particle energization in weakly collisional plasma turbulence, the collisionless dynamics leads to the continual transfer of energy back and forth between the electromagnetic field fluctuations and the particles. If the turbulent fluctuation is undamped, this energy transfer is oscillatory and reversible, contributing no net particle energization. For example, an \Alfven wave in the MHD limit $k \rho_i \ll 1$ is undamped, and involves an oscillatory transfer of energy between magnetic field energy and plasma bulk flow kinetic energy. If the energy transfer is not purely oscillatory, \emph{e.g.}, in the case of a resonant damping process, a portion of the energy transfer may contribute to a net particle energization, which we will define as the secular, or net, energy transfer. This secular energy transfer is manifested through the net increase of microscopic kinetic energy of plasma particles leading to perturbations of the VDF. In collisionless magnetic reconnection, as is studied in this paper, the dynamics are not typically oscillatory as the magnetic field is  reconfigured, but reversible kinetic energy transfer is still possible. Therefore, it is important to investigate the net energization of the particles during the evolution of a plasma undergoing magnetic reconnection.

In order to analyze the collisionless energy transfer in the first step of particle energization, we begin with the generalized Boltzmann equation,
\begin{equation}
\frac{\partial f_s}{\partial t} + \V{v} \cdot \V{\nabla} f_s + 
\frac{q_s}{m_s} \left[\V{E} + \frac{\V{v} \times \V{B}}{c}\right] \cdot \frac{\partial f_s}{\partial \V{v}} = \left(\frac{\partial f_s}{\partial t}\right)_{coll}.
\label{eq:boltz}
\end{equation}
This equation represents the evolution of the 3D-3V velocity distribution function $f_s\left(\V{r},\V{v},t\right)$ for a plasma species $s$. The species charge and mass are $q_s$ and $m_s$ respectively, $\V{v}$ is the velocity, $\V{E}$ and $\V{B}$ are the electric and magnetic fields, and $c$ is the speed of light, and the right-hand side represents the collision operator. Combining the Boltzmann equation for each plasma species together with Maxwell’s equations forms the closed set of Maxwell–Boltzmann equations that govern the nonlinear evolution of turbulent fluctuations in a magnetized kinetic plasma.

On the timescale of the energy transfer occurring in magnetic reconnection, the collisional term is negligible under typical conditions in space and astrophysical plasmas, so we may neglect the collision operator, recovering the Vlasov equation. The Vlasov equation describes the full phase-space dynamics of a collisionless magnetized plasma.  The ballistic or advection term, second on the left-hand side of \eqnref{eq:boltz}, represents the advection of particles. The third term is the classical Lorentz force term, which governs the self-consistent wave-particle interactions in the kinetic plasma. Therefore, we focus on the Lorentz term to characterize  the energization of particles in collisionless magnetic reconnection. 

\section{Methods}

\subsection{Field-particle correlation Technique}
Developed by \citet{Klein:2016a}, the field-particle correlation (FPC) technique produces a velocity-space representation of the phase-space energy density transfer in a kinetic plasma applicable to a Vlasov-Maxwell description for a collisionless plasma. The FPC technique was initially developed to separate oscillatory energy transfer during a physical process from any secular energy transfer, by taking an average over a a sufficiently long correlation intervals that the oscillatory transfer largely cancels out. This method has been used to identify mostly resonant processes leading to net positive particle energization occurring in: wave damping in heliospheric plasmas \cite{Klein:2016a,Howes:2017a,Howes:2017c}, broadband kinetic turbulence \cite{Klein:2017b,Chen:2018,TCLi:2019,Klein:2020,Horvath:2020}, strong \Alfven wave collisions \cite{Howes:2018}, collisionless shocks \cite{Juno:2021}, and laboratory evidence of the electron energization responsible for the aurora \cite{Schroeder:2021}.

Using the Vlasov equation for the evolution of the particle distribution function, we can define a new formulation that describes the phase-space energy density evolution for a given species $s$. Multiplying the Vlasov equation by kinetic energy $ m_s v^2 / 2$, we cast into a form 
\begin{equation}
    \frac{\partial w_{s}}{\partial t}  = -\V{v} \cdot \V{\nabla}  w_{s} - 
\frac{q_s}{m_s} \left[\V{E} + \frac{\V{v} \times \V{B}}{c}\right] \cdot \frac{\partial w_{s}}{\partial \V{v}},
\label{eq:vlasov-energy}
\end{equation}
which describes the rate of change of the phase-space energy density $w_{e,s} = m_s v^2 f_s / 2$. Using this description for the rate of change of phase-space energy density, we can then define a field-particle correlation that produces the velocity-space signature of energization at a single spatial location. 

For the application of this technique to data from our collisionless magnetic reconnection simulations using the Astrophysical Gyrokinetics Code \T{AstroGK} \cite{Numata:2010}, we note that the gyrokinetic distribution function $h_s(x,y,z,v_\perp,v_\parallel)$ \cite{Howes:2006} is related to
the total distribution function $f_s$ via 
\begin{equation}
f_s(\V{r}, \V{v}, t) = 
F_{0s}(v)\left( 1 - \frac{q_s \phi(\V{r},t)}{T_{0s}} \right)
+ {h_s}(\V{r},  v_\parallel, v_\perp,t).
\label{eqn:fullF}
\end{equation}
The parallel and perpendicular directions are with respect to the local equilibrium magnetic field $\V{B}_0$ up to $\Order(\epsilon^2)$ in the gyrokinetic ordering. \cite{Howes:2006} As a technical step, we transform from the gyrokinetic distribution
function $h_s$ to the complementary perturbed distribution function 
\begin{equation}
{g_s}(\V{r}, v_\parallel,v_\perp) = {h_s}(\V{r}, v_\parallel,v_\perp) 
- \frac{q_s F_{0s}}{T_{0s}} 
\left\langle 
\phi 
- \frac{\V{v}_\perp \cdot \V{A}_\perp}{c}
\right\rangle_{\V{R}_s},
\label{eqn:g+h}
\end{equation}
where $\langle ...\rangle$ is the gyroaveraging operator. \cite{Schekochihin:2009} The complementary distribution function $g_s$ describes perturbations to the background distribution in the frame of reference moving with the transverse oscillations of an \Alfven wave.  Field-particle correlations calculated using $h_s$ or $f_s$ yield qualitatively and quantitatively similar results to those computed with $g_s$. \cite{Klein:2017b} 

Below, we present the correlations between the complementary perturbed distribution function and the parallel electric field $E_\parallel$ at a single-point $\V{r}_0$
\begin{equation}
  C_{E_\parallel,s} (v_\parallel,v_\perp,t)= C\left(- q_s\frac{v_\parallel^2}{2}
  \frac{\partial g_s(\V{r}_0,v_\parallel,v_\perp,t)}{\partial
    v_\parallel},E_\parallel(\V{r}_0,t)\right).
   \label{eq:cepar_gs}
\end{equation}
More generally, the correlation can be modified if an oscillatory signal is present using a time sliding averaged window $\tau$; however, a laminar reconnection flow does not typically lead to a significant oscillatory energy transfer between the fields and particles during the main phase of reconnection.  Therefore, computing a time-average over a finite correlation interval is not necessary to cancel out a large oscillatory component. Therefore, we chose to simply evaluate the field-particle correlation instantaneously, taking the correlation interval $\tau=0$. 

The parallel electric field correlation defined in \eqnref{eq:cepar_gs} describes the phase-space energy transfer rate to species $s$ by $E_\parallel$ at a single point in space $\V{r}_0$ and is a three dimensional function in gyrotropic phase space and time, $(v_\parallel,v_\perp,t)$.  We present the correlation in several standard ways to aid visualization of the particle energization in velocity-space and time. A \emph{gyrotropic plot} of the correlation  $C_{E_\parallel,s} (v_\parallel,v_\perp)$ at a specific time $t_0$ shows how the rate of change of phase-space energy density varies in gyrotropic velocity space $(v_\parallel,v_\perp)$, as in \figref{fig:xpoint}(b).  We generally refer to the pattern of energization seen in the gyrotropic plot as the \emph{velocity-space signature} of the particle energization mechanism.  

Alternatively, we can integrate the correlation over the perpendicular velocity,
\begin{equation}
  C_{E_\parallel,s} (v_\parallel,t)= \int v_\perp d v_\perp
  C_{E_\parallel,s} (v_\parallel,v_\perp,t),
   \label{eq:cepar_reduced}
\end{equation}
to obtain the \emph{reduced parallel correlation}, $C_{E_\parallel,s} (v_\parallel,t)$.  
A \emph{timestack plot} presents this reduced parallel correlation as a function of  $v_\parallel$ and time, which is particularly useful to explore the rate energization of particles over the course of the main phase of magnetic reconnection in our simulations, as in the main panel of \figref{fig:xpoint}(c).

Integrating the reduced parallel correlation over time yields
\begin{equation}
  C_{E_\parallel,s} (v_\parallel)= \int  dt
  C_{E_\parallel,s} (v_\parallel,v_\perp,t),
   \label{eq:cepar_reduced_vpar}
\end{equation}
a simple one-dimensional representation of the rate of energization of particles as a function of $v_\parallel$ over the course of the simulation, as in the lower panel of  \figref{fig:xpoint}(c). This visualization facilitates the identification of the bipolar signatures that are indicative of collisionless resonant energization mechanisms, such as Landau damping \citep{Chen:2018,Horvath:2020}.

Alternatively, one can integrate the reduced parallel correlation over $v_\parallel$ to obtain the rate of particle energization at the single spatial point $\V{r}_0$ as a function of time, given by 
\begin{equation}
  \left(\frac{\partial  W_s(t)}{\partial t}\right)_{E_\parallel} \equiv \int  dv_\parallel
  C_{E_\parallel,s} (v_\parallel,v_\perp,t) = j_{\parallel,s}(\V{r}_0,t) E_{\parallel}(\V{r}_0,t),
    \label{eqn:corrworkequiv}
\end{equation}
as in the left-hand panel of \figref{fig:xpoint}(c).  Note that this form shows that, when integrated over all velocity space, the parallel electric field correlation simply yields the rate of work done by the parallel electric field on the particle species  $s$ at position $\V{r}_0$ versus time.

\subsection{Simulation} 
In this paper, we analyze 2D magnetic reconnection simulations ($d/dz = 0$) with a strong out-of-plane guide field. The domain consists of a doubly-periodic slab geometry with an in-plane reconnection field. To solve the fully electromagnetic gyrokinetic equations for the electrons and ions, \T{AstroGK} employs a pseudo-spectral algorithm for the spatial coordinates $(x,y)$, and Gaussian quadrature for velocity space integrals. The velocity grid is discretized into energy $E_s = m_s v^2/2$ and pitch angle $\lambda = v_{\perp}^2 / (B_{z0} v^2)$ values, where $B_{z0}$ is the constant, background (guide) magnetic field. Derivatives of velocity space in the collision operator are estimated using a first-order finite difference scheme on an unequally spaced grid according to the quadrature rules in \citet{Barnes:2009}. 
We perform the same simulations as in \citet{Numata:2015} with a fixed collisionality $\nu_{ei} = \nu_{ee} =  \nu_{ii} = 1.0 \times 10^{-4}$ for all simulations. The simulation is initialized with an unstable tearing mode for the in-plane magnetic field configuration as in \citet{Numata:2010, Numata:2011}. The equilibrium total magnetic field is given by 
\begin{equation}
    \V{B} = B_{z0} \zhat + B_{y}^{eq}(x) \yhat, \quad B_{y}^{eq} / B_{z0} \sim \epsilon \ll 1,
\end{equation}
where $B_{y}^{eq}$ is the in-plane, reconnecting component, with a maximum value $B_y^{max} = 1$, determined from the parallel vector potential by $B_{y}^{eq}(x) = -\partial A_{\parallel}^{eq} / \partial x$, $\epsilon$ is the gyrokinetic epsilon --- a small expansion parameter defining scale separation in gyrokinetics (see \citet{Howes:2006}). The background Maxwellian electron distribution is perturbed with a perturbation of the from $\delta f_e \varpropto V_{\parallel} f_{e0}$. To support the modified distribution function, the vector potential is defined as follows using a shape function $S_{h}(x)$ to enforce periodicity (see \citet{Numata:2010}) such that 
\begin{equation}
    A_{\parallel}^{eq}(x) = A_{\parallel 0}^{eq} \cosh{\left(\frac{x - L_x/2}{a}\right)} S_{h}(x).
\end{equation}
$A_{\parallel}^{eq}$ arises from the parallel electron current that must satisfy Amp\`ere's law. The dimensions of the simulation determine the scale lengths, with equilibrium current width $a$ and $L_x$ the scale of the box in the $x$-direction, where $L_x/a = 3.2\pi$. For the $y$-direction, the width of the box is $L_y/a = 2.5\pi$. The tearing mode is imposed by a small sinusoidal perturbation to the equilibrium magnetic field, so that $\Tilde{A}_{\parallel} \varpropto \cos{(k_y y)}$ with wave number $k_y a = 2\pi a/L_y = 0.8$, which yields $\Delta^{\prime} a \approx 23.2$ for the tearing instability parameter. The plasma considered is quasi-neutral, so that $n_{0i} = n_{0e} = n_0$, with singly charged ions $q_i = -q_e = e$.

The scale of the system is determined by the equilibrium magnetic field. Thus, we normalize time by the \Alfven time $\tau_A \equiv a/V_A$, where $V_A \equiv B_y^{max} / \sqrt{4 \pi n_0 m_i}$ is the in-plane \Alfven velocity corresponding to the maximum initial $B_y^{eq}$. 

Additional fundamental parameters define the physical scales within the simulation: The mass ratio, $\mu = m_e/m_i$, the equilibrium plasma temperature ratio $\tau \equiv T_{0i}/T_{0e}$, the electron plasma beta, $\beta_e \equiv n_o T_{0e}/ (B_{z0}^2/8 \pi)$, and the ratio of ion sound Larmor radius to the equilibrium scale length a (i.e. width of the current sheet), $\rho_{se}/a \equiv c_{se} / (\Omega_{ci} a)$. The ion sound speed for cold ions is $c_{se} = \sqrt{T_{0e/m_i}}$, and the ion cyclotron frequency is $\Omega_{ci} = e B_{z0}/(m_i c)$. The following parameters are fixed for all simulations throughout this paper: 
\begin{align}
    \rho_{se}/a &= 0.25/\sqrt{2} & \mu &= 0.01 & \tau &= 1.
\end{align}
These scale parameters require $\rho_i / a = 0.25$, $\rho_e/a = 0.025$, and $\beta_i = \beta_e$. Five simulations are performed using varying values of $\beta_i = (0.01,0.03,0.1,0.3,1.0)$. 

In these simulations, the electron current layer width $\delta_{CS,e}$ decreases with increasing $\beta_e$, such that the electron Larmor radius $\rho_e = \sigma^{1/2} \rho_{se} \sqrt{2}$ approaches the electron skin depth $d_e = \beta_{e}^{-1/2} \sigma^{1/2} \rho_{se} \sqrt{2}$.  \citet{Numata:2015} demonstrate with linear simulations in the collisionless regime, the frozen-flux condition is broken by electron inertia for small $\beta_e$. When $\beta_e$ is greater than unity, $\rho_e$ becomes larger than $d_e$, such that electron FLR effects, rather than electron inertia, lead to field line breaking. By varying the collisionality, they also find in the linear regime small ($\nu_e \tau_A \lesssim 1 \times 10^{-3}$), but finite collisionality results in asymptotic growth rates and current sheet width.

\section{Results}

\subsection{Partitioning of Energization by Species}
The energy budget for the simulations shown in \figref{fig:rxn_energy} of Appendix \secref{sec:rxn_energy} confirms the energy released from the in-plane magnetic field flows into the electrons. This result is consistent with previous investigations. \cite{Egedal:2012, Dahlin:2014} As $\beta_i$ increases, a smaller proportion of the magnetic energy is transferred to the electrons and more energy is transferred to the ions, with near equipartition between electrons and ions at $\beta_i = 1$. This change is due to the electron flow decreasing in the reconnection region, and subsequently a slower reconnection onset and smaller reconnection rate. The advection rate of the magnetic field lines decreases as $\beta_i$ increases in a self-consistent manner with plasma pressure. In addition, the total fraction of the initial in-plane magnetic energy transferred to both species is lower, indicating that the energization is inhibited. This decrease occurs consistently with the notion that a smaller $E_\parallel$ is needed to support the outflow pressure balance. In this paper, we focus strictly on investigating the electron energization, leaving the energization of the ions to be explored in future work.
 
\subsection{Locations of Electron Energization}
The electron energization as a function of the position in the $(x,y)$ plane is given by the work done on the electrons by the electric field, $\V{j}_e \cdot \V{E}$.  Our gyrokinetic simulations of collisionless magnetic reconnection are valid in the limits of strong guide field $B_{z0}/ B_{y}^{eq} \sim \epsilon^{-1} \gg 1$ and of the gyrokinetic approximation with $k_\perp \gg k_\parallel$, where the parallel direction is along the guide field in the out-of-plane, $z$ direction. In all of our simulations, the summed contributions to the electron energization by the in-plane (perpendicular) components of the electric field are much smaller than that by the out-of-plane component, $j_{x,e} E_x + j_{y,e} E_y \ll j_{\parallel,e} E_\parallel$, as expected in the gyrokinetic limit.  In the gyrokinetic approximation, the quasi-neutrality condition \citep{Howes:2006} dictates that  $\V{\nabla} \cdot \V{j} = 0$, so the limit $k_\perp \gg k_\parallel$ implies that $\V{k}_\perp \cdot \V{j}_\perp =0$  to lowest order in the gyrokinetic expansion parameter $\epsilon$.  Since the perpendicular electric field in the same $k_\perp \gg k_\parallel$ limit scales as $\V{k}_\perp \phi$, where $\phi$ is the electrostatic potential, then the work done by the perpendicular electric field $\V{j}_{\perp,e} \cdot \V{E}_\perp$  scales as $\V{j}_{\perp,e} \cdot \V{k}_\perp \phi \simeq 0$ to lowest order in $\epsilon$.  Therefore, in our analysis here we focus strictly on the parallel contribution to the rate of electron energization, $j_{\parallel,e} E_\parallel$.

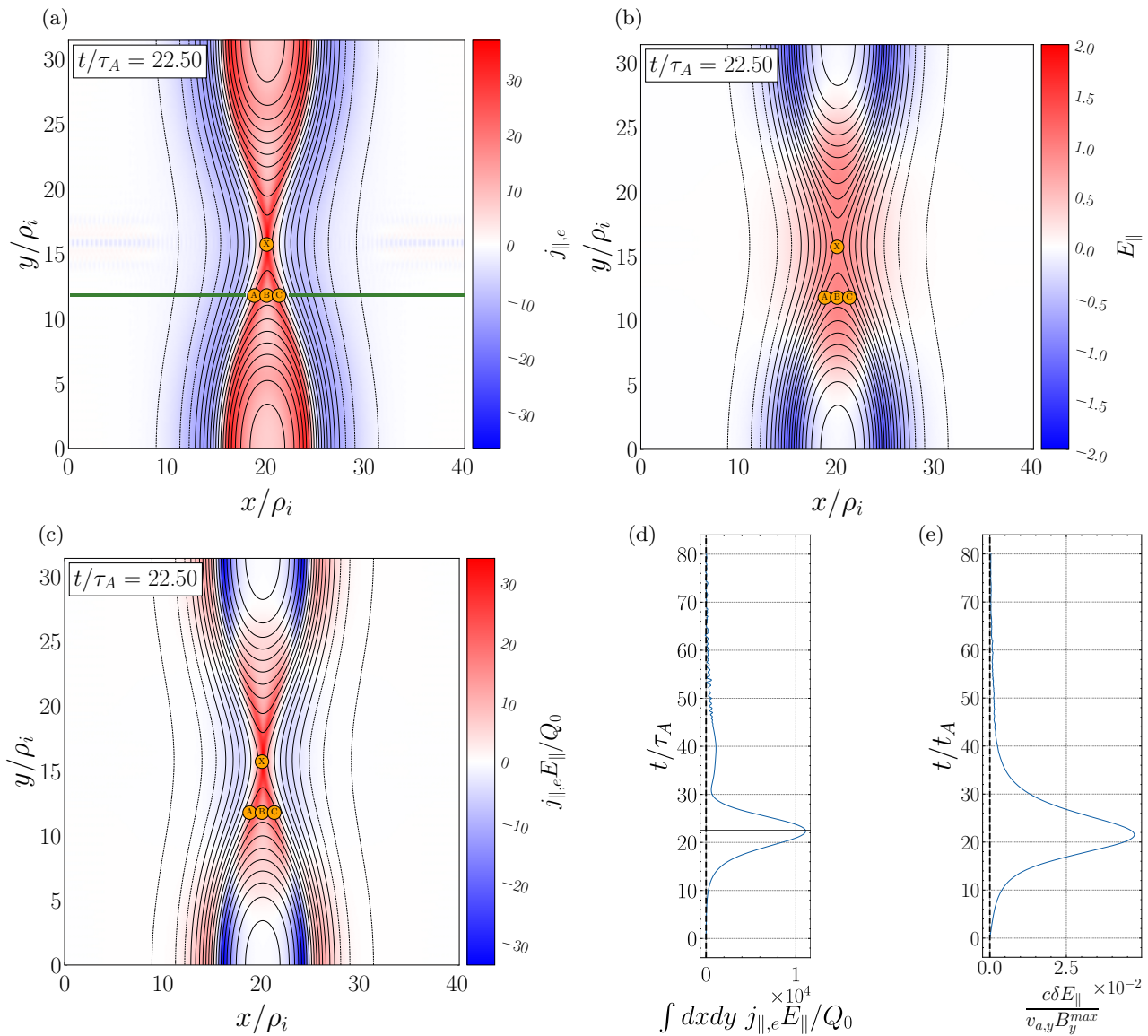
\begin{figure*}
   \setlength{\unitlength}{1in}
   \begin{picture}(3.3,3.0)
       \put(0,0){\includegraphics[width=3.3in]{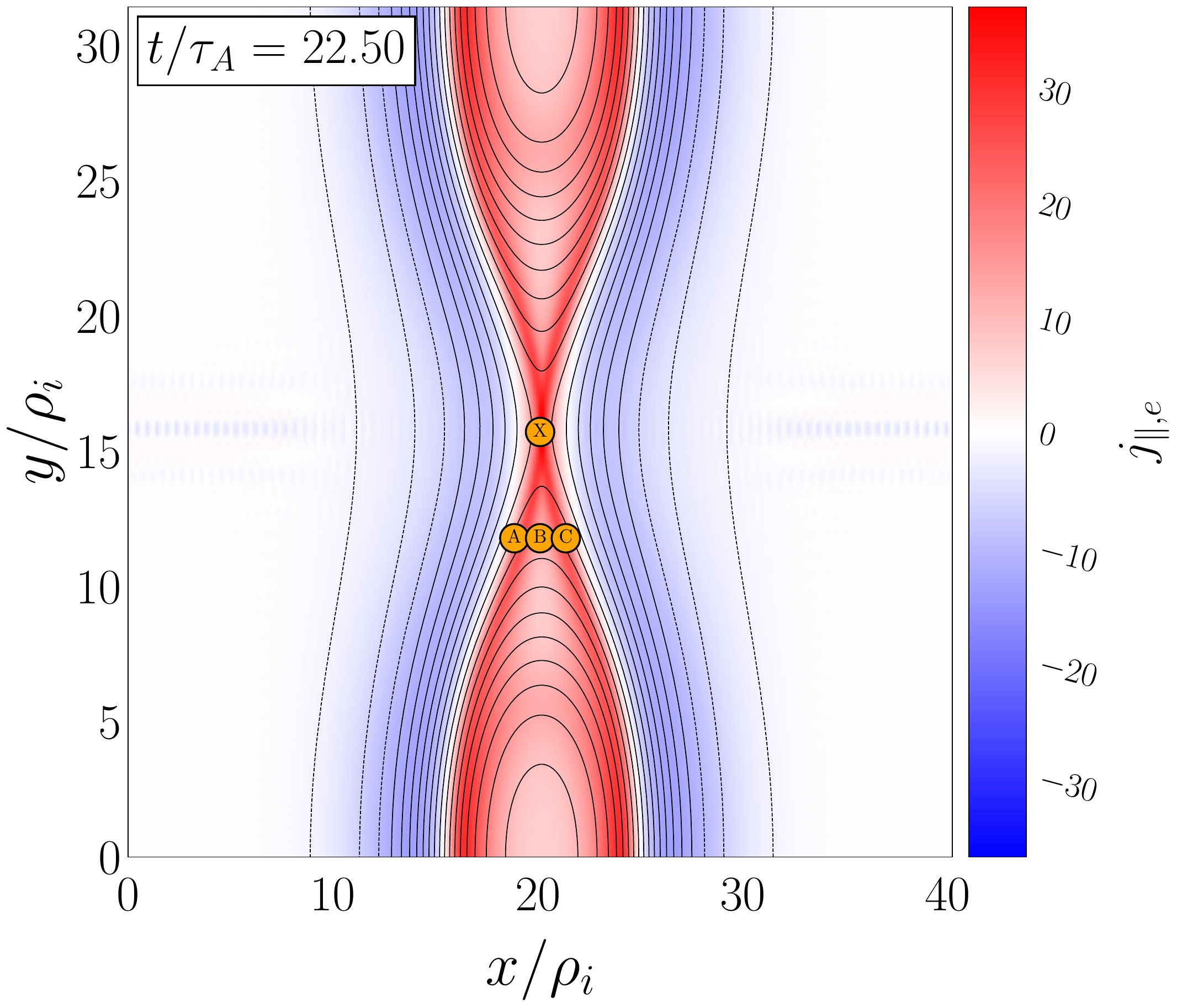}}
       \put(0.2,2.9){(a)}
       \linethickness{1.25pt}
       \put(0.325,1.32){ \color{OliveGreen} \line(1,0){1.025}}
       \put(1.6,1.32){ \color{OliveGreen} \line(1,0){1.025}}
   \end{picture}
   \begin{picture}(3.3,3.0)
       \put(0,0){\includegraphics[width=3.3in]{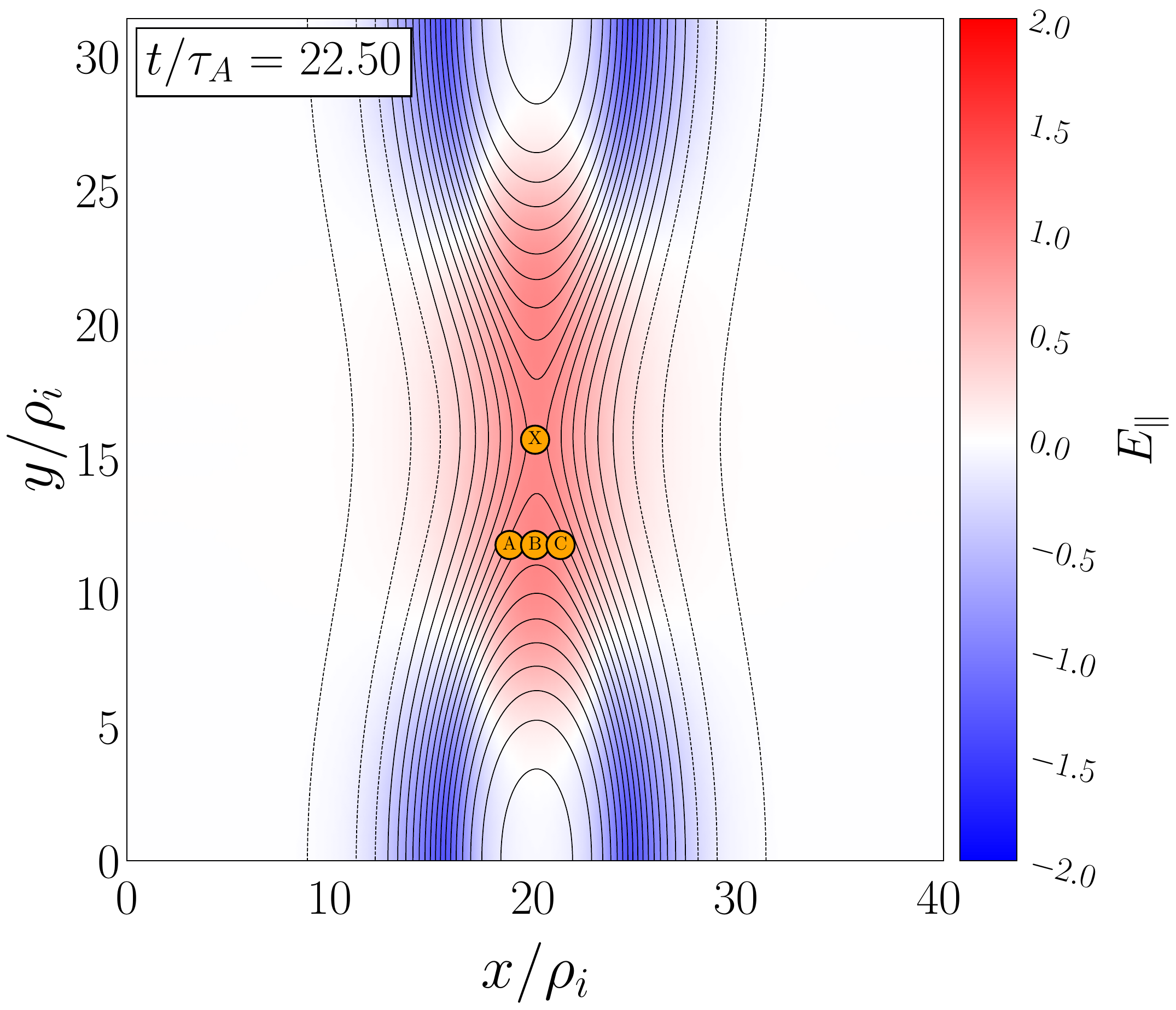}}
       \put(0.2,2.9){(b)}
   \end{picture} 
    \\
     \begin{picture}(3.3,3.0)
       \put(0,0){\includegraphics[width=3.3in]{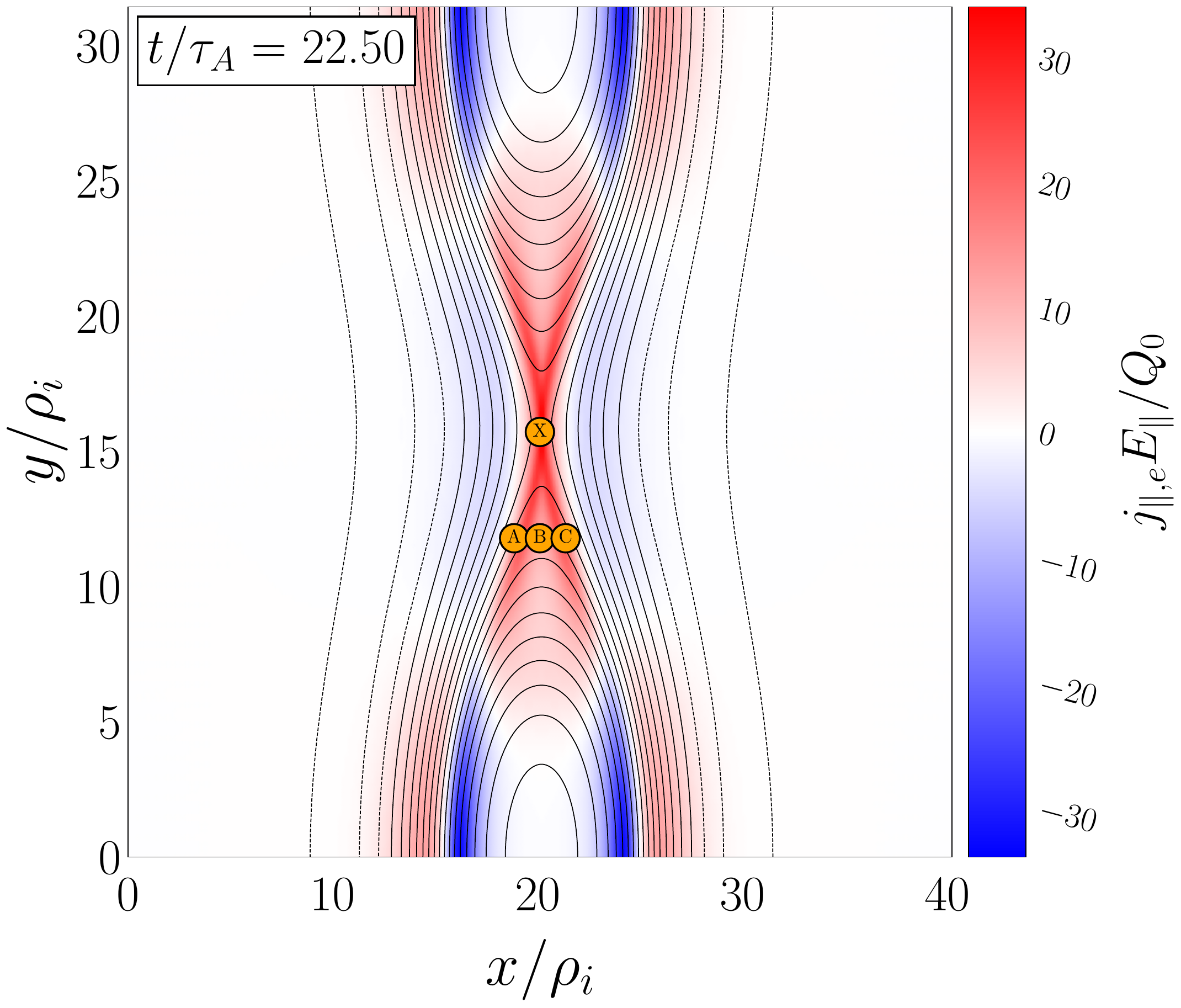}}
       \put(0.2,2.9){(c)}
   \end{picture}
   \begin{picture}(1.65,3.0)
       \put(0.35,0){ \includegraphics[height=2.95in]{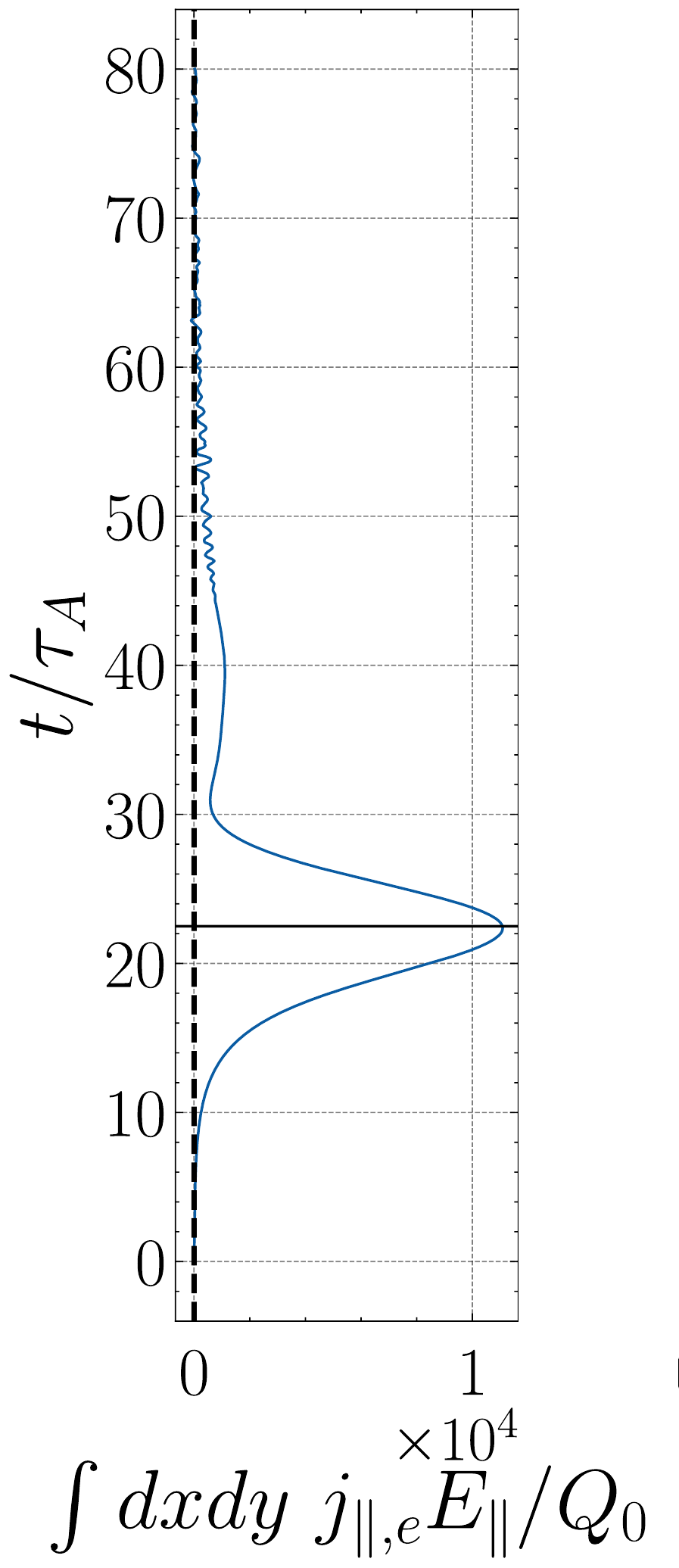}}
       \put(0.3,2.9){(d)}
   \end{picture} 
    \begin{picture}(1.65,3.0)
       \put(0.35,0){\includegraphics[height=2.95in]{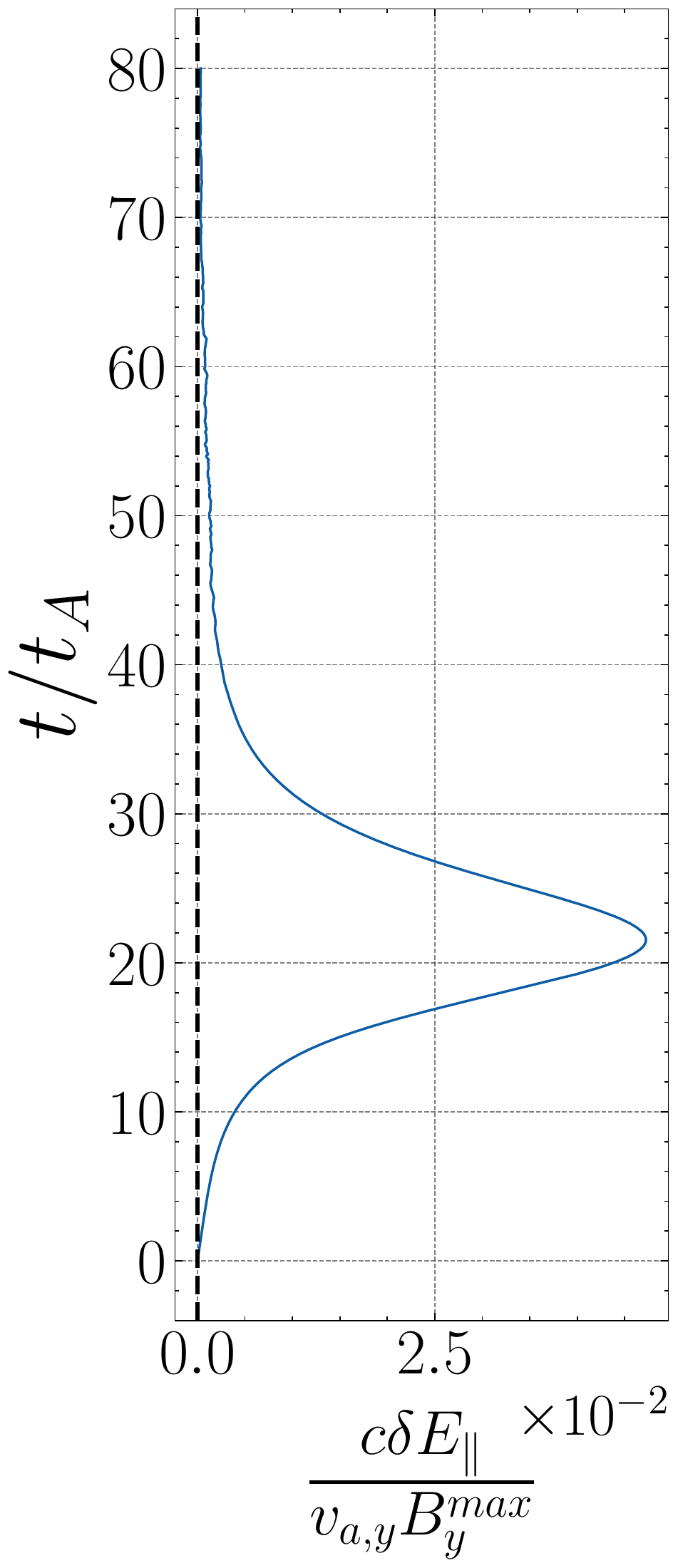}}
       \put(0.3,2.9){(e)}
   \end{picture} 
    \caption{Plots for the $\beta_i = 0.01$ simulation of the reconnecting plane displaying (a) electron $j_\parallel E_\parallel$c, (b) $E_\parallel$, and (c) electron $j_\parallel$ at the time of maximum reconnection rate $t/\tau_A = 22.5$. (d) Spatially integrated $j_\parallel E_\parallel$ for the entire simulation interval with horizontal black line at maximum reconnection rate. (e) Normalized reconnection rate for the entire simulation interval.} 
    \label{fig:coordinate2d_panel}
\end{figure*}

In \figref{fig:coordinate2d_panel}, for the $\beta_i = 0.01$ simulation, we plot the spatial distribution over the $(x,y)$ plane of (a) the parallel electron current $j_{\parallel,e}$, (b) the parallel electric field $E_{\parallel}$, and (c) the work done on the electrons due to the reconnection electric field $j_{\parallel,e} E_{\parallel} / Q_0$, where we normalize by the characteristic heating rate per unit volume, $Q_0 = (n_{0i} T_{0i} v_{ti} / L_\parallel)(\pi/8)(L_\perp/L_{\parallel})^2$.  We plot these quantities at the time $t/\tau_A = 22.5$ of the maximum  spatially-integrated electron energization rate $\int dx dy j_{\parallel,e} E_{\parallel} / Q_0$, plotted in \figref{fig:coordinate2d_panel}(d). The reconnection rate, estimated using the magnitude of the reconnecting (parallel) electric field $E_\parallel$ at the x-point in the simulation $c E_\parallel/(v_{A,y} B^{max}_y)$, is plotted in \figref{fig:coordinate2d_panel}(e), and its peak roughly coincides in time with the spatially-integrated electron energization rate during the main phase of reconnection, roughly spanning $10 \lesssim t/\tau_A \lesssim 30$.   Note that, for the initial Porcelli equilibrium \cite{Porcelli:2002} employed in these simulations, there is a limited amount of upstream magnetic flux, so the main phase of reconnection eventually ceases once a majority of the initial flux has reconnected.

In \figref{fig:coordinate2d_panel}(a), the parallel electron current in the $+z$ direction peaks at the x-point and along the separatrices. The parallel electric field which drives the reconnection flow is fairly uniform throughout the ion diffusion region (approximately spanning the range $17< x/\rho_i<23$ and $7< y/\rho_i<25$), as shown in \figref{fig:coordinate2d_panel}(b).  The rate of change of electron energy density is given by the product of these two quantities, and we find that the positive electron energization occurs dominantly over a relatively small region at the x-point and along the separatrices within the ion diffusion region, as indicated by the bright red regions in \figref{fig:coordinate2d_panel}(c).  Since the field-particle correlation technique is applied at specific spatial locations to determine the nature of the mechanisms for particle energization at those positions, we choose the following positions to investigate the electron energization in this study: (i) the x-point, denoted by point ``X,'' and (ii) three positions as you traverse horizontally (green line) through the lower exhaust region, denoted by points ``A,'' ``B,'' and ``C.''  Below, we perform a field-particle correlation analysis at each of these points to identify the velocity-space signatures of electron energization in collisionless magnetic reconnection with a strong guide field.

Note that our finding that the electron energization is dominated by $j_{\parallel,e} E_{\parallel}$ is consistent with previous investigations of magnetic reconnection in the moderate to strong guide field limit \cite{Dahlin:2014, Munoz:2015, Dahlin:2016}.  The energization of electrons in the out-of-plane direction, parallel to the guide field to lowest order in $\epsilon$, is qualitatively different from magnetic reconnection in the small to zero guide field limit, where the local parallel direction is mostly along the in-plane reconnecting field.  We see no first order Fermi acceleration \cite{Drake:2006,Dahlin:2014} due to the gyrokinetic ordering, so an analysis of that source of acceleration is neglected here.

\subsection{Energization at x-point}
\begin{figure}
   \setlength{\unitlength}{1in}
   \begin{picture}(3.3,2.1)
       \put(0,0){\includegraphics[width=3.3in]{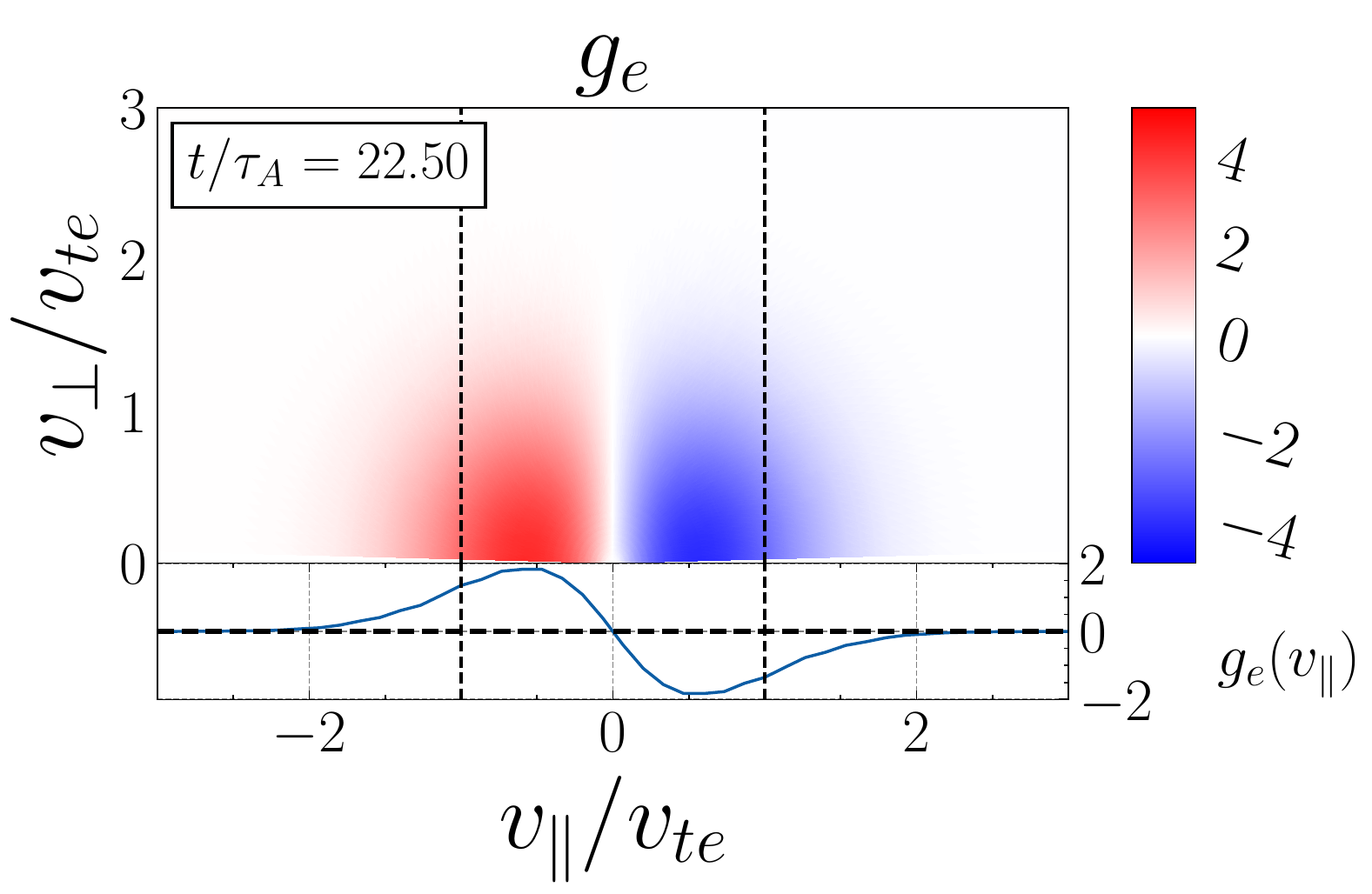}}
       \put(0.1,2.0){(a)}
   \end{picture} \\
   \begin{picture}(3.3,2.1)
       \put(0,0){\includegraphics[width=3.3in]{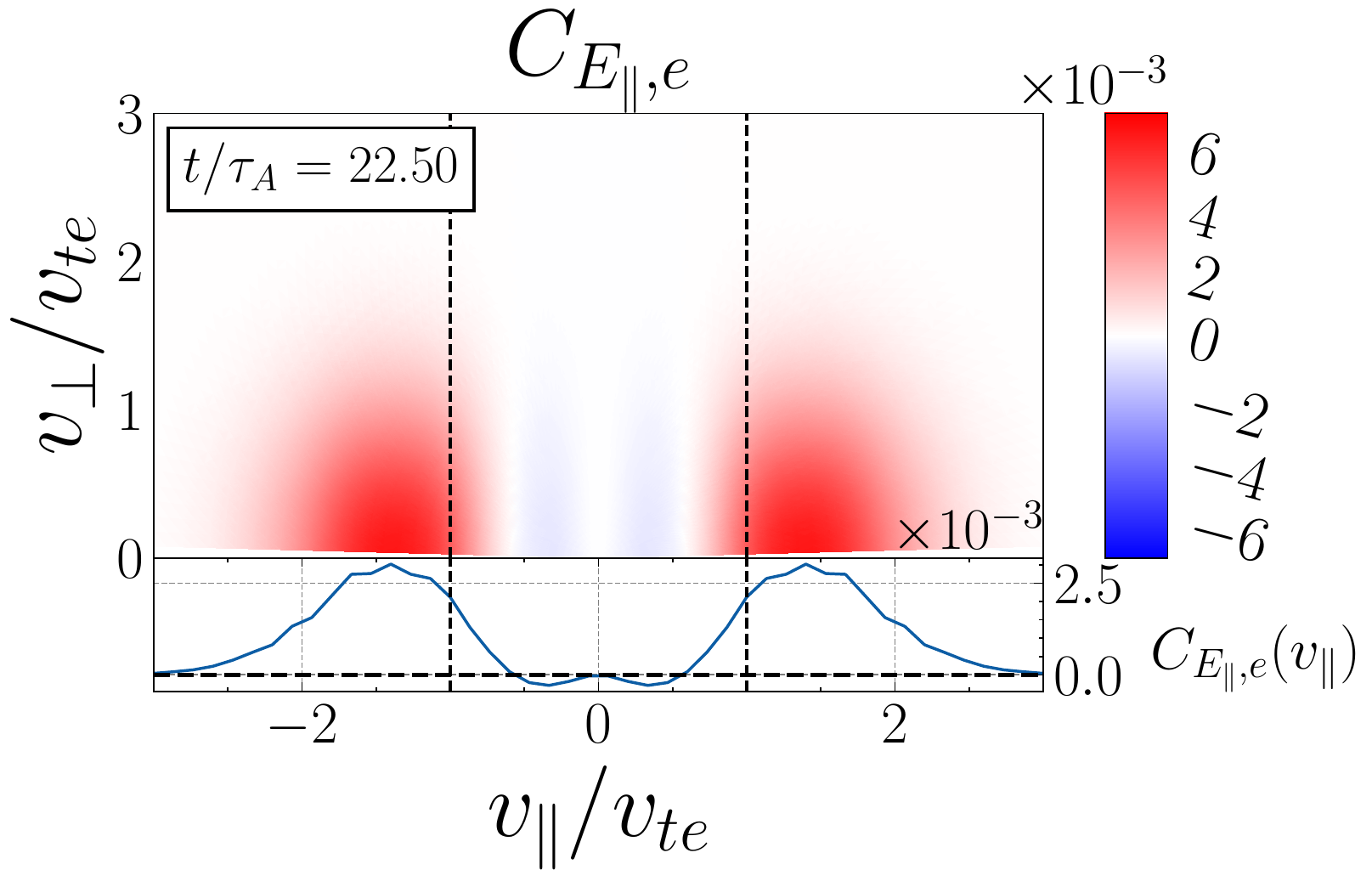}}
       \put(0.1,2.0){(b)}
   \end{picture} \\
   \begin{picture}(3.3,2.2)
       \put(0,0){\includegraphics[width=3.3in]{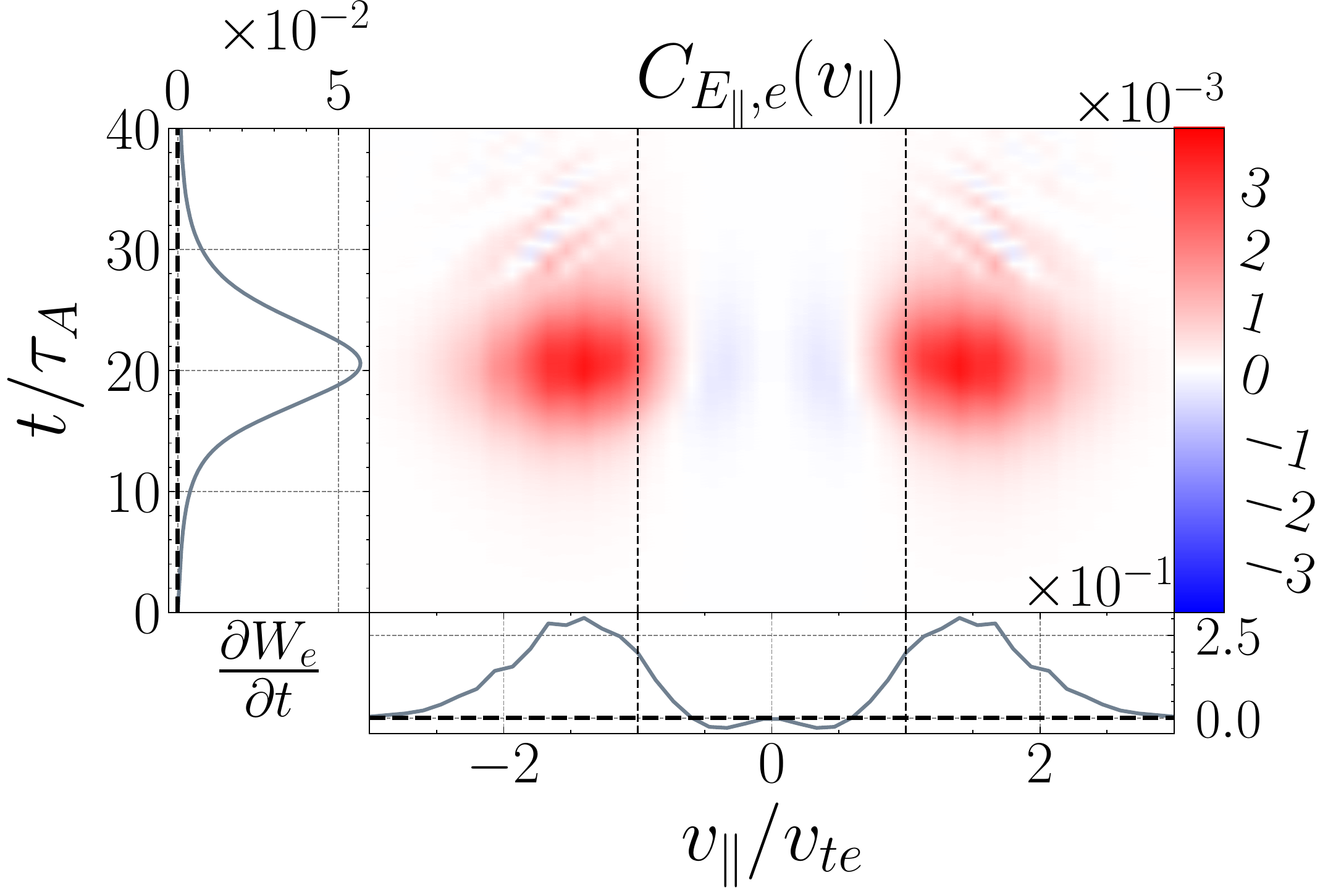}}
       \put(0.1,2.15){(c)}
   \end{picture} 
    \caption[x-point velocity space field particle correlations]{(a) A gyrotropic plot of the complementary perturbed electron distribution function $g_e(v_\parallel,v_\perp)$ at the time of the peak energization rate $t/\tau_A=22.5$ at the x-point, $(x/\rho_i, y/\rho_i) = (20.1, 15.7)$.  The reduced parallel perturbed distribution $g_e(v_\parallel)$, obtained by integrating over $v_\perp$, is shown in the lower panel. (b) The gyrotropic parallel electric field correlation $C_{E_\parallel}(v_\parallel,v_\perp)$ at the same time and position, along with the reduced parallel correlation $C_{E_\parallel}(v_\parallel)$ in the lower panel. The vertical dashed black lines indicate the \Alfven velocity $\pm v_{A,z}/v_{te}$ in the parallel direction. (c) A timestack plot of the reduced parallel correlation over time $C_{E_\parallel}(v_\parallel,t)$, with the net electron energization rate vs.~time in the left panel and the time-integrated reduced parallel energization rate over the full interval shown in the lower panel, demonstrating that the energization at the peak time $t/\tau_A=22.5$ is consistent with the time-integrated energization at the x-point.
    \label{fig:xpoint}}
\end{figure}
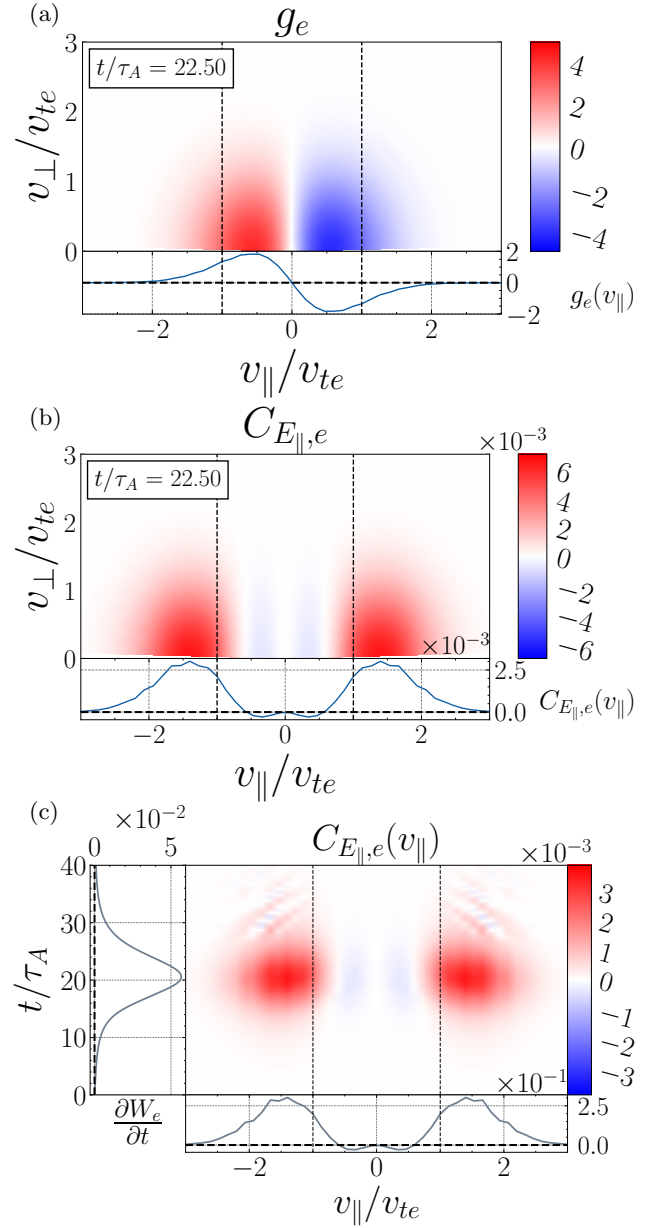

To investigate the energization of the electrons at the x-point of the reconnection geometry, located at $(x/\rho_i, y/\rho_i) = (20.1, 15.7)$, we focus initially on the perturbations to the electron velocity distribution function and the parallel electric field at the time $t/\tau_A=22.5$ when the electron energization rate $j_{\parallel,e} E_{\parallel}$ peaks at the x-point.  In \figref{fig:xpoint}(a), we plot the complementary perturbed electron distribution function $g_e(v_\parallel,v_\perp)$ over gyrotropic velocity space, along with the reduced parallel perturbed distribution $g_e(v_\parallel)$, obtained by integrating over $v_\perp$, in the lower panel. This perturbed velocity distribution leads to the parallel electron current $j_{\parallel,e}$ needed to sustain the change in the $B_y$ component of the magnetic field across the midplane, as seen in \figref{fig:coordinate2d_panel}.

To explore the electron energization at the x-point, we use $g_e(v_\parallel,v_\perp)$ and the parallel electric field $E_\parallel$ to compute the field-particle correlation in gyrotropic velocity space $C_{E_\parallel,e}(v_\parallel,v_\perp)$, given by \eqnref{eq:cepar_gs} and plotted in \figref{fig:xpoint}(b).  In this figure, we see a symmetric (about $v_\parallel)$ increase in the phase-space energy density $w_e$ of the electrons, dominantly occurring over the parallel velocity range $1 \lesssim |v_\parallel|/v_{te} \lesssim 2$, as made clear by the lower panel where the reduced parallel correlation $C_{E_\parallel}(v_\parallel)$, computed by integrating over $v_\perp$, is plotted. The vertical dashed black lines indicate the \Alfven velocity $v_{A,z} = B_{z0} / \sqrt{4 \pi n_0 m_i} $ in the parallel direction, $\pm v_{A,z}/v_{te}$, for $\beta_i=0.01$ and mass ratio $m_i/m_e=100$. The gyrotropic correlation  $C_{E_\parallel,e}(v_\parallel,v_\perp)$ in panel (b) is the \emph{velocity-space signature} of the electron energization at the x-point in this simulation of collisionless magnetic reconnection in the strong-guide-field limit.

To probe how this electron energization as a function of $v_\parallel$ evolves over time at the x-point, we plot in \figref{fig:xpoint}(c) a timestack plot of the reduced parallel correlation $C_{E_\parallel}(v_\parallel,t)$ over time.  The vertical panel on the left presents the net electron energization rate due to $E_\parallel$, $\partial W_e(t)/\partial t = \int d{v_\parallel} C_{E_\parallel}(v_\parallel,t)$ as a function of time. Note that the majority of the electron energization at the x-point occurs over the interval $10 \le t/\tau_A \le 30$, with the peak energization rate occurring around $t/\tau_A=22.5$.  The horizontal panel below presents the time-integrated reduced parallel energization rate over the full interval shown in the lower panel, $\int dt C_{E_\parallel}(v_\parallel,t)$.  By comparing this time-integrated energization rate at the x-point to the energization rate at the peak time $t/\tau_A=22.5$, shown in the lower panel of (b),  we see that both the peak and time-integrated energization rates have a nearly identical dependence on $v_\parallel$.

\begin{figure}
    \centering
    \includegraphics[width=2.8in]{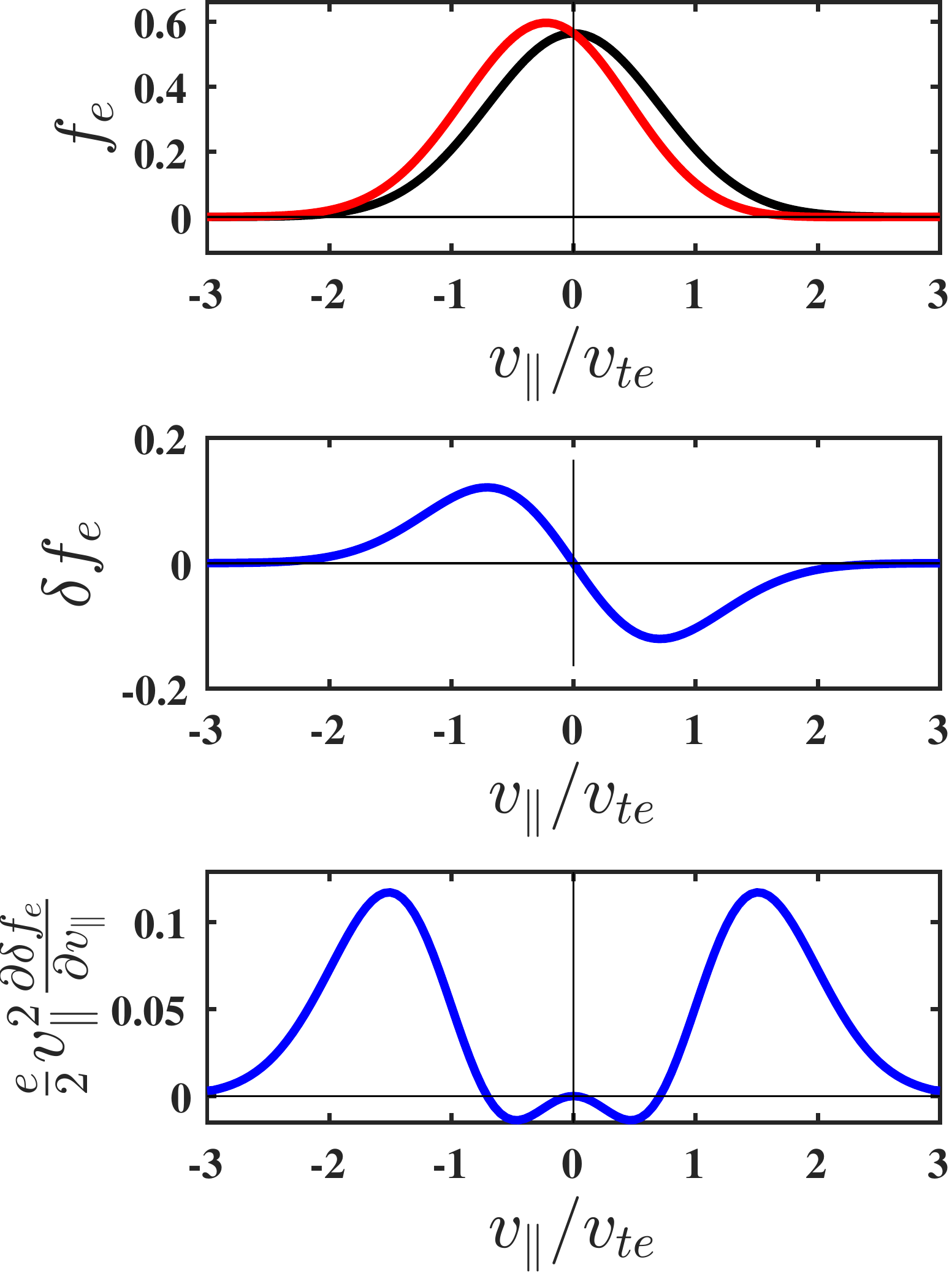}
    \caption[Simple model of modified distribution function changing field-particle correlation signature]{(a) A model form of the total electron parallel velocity distribution (red) with the original Maxwellian equilibrium electron parallel velocity distribution (black). (b) The model perturbed electron velocity distribution function as in \eqnref{eq:deltafe}. (c) The corresponding reduced correlation representing the phase-space density signature of the perturbed electron velocity distribution function.}
    \label{fig:model_xpoint}
\end{figure}

A simple model can be constructed that explains the qualitative and quantitative features of the reduced velocity-space signature observed in the lower panel of \figref{fig:xpoint}(b). The perturbed distribution function plotted in \figref{fig:xpoint}(a) is well approximated by a form 
\begin{equation}
\delta f_e (v_\parallel) = 2 \frac{U_{\parallel,e}}{v_{te}}\frac{v_{\parallel}}{v_{te}} f_{0,e} (v_\parallel),
\label{eq:deltafe}
\end{equation}
where the integration over this perturbation leads to zero density perturbation and a parallel flow given by the parameter $U_{\parallel,e}$. The total electron velocity distribution is given by $f_e(v_\parallel)= f_{0,e}(v_\parallel) +  \delta f_e(v_\parallel)$, where $f_{0,e}(v_\parallel)$ is the equilibrium Maxwellian parallel velocity distribution.  In \figref{fig:model_xpoint}(a), we plot the equilibrium electron parallel velocity distribution $f_{0,e}(v_\parallel)$ (black) and the total parallel electron velocity distribution $f_e(v_\parallel)$ (red).  Panel (b) shows the perturbed electron velocity distribution function $\delta f_e(v_\parallel)$ (blue) with parameter $U_{\parallel,e}/v_{te}=-0.25$.  The factor of the reduced parallel correlation $C_{E_\parallel,e}(v_\parallel)$ that depends on the distribution function, $+e v_\parallel^2/2 (\partial \delta f_e/\partial v_\parallel)$, where we have substituted the electron charge $q_e=-e$, is plotted in \figref{fig:model_xpoint}(c).  This form of the reduced parallel correlation shows excellent agreement with the analysis of the simulations in the lower panels of  \figref{fig:xpoint}(b) and (c), where there is a small loss of phase-space energy density at $|v_{\parallel}|/v_{te} \le 1/\sqrt{2}$, and the bulk of the increase in the phase-space energy density occurs in the range $1 \lesssim |v_\parallel|/v_{te} \lesssim 2$.  Thus, the velocity-space signature of the electron energization at the x-point is simply due to the bulk acceleration of the electrons in the $-z$ direction by the parallel electric field $E_\parallel$, where the parallel electron flow $U_{\parallel,e}$ supports the parallel current arising due to the odd (in $v_\parallel$) perturbation of $\delta f_e(v_\parallel)$, with a form well modeled by \eqnref{eq:deltafe}.  Thus, this velocity-space signature represents the bulk acceleration of the electrons in the out-of-plane direction by $E_\parallel$, with a net electron energization rate at the x-point simply given by  $j_{\parallel,e} E_{\parallel} = \int dv_\parallel C_{E_\parallel,e}(v_\parallel)$.

In summary, for collisionless magnetic reconnection in the strong-guide-field limit, the electron energization at the x-point is dominated by bulk acceleration of the electrons by $E_\parallel$. The particular form of the distribution function arising in the simulation yields a positive rate of electron energization with a signature that is symmetric about $v_\parallel=0$ and that peaks over the velocity range $1 \lesssim |v_\parallel|/v_{te} \lesssim 2$.  Note that this is \emph{not} a resonant acceleration of the electrons, as would be expected for collisionless damping via the Landau resonance, but instead it is a bulk acceleration of the electrons, in agreement with previous findings on electron energization by $E_\parallel$ in the strong-guide-field limit of collisionless magnetic reconnection
\cite{Egedal:2012,Dahlin:2014,Dahlin:2015,Dahlin:2016}.

\subsection{Energization in Exhaust}
\begin{figure*}
    \centering
    \setlength{\unitlength}{1in}
    \begin{picture}(5.0,2.5)
       \put(0.1,0.1){\includegraphics[width=5in]{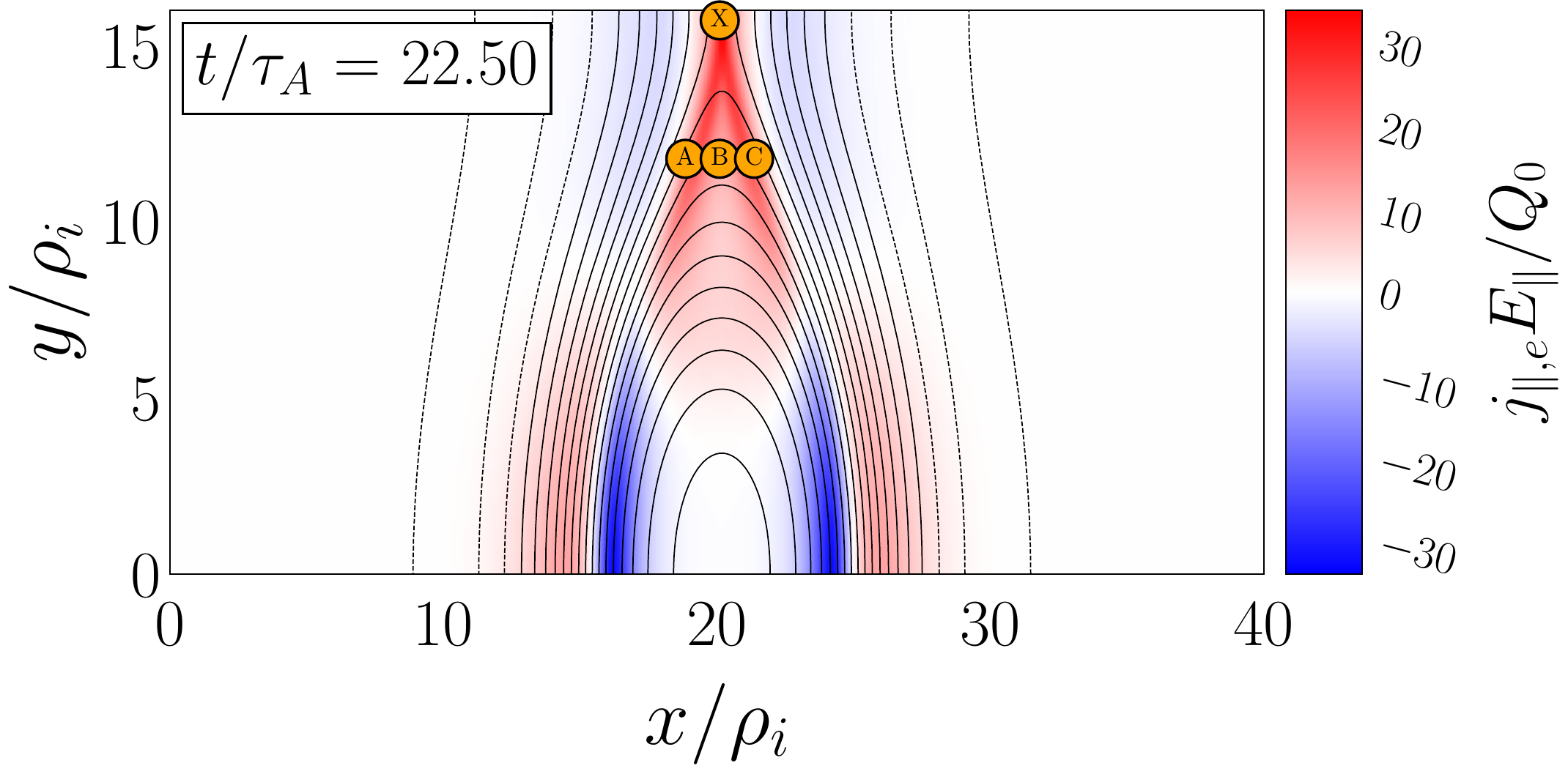}}
       \put(0.1,2.4){(a)}
    \end{picture} \\
    \begin{picture}(2.2,1.5)
       \put(0,0){\includegraphics[width=2.2in]{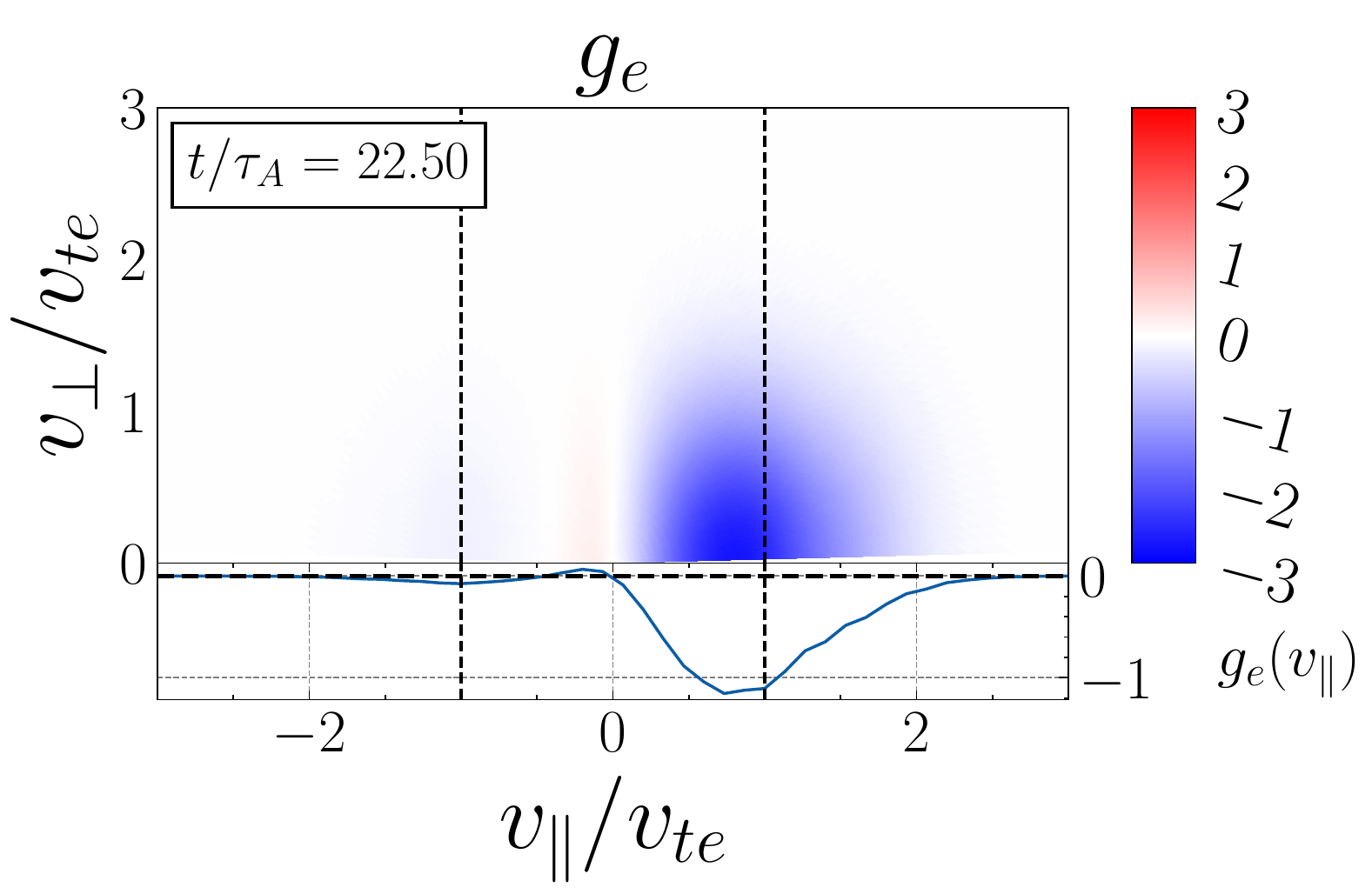}}
       \put(0.1,1.4){(b)}
       \put(0.9,1.5){\large \textcircled{\small \raisebox{-0.9pt}{A}}}
    \end{picture}
    \begin{picture}(2.2,1.5)
       \put(0,0){\includegraphics[width=2.2in]{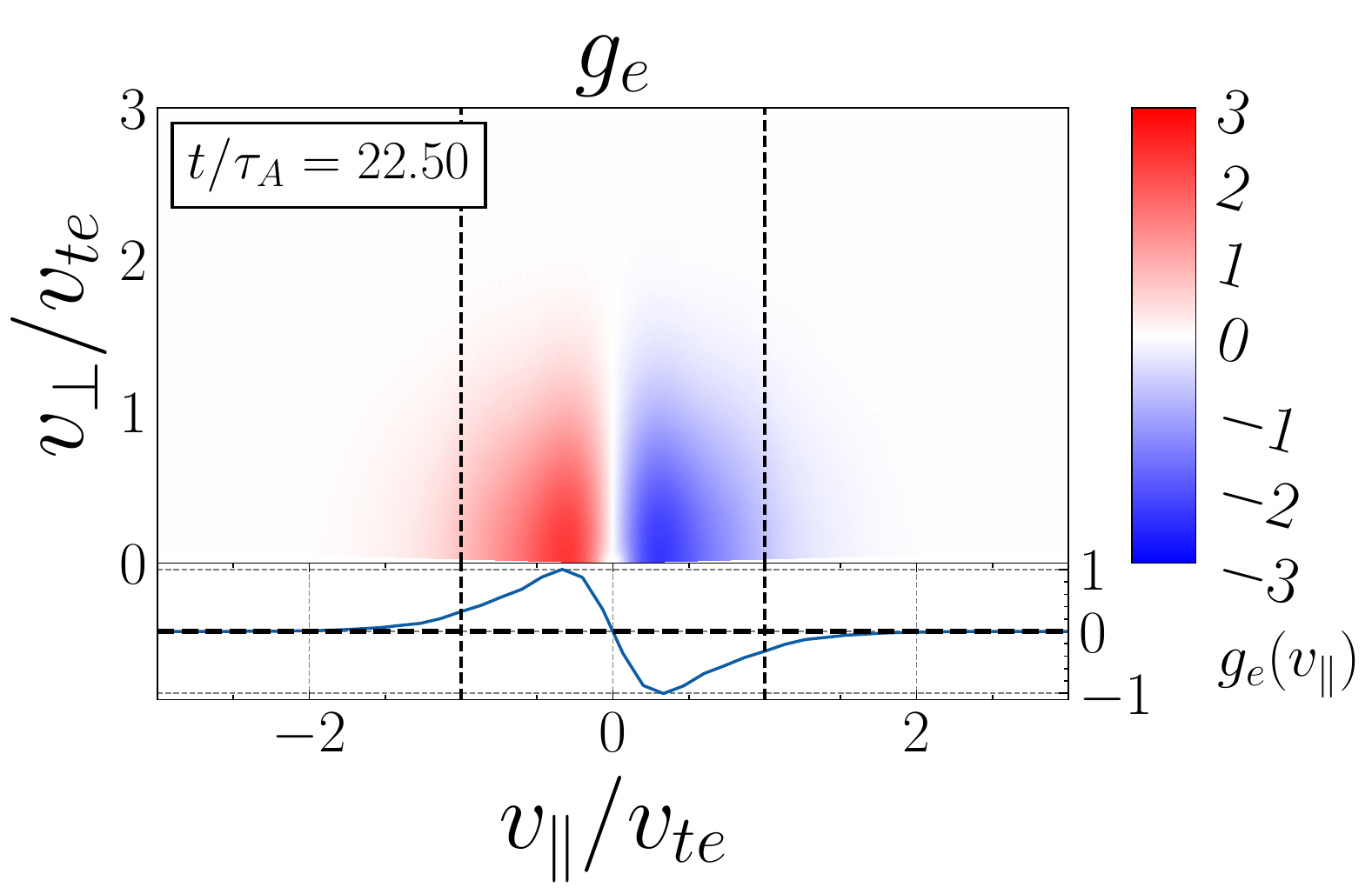}}
       \put(0.1,1.4){(c)}
       \put(0.9,1.5){\large \textcircled{\small \raisebox{-0.9pt}{B}}}
    \end{picture}
    \begin{picture}(2.2,1.5)
       \put(0,0){\includegraphics[width=2.2in]{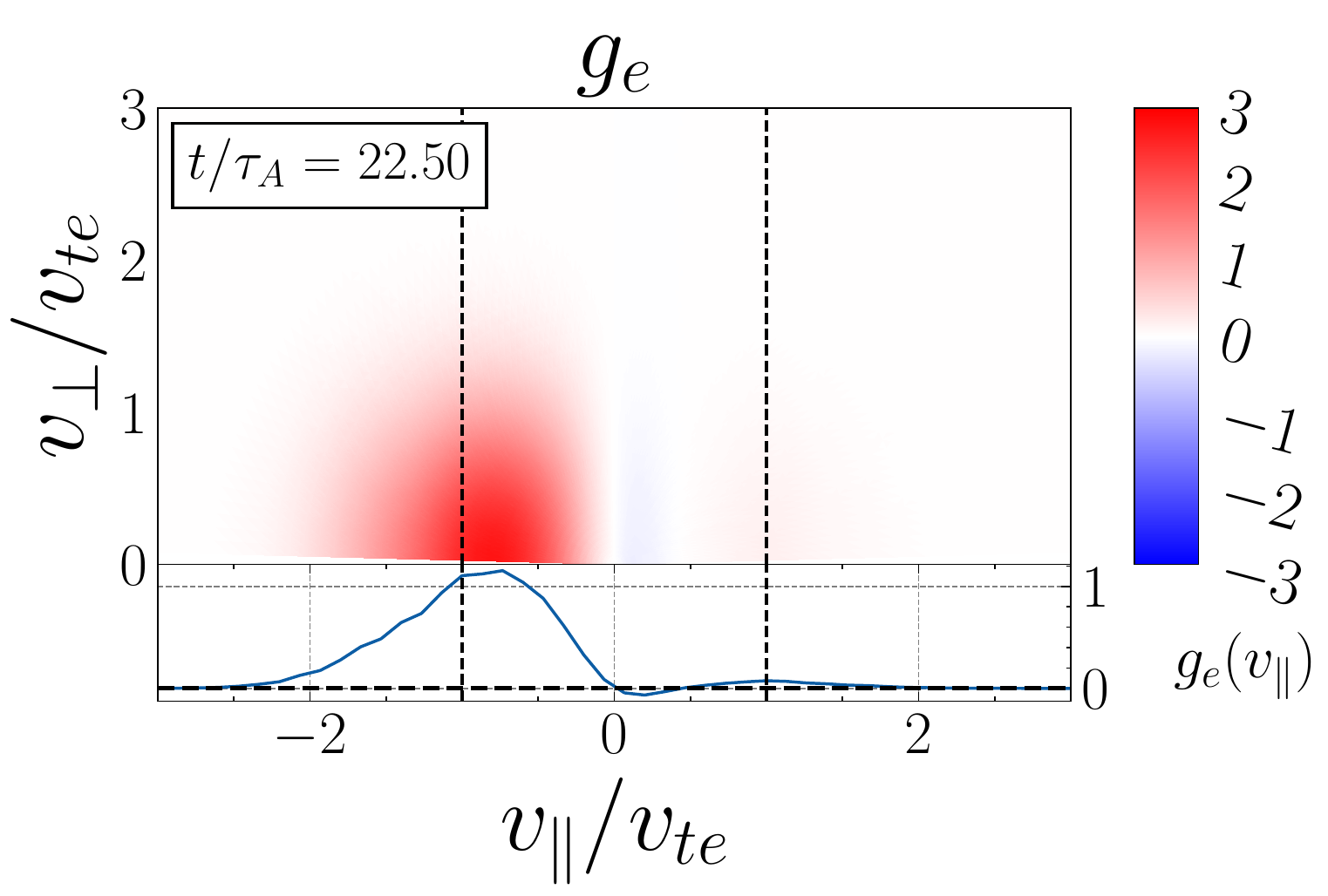}}
       \put(0.1,1.4){(d)}
       \put(0.9,1.5){\large \textcircled{\small \raisebox{-0.9pt}{C}}}
    \end{picture} \\
    \begin{picture}(2.2,1.6)
       \put(0,0){\includegraphics[width=2.2in]{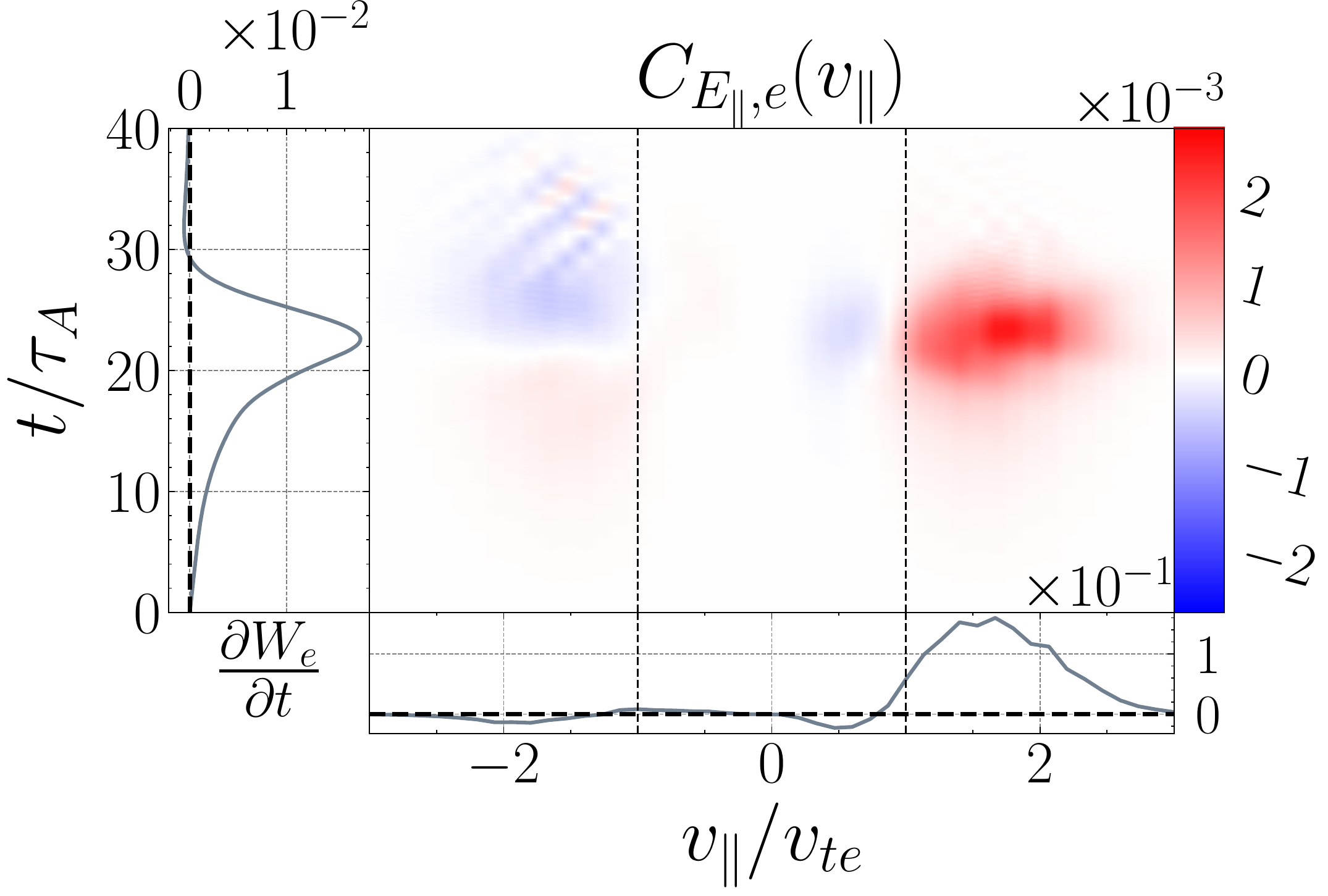}}
       \put(0.1,1.5){(e)}
    \end{picture}
    \begin{picture}(2.2,1.6)
       \put(0,0){\includegraphics[width=2.2in]{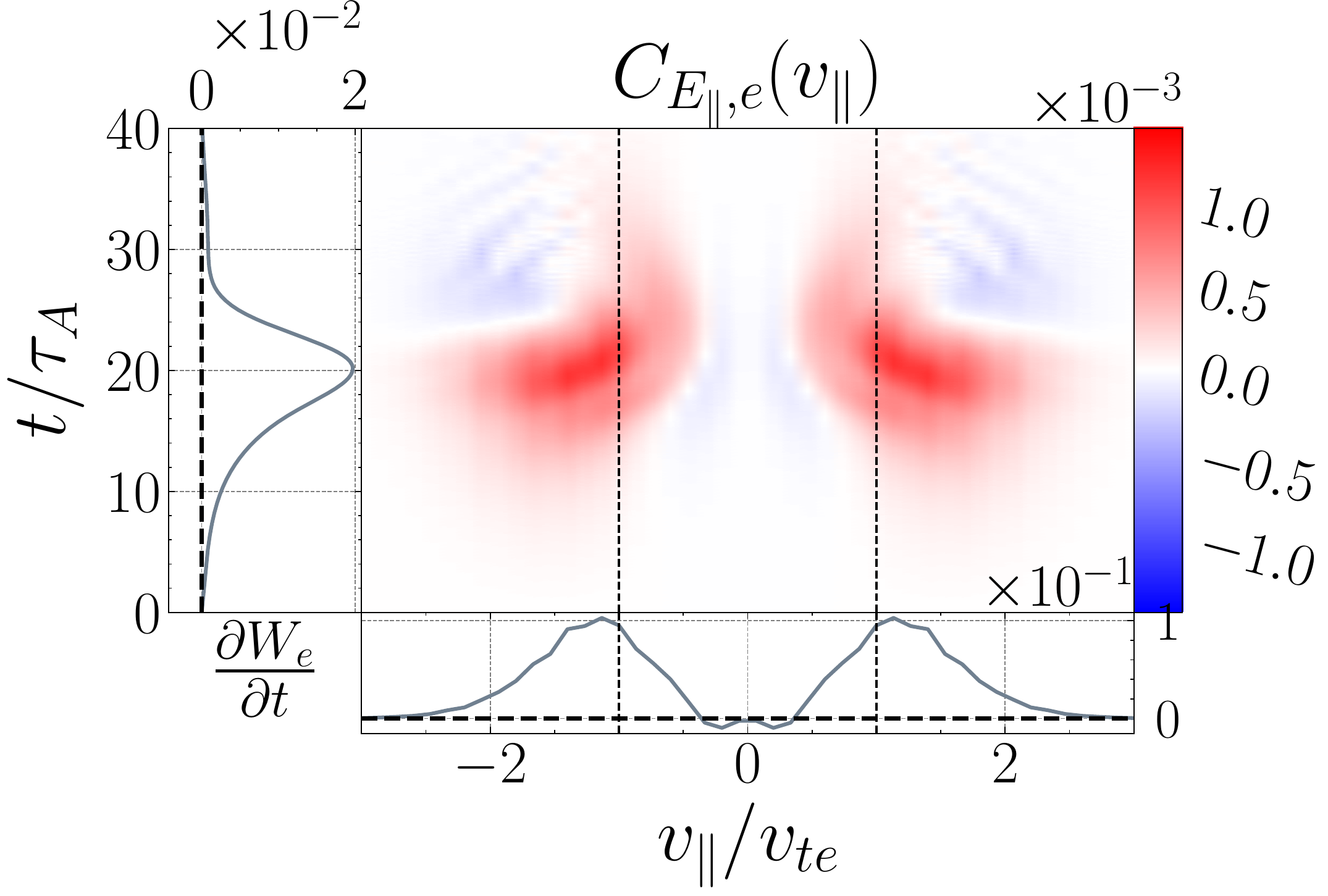}}
       \put(0.1,1.5){(f)}
    \end{picture}
    \begin{picture}(2.2,1.6)
       \put(0,0){\includegraphics[width=2.2in]{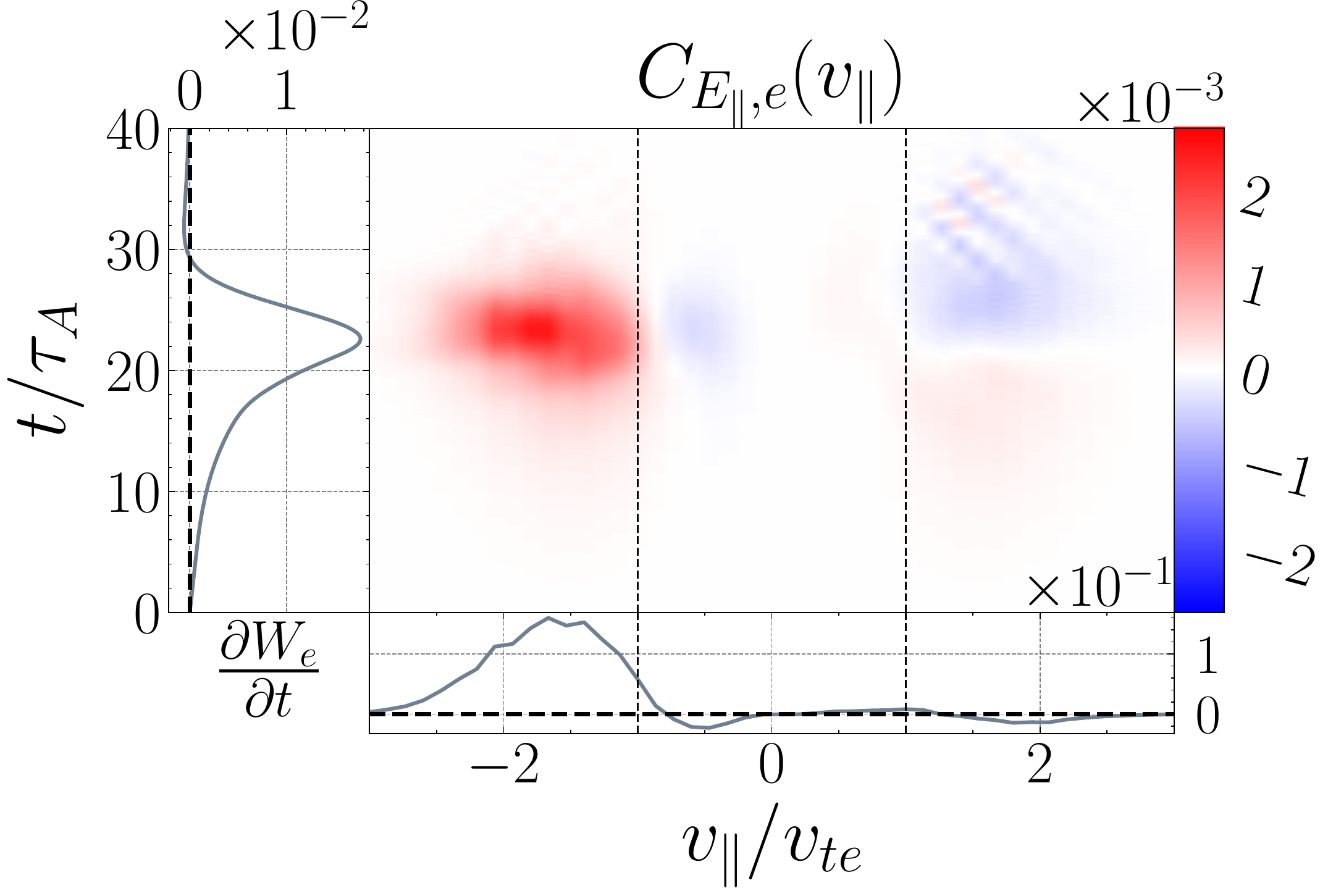}}
       \put(0.1,1.5){(g)}
    \end{picture}
    \caption[Velocity space signatures in the reconnection exhaust]{(a): Lower half of $\beta_i = 0.01$ simulation of net electromagnetic work with selected diagnostic probe positions given by A, B, C located at $\mathbf{r}/\rho_i = (x/\rho_i, y/\rho_i)$ where $\mathbf{r_A}/\rho_i = (18.9, 11.8)$, $\mathbf{r_B}/\rho_i = (20.1, 11.8)$, $\mathbf{r_C}/\rho_i = (21.4, 11.8)$ at the time of the peak energization rate $t/\tau_A=22.5$. Middle Row: Complementary perturbed gyrokinetic distribution function $g_e(v_{\parallel},v_{\perp})$ at the same time and at each diagnostic probe position in the exhaust point A: (b), point B: (c), point C: (d), with $g_e(v_\parallel)$ shown in the lower panel of each. Bottom Row: Timestack plots of the reduced parallel correlation over time $C_{E_\parallel,e}(v_\parallel,t)$, with the net electron energization rate vs.~time in the left panel and the time-integrated reduced parallel energization rate $\int C_{E_\parallel,e}(v_\parallel,t) dt$  over the full interval shown in the lower panel, at $\mathbf{r_A}$ (e), $\mathbf{r_B}$ (f), $\mathbf{r_C}$ (g). The vertical dashed black lines again indicate the \Alfven velocity, $\pm v_{A,z}/v_{te}$.}
    \label{fig:exhaust_composite}
\end{figure*}
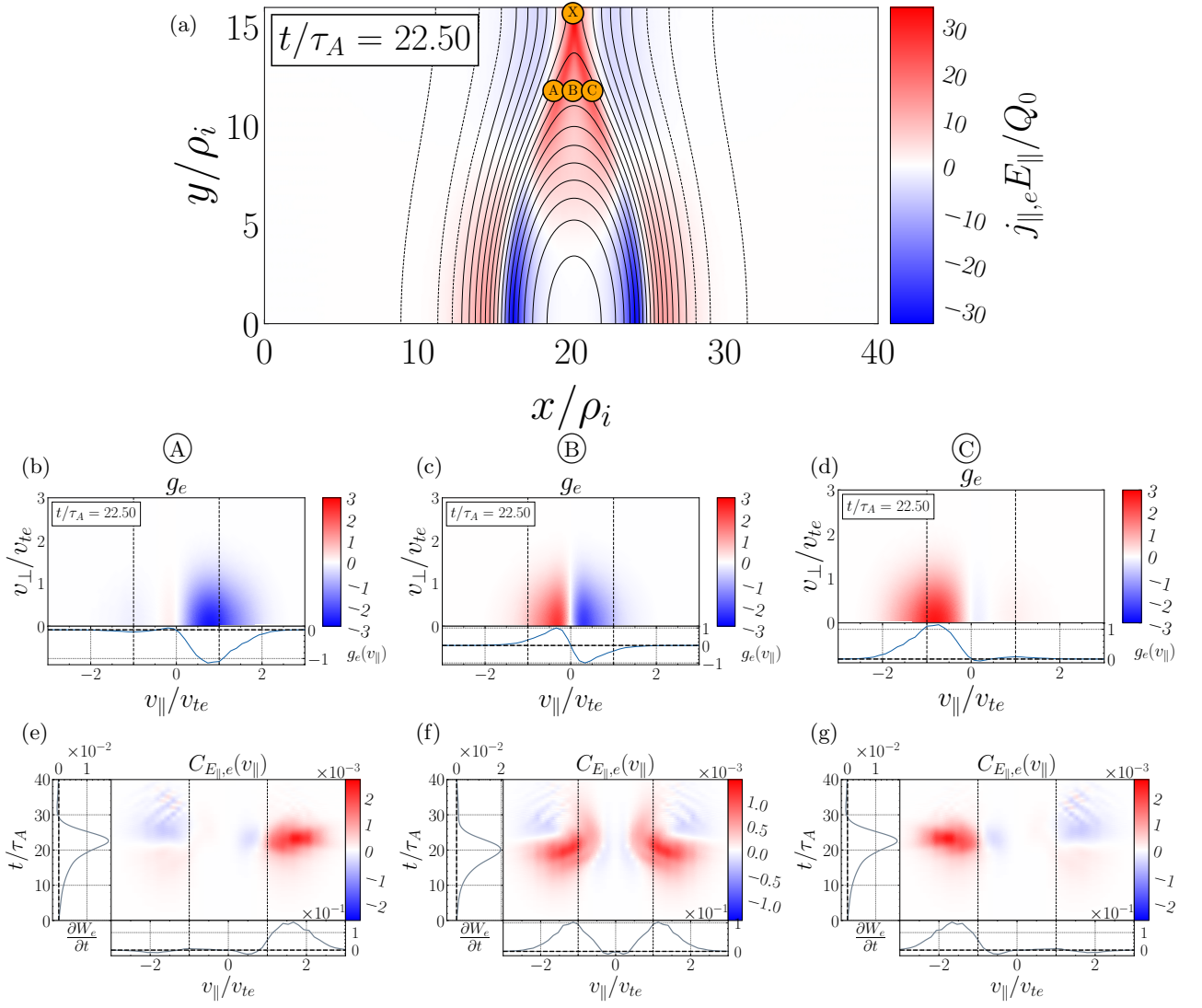

In the same manner as for the x-point, we now select three points located at $\mathbf{r}/\rho_i = (x/\rho_i, y/\rho_i)$, where $\mathbf{r_A}/\rho_i = (18.9, 11.8)$, $\mathbf{r_B}/\rho_i = (20.1, 11.8)$, $\mathbf{r_C}/\rho_i = (21.4, 11.8)$ to investigate the particle energization within the exhaust. As before, we begin with analyzing the perturbations to the electron velocity distribution function and electric field at the time $t/\tau_A=22.5$. In \figref{fig:exhaust_composite}(a) we show the total spatial electron energization rate $j_{\parallel,e} E_{\parallel}$ again for reference. The points $\mathbf{r_A}$ and $\mathbf{r_C}$, on either side of the midplane, are located just inside of the separatrix boundary at time $t/\tau_A=22.5$. Each column in the lower two rows of \figref{fig:exhaust_composite} corresponds to labels A, B, and C in the exhaust of panel (a) from left to right, respectively. 

At point $\mathbf{r_B}$, directly downstream from the x-point, along the midplane, we see a qualitatively similar perturbed complementary electron velocity distribution function to that at the x-point, shown in \figref{fig:exhaust_composite}(c), with the reduced parallel perturbed distribution shown in the lower panel. The odd perturbation in $v_{\parallel}$ is more confined in parallel velocity than at the x-point, where the bulk of the perturbed electron VDF is contained at $|v_{\parallel}/v_{te}| < 1$. Contrasting this point on the midplane in \figref{fig:exhaust_composite}(c) and \figref{fig:exhaust_composite}(f) to the points near the separatrix on either side, there is an asymmetric signature of $g_e(v_\parallel,v_\perp)$. The signatures are mirrored across the midplane in magnitude, but are oppositely signed such that $g_{e,A} (v_{\parallel}, v_{\perp}) = -g_{e,C} (-v_{\parallel}, v_{\perp})$. This symmetry means that the first moment of the distribution yields an identical value of the parallel current at both points. The corresponding reduced distribution function $g_e(v_\parallel)$ is shown as before in the lower panels of (b), (c), and (d). 

To analyze the particle energization within the exhaust, we plot the reduced field-particle correlation $C_{E_\parallel}(v_\parallel,t)$ timestack shown in the bottom row of \figref{fig:exhaust_composite}. In the left vertical panel for each point in the exhaust, we see the net energization rate $\partial W_e(t)/\partial t$ peaks at a similar time of $t/\tau_A=22.5$ for either side of the midplane (e) and (g). At the midplane in the exhaust (f), the net energization rate peaks slightly earlier at $t/\tau_A \approx 20.0$. In the lower horizontal panel for each point, we again plot the time-integrated reduced parallel energization rate over the full interval. At the midplane in the exhaust, we again find a symmetric (about $v_{\parallel}$) increase in phase-space energy density $w_e$ of the electrons. The energization extends over a range that is slightly narrower than at the x-point, $0.5 \lesssim |v_{\parallel}|/v_{te} \lesssim 2.0$, and is centered close to $|v_{\parallel}|/v_{te} = 1$. On either side of the midplane, there is an asymmetric increase in phase-space energy density over the velocity range $0.7 \lesssim |v_{\parallel}|/v_{te} \lesssim 3$. At point $\mathbf{r_A}$, the electrons travelling in the parallel direction $v_\parallel > 0$ are preferentially accelerated, gaining phase-space energy density. At point $\mathbf{r_C}$, on the other hand, the electrons travelling in the anti-parallel direction are preferentially gaining phase-space energy density. On either side of the midplane, there is a clear cutoff in the dominant energization signature at $|v_{\parallel}|/v_{te} = 1$. Panels (e), (f), and (g) in \figref{fig:exhaust_composite} show the characteristic velocity-space signatures of electron energization in the exhaust region of strong-guide-field magnetic reconnection, a key result of this investigation. The parallel velocity ranges for the electron energization are apparent in the lower panels of (e), (f) and (g), showing the time-integrated reduced correlation $\int dt C_{E_\parallel}(v_\parallel,t)$. 

\begin{figure*}
   \centering
   \setlength{\unitlength}{1in}
   \begin{picture}(2.2,2.8)
       \put(0,0){ \includegraphics[height=2.65in]{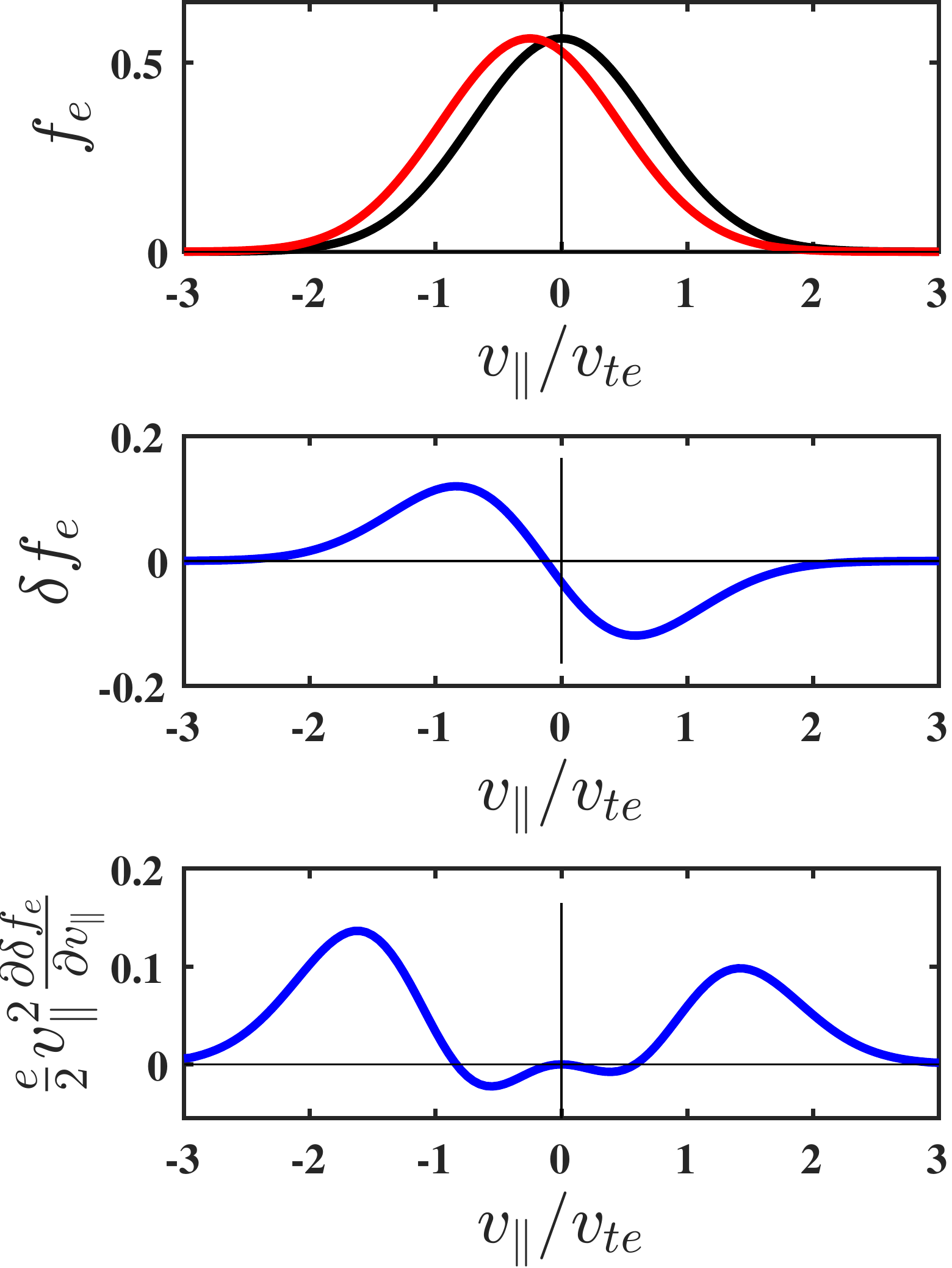}}
       \put(0.2,2.7){(a)}
    \end{picture}
     \begin{picture}(2.2,2.8)
       \put(0,0){ \includegraphics[height=2.65in]{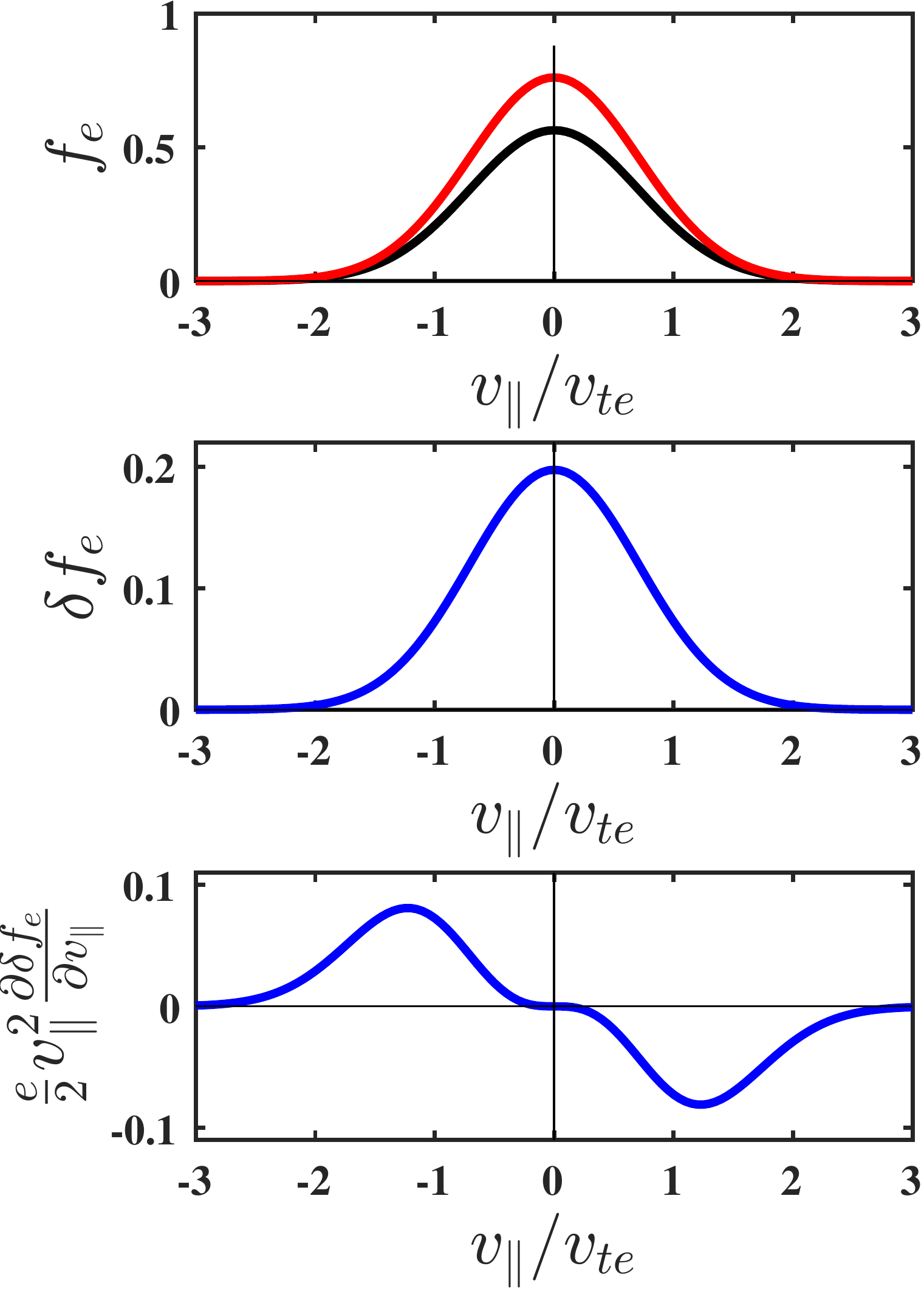}}
       \put(0.2,2.7){(b)}
    \end{picture}
     \begin{picture}(2.2,2.8)
       \put(0,0){\includegraphics[height=2.65in]{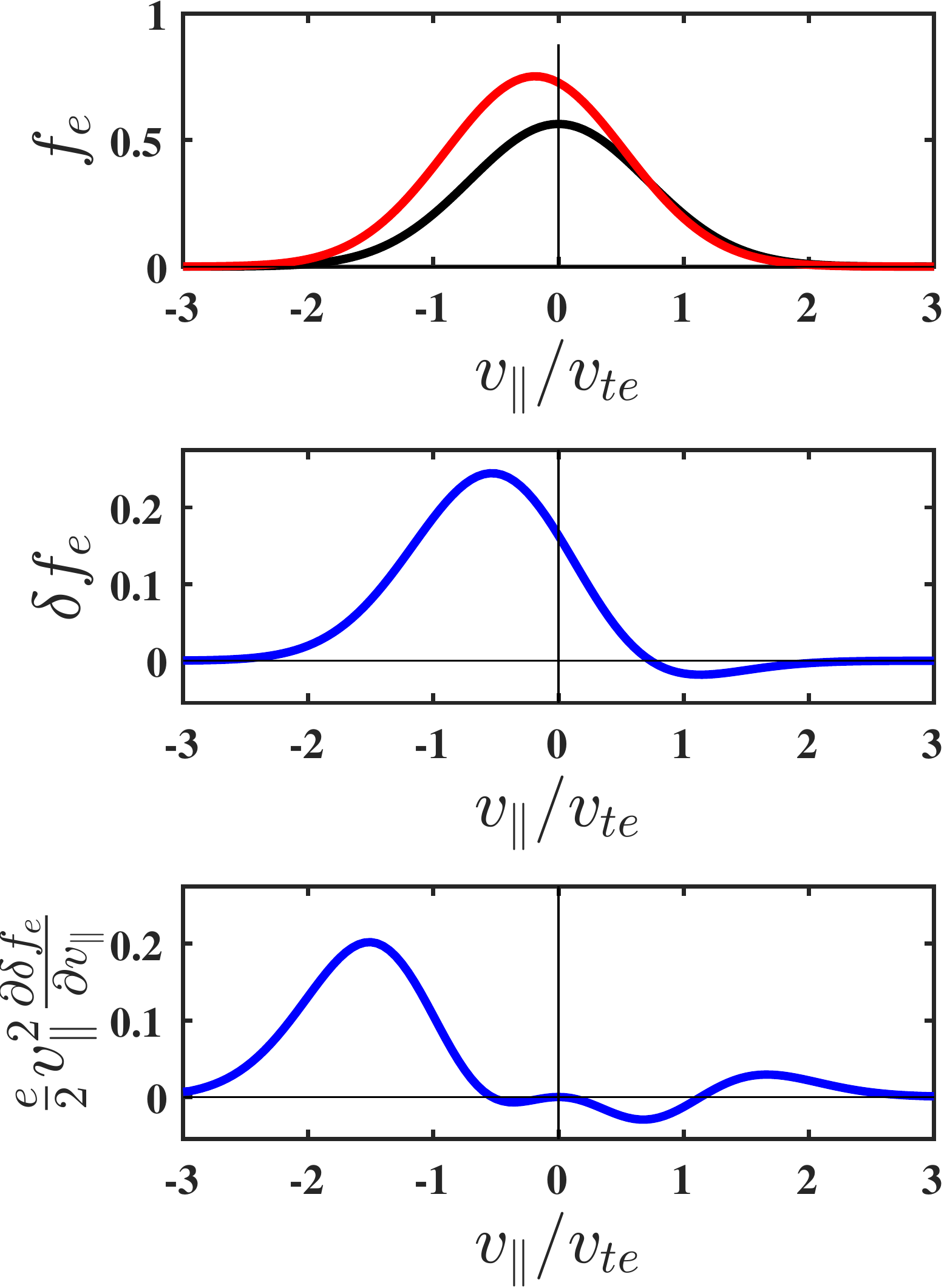}}
       \put(0.2,2.7){(c)}
    \end{picture}
   \caption[Simple model of modified distribution function changing field-particle correlation signature]{ Simple model of (a) shifted Maxwellian $U_{\parallel,e}/v_{te} = -0.25$ (b) density perturbation $\delta n_e/n_{0e}=+0.35$ (c) combined shifted Maxwellian with a density perturbation.  Top row: Comparison of the total distribution function (red) with equilibrium Maxwellian distribution function (black). Middle row: Perturbed distribution function. Lower row: Corresponding phase-space energy density signature of the perturbed distribution function. The signature here is similar qualitatively to that of the reduced correlation for point $\mathbf{r_C}$.}
    \label{fig:model_exhaust}
\end{figure*}
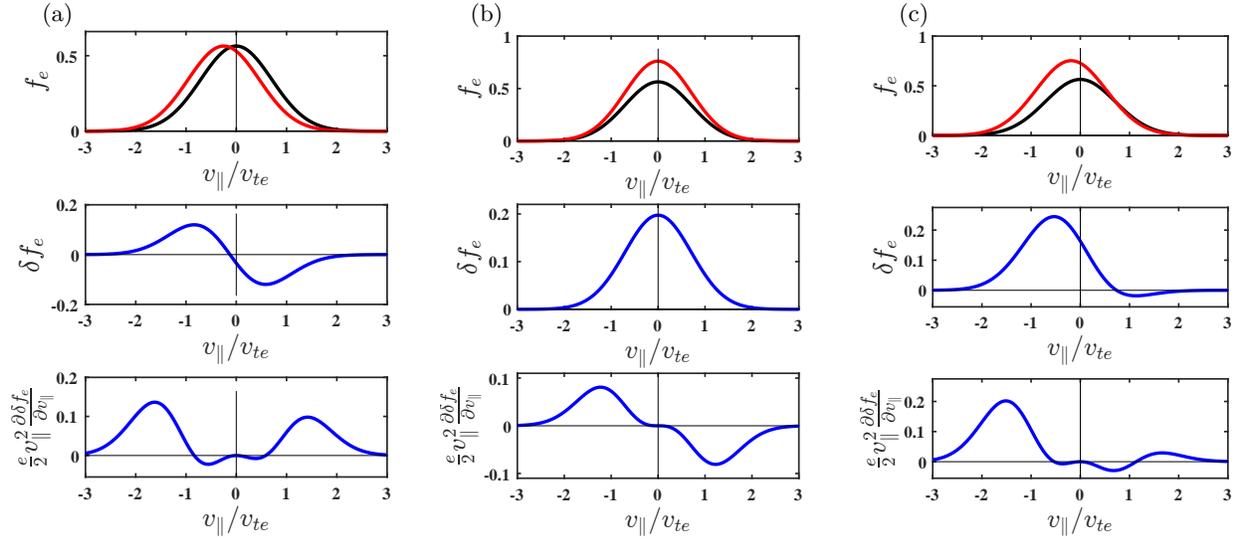

The asymmetry in the perturbed distribution function and reduced parallel energization rate on either side of the midplane within the exhaust is at first surprising, given the symmetric signature of $j_{\parallel,e} E_{\parallel}$ both in magnitude and sign. We hypothesize here that the asymmetric velocity-space signatures in the exhaust region, shown in \figref{fig:model_exhaust}(e) and (g), can be explained by the combination of a parallel electron flow with an electron density perturbation. If we look at the density perturbation in the $(x,y)$ plane, shown in \figref{fig:exhaust_trajectory}(a) of Appendix \ref{sec:exhaust_trajectory}, we see in the lower half of the simulation plane there is a decrease in electron density to the left of the midplane and an increase in density to the right of the midplane along each lower separatrix arm. A similar quadrupolar density pattern is well-known from previous Hall-MHD and two-fluid simulations. \cite{Uzdensky:2006, Birn:2007}  The perturbed electron velocity distributions shown in \figref{fig:exhaust_composite}(b) and (d) are consistent with the density perturbations shown in \figref{fig:exhaust_trajectory}(a). Since the velocity-space signature of energization depends on the details of the electron velocity distribution, it is expected that this density perturbation will influence the form of the observed velocity-space signature.

To demonstrate that a parallel electron flow combined with a density perturbation can indeed generate the asymmetric velocity-space signatures of electron energization seen in \figref{fig:exhaust_composite}(e) and (g), we present a simple model with either (i) a shifted Maxwellian distribution for a bulk parallel electron flow $U_{\parallel,e}$, given by 
\begin{equation}
f_e (v_\parallel) = \frac{n_{0e}}{\pi^{1/2} v_{te}} e^{-(v_\parallel- U_{\parallel,e})^2/v_{te}^2},
\label{eq:shifted}
\end{equation}
or (ii) an electron density perturbation $\delta n_e$
\begin{equation}
f_e (v_\parallel) = \frac{(n_{0e}+ \delta n_e)}{\pi^{1/2} v_{te}} e^{-v_\parallel^2/v_{te}^2},
\label{eq:delta_ne}
\end{equation}
or (iii) a linear combination of the deviations from a  Maxwellian distribution for both a shifted Maxwellian with flow $U_{\parallel,e}$ and an electron density perturbation $\delta n_e$. We emphasize that the precise quantitative form in velocity space of the net parallel flow and density perturbation is not critical; what is important is the general concept that the sum of a parallel flow with a density perturbation qualitatively leads to asymmetric signatures as seen in \figref{fig:exhaust_composite}(g).  For each simple model, we plot a column in \figref{fig:model_exhaust} with the equilibrium parallel electron velocity distribution (black) and total parallel electron velocity distribution (red) in the top row, the perturbed velocity distribution in the middle row, and the form of the velocity dependence $+e v_\parallel^2/2 (\partial \delta f_e/\partial v_\parallel)$ of the field-particle correlation $C_{E_\parallel}(v_\parallel)$ in the bottom row. 

In the left column (a), we plot the case for a shifted Maxwellian with bulk parallel electron flow $U_{\parallel,e}/v_{te} = -0.25$, showing that it results in an electron energization signature that is asymmetric in $v_\parallel$, with a larger signature of energization in the same direction as the bulk parallel flow. In the middle column (b), we plot the case for a density perturbation with $\delta n_e/n_{0e}=+0.35$, showing this density perturbation alone leads to an energization signature that is odd in $v_\parallel$, meaning there is zero net energization of the electrons by the parallel electric field due to a density perturbation when integrated over velocity. In the right column (c), we plot the case with a superposition of the shifted Maxwellian with $U_{\parallel,e}/v_{te} = -0.25$ and the density perturbation with  $\delta n_e/n_{0e}=+0.35$. The combination of the flow and density perturbations leads to an energization signature as a function of $v_\parallel$ that is qualitatively similar to that seen at point $\V{r}_C$ in \figref{fig:exhaust_composite}(g).  

Although the detailed perturbed electron velocity distributions arising through the evolution of the simulation show modest quantitative differences from the forms used in this simple model, this example demonstrates that the combination of a parallel flow and a density perturbation can indeed lead to the asymmetric signatures of electron energization in the exhaust region of collisionless magnetic reconnection in the strong-guide-field limit seen in \figref{fig:exhaust_composite}. 

In summary, the electron energization in the exhaust region of collisionless magnetic reconnection in the strong-guide-field limit is caused by a bulk acceleration of the electrons by the parallel component of the electric field $E_\parallel$.  Although, along the separatrices away from the midplane in the exhaust, we find an asymmetric signature of electron energization, since a density perturbation leads to zero net electron energization when integrated over $v_\parallel$ (as seen in the lower middle panel of \figref{fig:exhaust_composite}), the net electron energization is simply due to this bulk acceleration of the out-of-plane electron flow by the reconnection electric field.  This asymmetric signature about $v_{\parallel}=0$ is indicative of the spatial location in the exhaust where the diagnostic is sampling velocity space distributions. On each side of the midplane, we see equal magnitudes of electron energization (increase in phase-space energy density). The asymmetry in the signature of electron energization motivates the possibility to identify observationally, the physics of electron energization by collisionless magnetic reconnection, using only single-point measurements of the electromagnetic fields and electron velocity distributions. This simulation demonstrates a characteristic velocity-space signature for electron acceleration through bulk parallel acceleration, by the reconnection electric field in the exhaust of collisionless magnetic reconnection in the strong-guide-field limit. Further evidence that this energization is \emph{not} a resonant acceleration of electrons, is provided by investigating the variation of the electron energization signatures with the plasma beta parameter.

\subsection{Variation of Energization with Plasma $\beta_i$}
The general qualitative picture of the energization rate for the remaining simulations is consistent as $\beta_i$ increases from $0.03 \le \beta_i \le 1.0$. The energization of electrons for all five simulations is largely dominated in the electron diffusion region (EDR) around the x-point and into the exhaust along the separatrices as in \figref{fig:coordinate2d_panel}(c). The overall reconnection geometry persists as $\beta_i$ varies from 0.01 to 1, with similar energization signatures. However, there are some differences in the dynamic evolution of the reconnecting field. In the $\beta_i = 1$ case, there is a clear development of a secondary island at the original x-point. This secondary island is formed as a consequence of the plasmoid instability. \cite{Loureiro:2007} As noted by \citet{Numata:2015}, this secondary island eventually moves in the -y direction due to numerical noise, and secondary reconnection commences, which allows renewed particle energization and plasma heating late in the simulation at a lower magnitude. 

\begin{figure*}   
   \setlength{\unitlength}{1in}
   \begin{picture}(3.3,3.0)
       \put(0,0){\includegraphics[width=3.3in]{figs/rxnb001a_JZ_S2_nc001_t0180.pdf}}
       \put(0.2,2.9){(a)}
       \put(1.3,2.9){$\beta_i = 0.01$}
   \end{picture}
   \begin{picture}(3.3,3.0)
       \put(0,0){\includegraphics[width=3.3in]{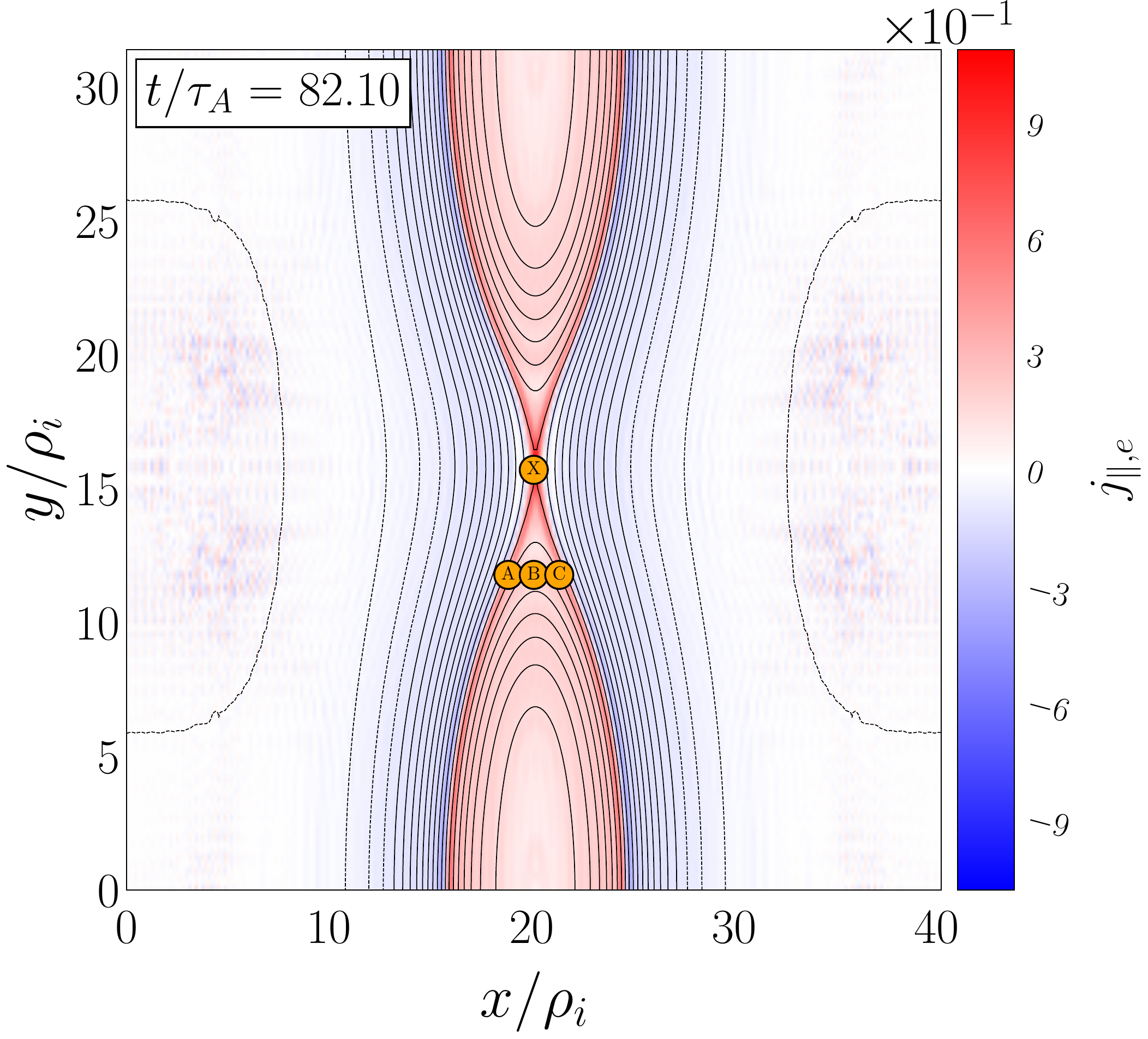}}
       \put(0.2,2.9){(b)}
       \put(1.3,2.9){$\beta_i = 1$}
   \end{picture}
\caption[Current Sheet Thickness Comparison]{\label{fig:current_sheet_comparison} Snapshot of parallel current at maximum reconnection rate which occurs (a) at $t/\tau_A = 22.5$ for $\beta_i=0.01$, and (b) at $t/\tau_A = 82.10$ for $\beta_i = 1$, illustrating a thinning of the current sheet along the separatrices at higher $\beta_i$.}
\end{figure*}

As $\beta_i$ increases, we see a thinning of the current sheet supporting the reconnection process, as shown in \figref{fig:current_sheet_comparison}(b) for $\beta_i = 1$. The opening angle of the exhaust decreases slightly with higher $\beta_i$, and subsequently, the separatrix arms move closer to the midplane. In addition to these qualitative changes to the reconnection geometry and associated current sheets, the magnitude of both the current sheet and self-consistent $E_\parallel$ decrease in magnitude with increased $\beta_i$.

\begin{figure}   
    \includegraphics[width=3.3in]{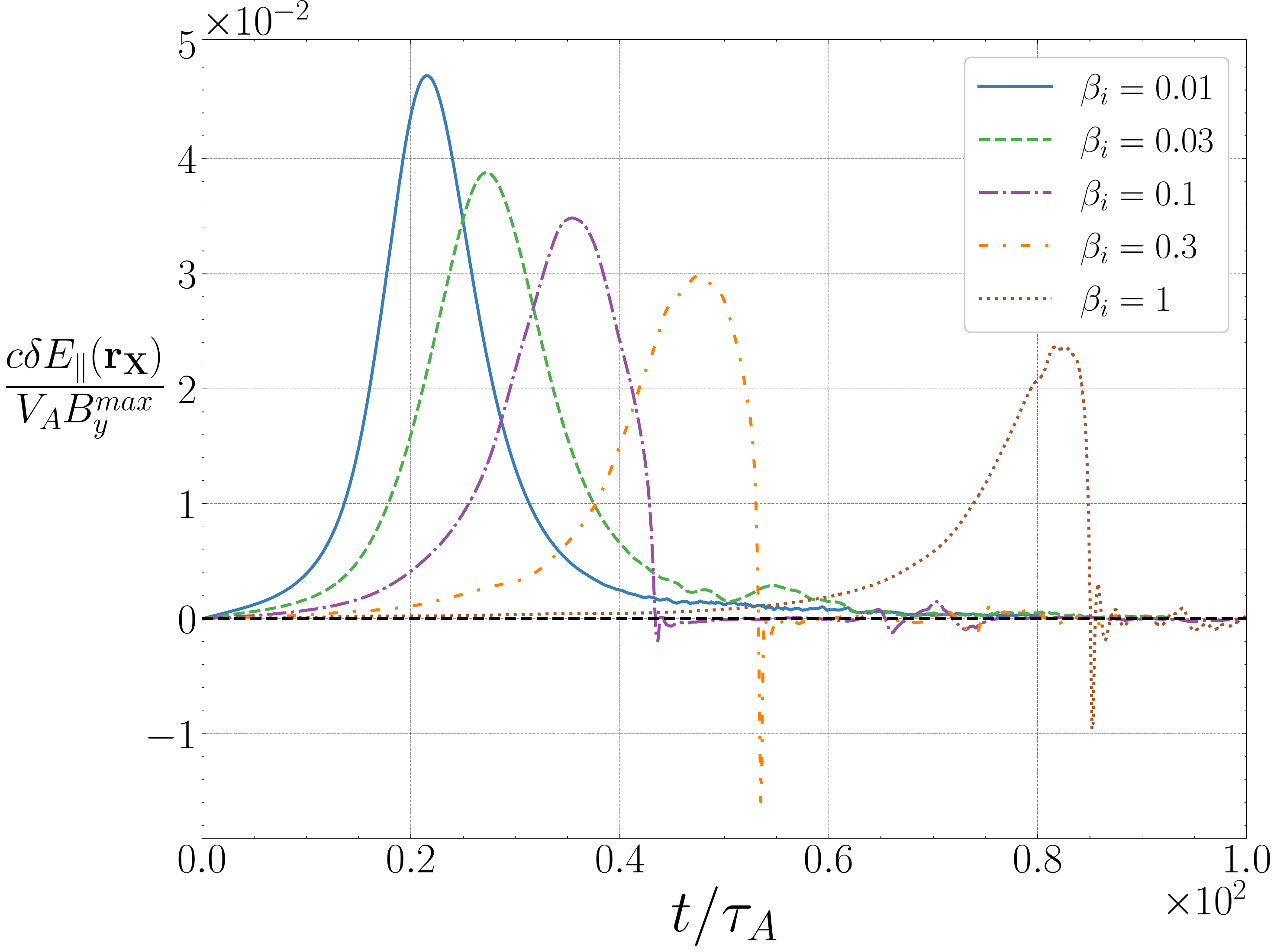}
    \caption[Reconnection rates for all simulations]{\label{fig:reconnection_rate_all} Normalized reconnection rate for the five simulations using the out-of-plane reconnecting electric field at the center of the simulation with $\beta_i = 0.01$ blue, $\beta_i = 0.03$ green, $\beta_i = 0.1$ yellow, $\beta_i = 0.3$ red, $\beta_i = 1$ purple. The reversal of the electric field observed for $\beta_i \ge 0.1$ is evidence of a conversion at the x-point to an O-point, where the current sheet becomes unstable due to the formation of a plasmoid. At this point in time, $E_\parallel$ ceases to represent the reconnection rate for the central x-point.}
\end{figure}

In \figref{fig:reconnection_rate_all}, we show the normalized reconnection rate $c E_\parallel(\V{r_X})/V_{A} B^{max}_y$  for all five $\beta_i$ cases. The time evolution of the out-of-plane electric field at the x-point $(x/\rho_i,y\rho_i) = (L_x/2, L_y/2)$ is used as the measure of the reconnection rate. The peak reconnection rate decreases in magnitude as $\beta_i$ increase and occurs later in time as the tearing instability develops more slowly at higher $\beta_i$. In the higher $\beta_i$ runs, there is a steep drop in the electric field just after the reconnection rate peaks, and the field eventually reverses sign at the x-point. This is a consequence of the formation of the plasmoid instability \cite{Loureiro:2007, Loureiro:2013a}, leading to  the conversion from an x-point to an o-point at the center of the reconnecting geometry. 

At the x-point, the energization signatures stay qualitatively consistent in shape as $\beta_i$ increases, with the energization abruptly ceasing when the plasmoid instability causes the x-point to convert to an o-point in the $\beta_i \ge 0.1$ simulations. The magnitude of energization for electrons decreases with increasing $\beta_i$, consistent with the lower magnitudes of $j_{\parallel,e}$ and $E_{\parallel}$ with increasing $\beta_i$, and also consistent with the decreasing conversion of energy as shown in \figref{fig:rxn_energy}.

\begin{figure*}
    \setlength{\unitlength}{1in}
    \begin{picture}(6.6,2.3)
        \put(0.05,0){\includegraphics[width=3.2in]{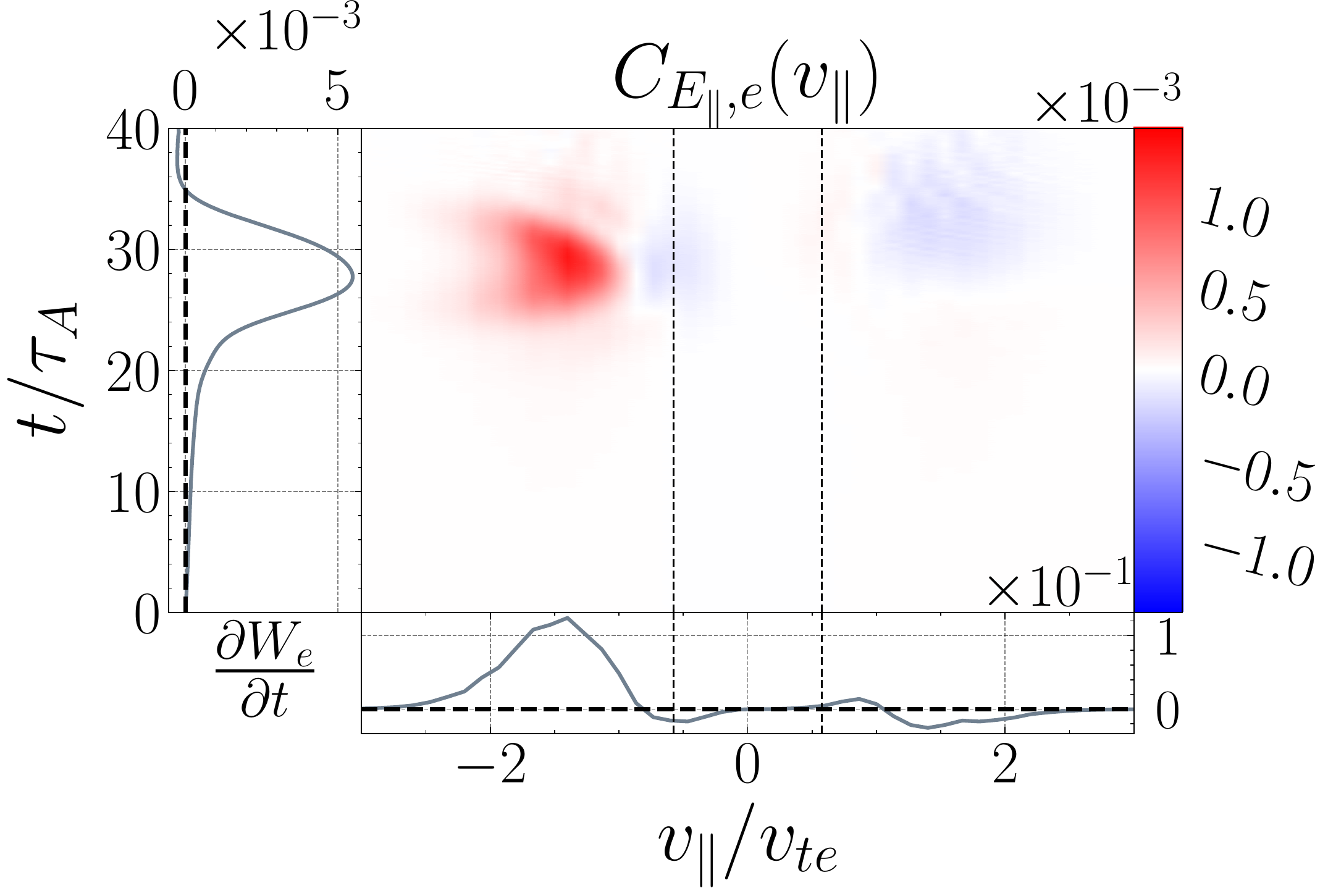}}
        \put(0.1,2.2){(a) $\beta_i = 0.03$}
        \put(3.3,0){\includegraphics[width=3.2in]{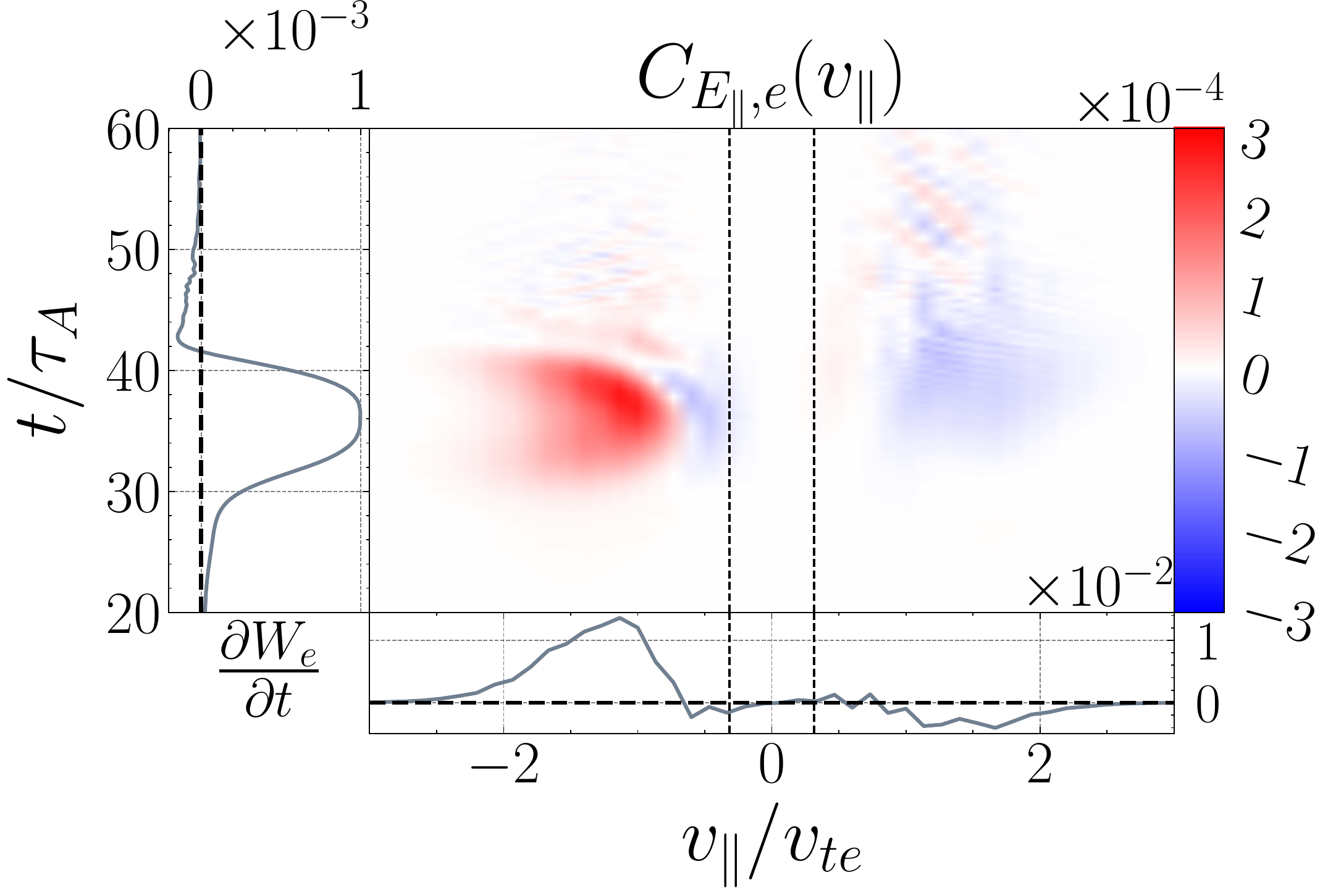}}
        \put(3.2,2.2){(b) $\beta_i = 0.1$}
    \end{picture}
    \\
    \begin{picture}(6.6,2.3)
        \put(0.05,0){\includegraphics[width=3.2in]{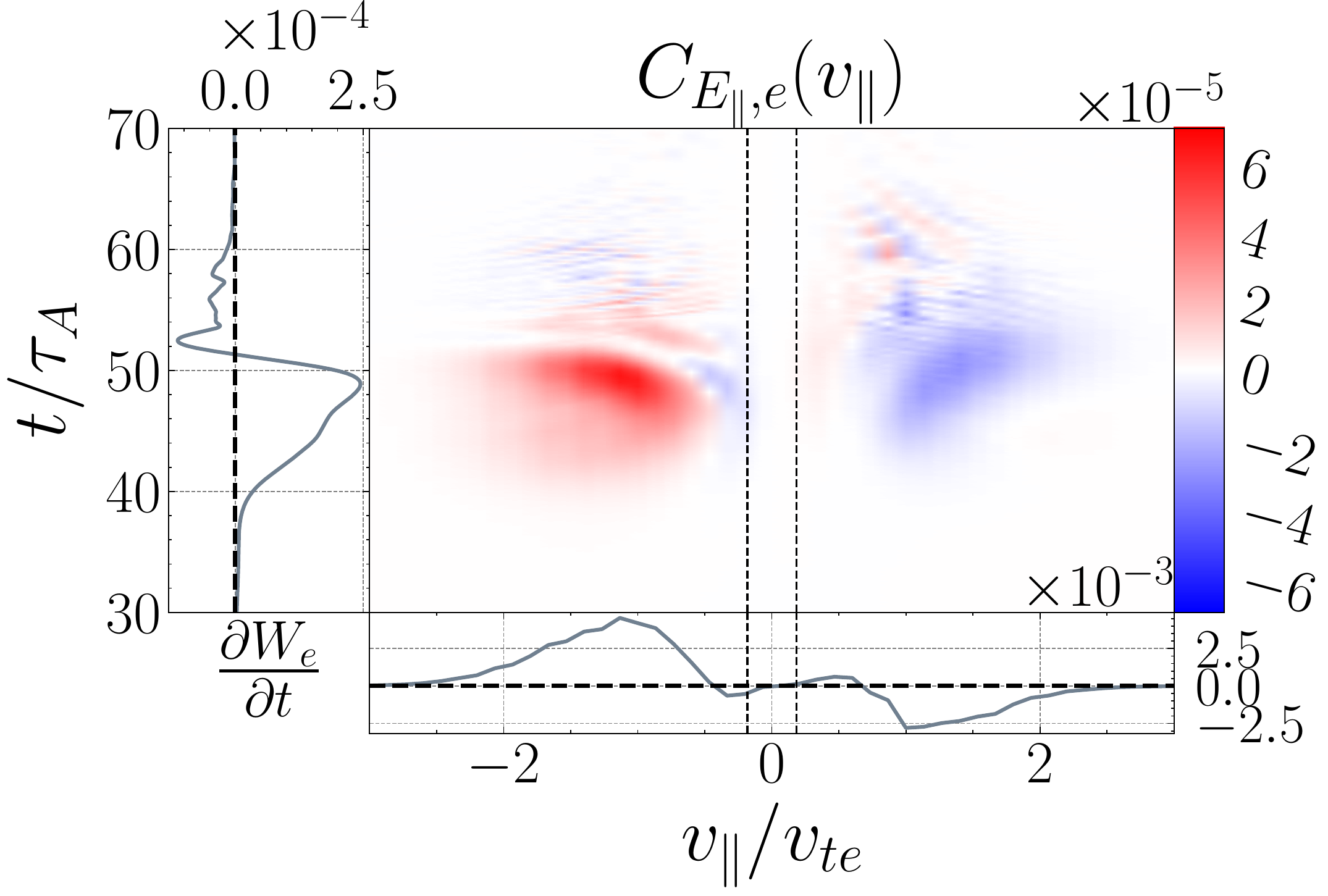}}
        \put(0.1,2.2){(c) $\beta_i = 0.3$}
        \put(3.3,0){\includegraphics[width=3.2in]{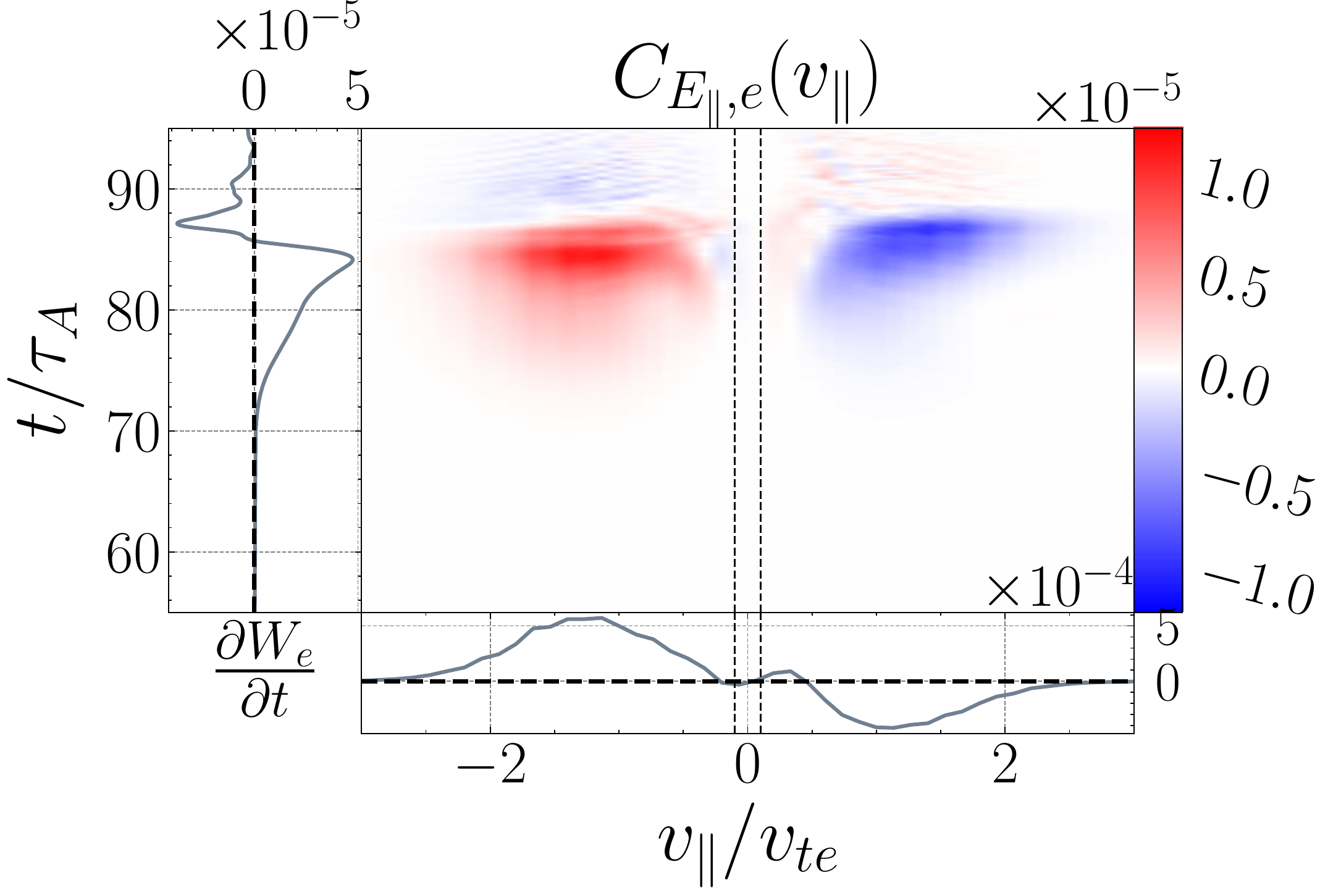}}
        \put(3.2,2.2){(d) $\beta_i = 1$}
    \end{picture}
    \caption[Timestack Plots for $\beta_e = 0.03,0.1,0.3,1.0$]{  \label{fig:exhaust_betas_timestack_panel}Timestack plots of $C_{E_{\parallel},e}(v_{\parallel},t)$ for point $\mathbf{r_C}$ (right of the mid-line)  in the exhaust for (a) $\beta=0.03$, (b) $\beta=0.1$, (c) $\beta=0.3$, (d) $\beta=1$. The higher $\beta_i$ simulations develop a loss of phase-space energy density to form a quadrupolar pattern in $v_\parallel$ of particle energization. The vertical black lines again indicate the \Alfven velocity, $\pm v_{A,z}/v_{te}$.}
\end{figure*}

In the exhaust, there is markedly more variation in the electron energization signatures as $\beta_i$ increases. We show the reduced correlation timestack plots in the exhaust for point $\mathbf{r_C}$ in \figref{fig:exhaust_betas_timestack_panel} for $\beta_i=$ (a) 0.03, (b) 0.1, (c) 0.3, and (d) 1. As $\beta_i$ increases, the asymmetric electron energization rate signature develops a pronounced loss of energy for $v_\parallel >0$ at point $\mathbf{r_C}$. 

We suggest that this development of a loss of phase space energy density in the exhaust at $\V{r}_C$ for $v_\parallel >0$, with increasing $\beta_i$, is due to an incomplete cancellation of the contributions to the rate of electron energization from the parallel electron flow and the electron density perturbation.  For the $\beta_i=0.01$ simulation, we show in Fig 5 that (a) the positive energization at  $v_\parallel >0$ due to the parallel flow $U_{\parallel, e}<0$ is almost exactly canceled out by (b) the negative energization at $v_\parallel >0$ due to the electron density perturbation $\delta n_e >0$, leading to (c) little net energization of electrons at $v_\parallel >0$.  If the magnitude of the net parallel flow $U_{\parallel, e}$ decreases more rapidly with increasing $\beta_i$ than the magnitude of the density perturbation $\delta n_e$ decreases with increasing $\beta_i$,  then the sum of these two contributions will not cancel out, but rather will lead to a net loss of phase-space energy density at $v_\parallel >0$.  This picture appears to be consistent with the evolution of the energization signatures in the timestack plots shown in Fig 8.

\begin{figure}   
    \includegraphics[width=3.3in]{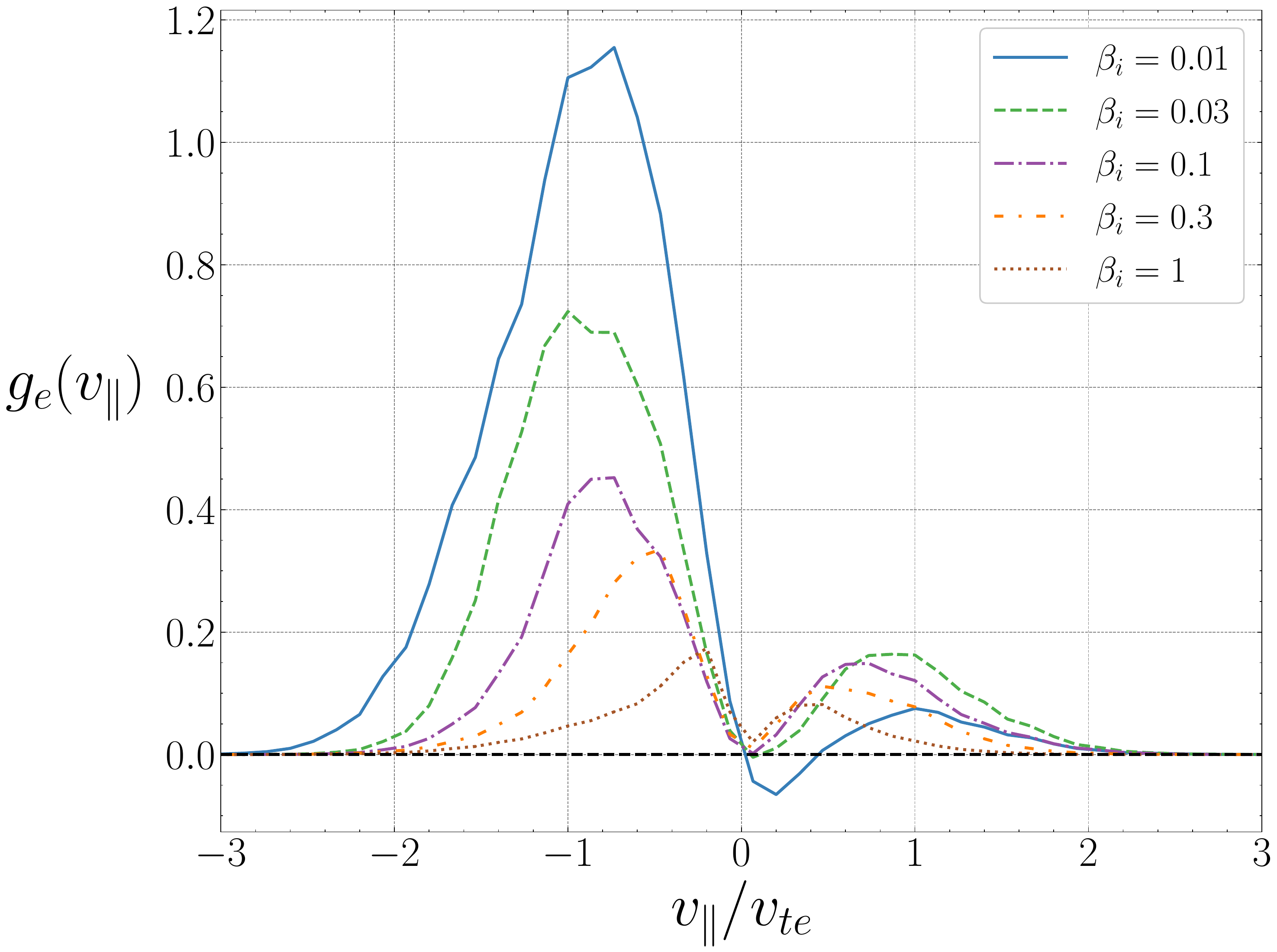}
    \caption[Perturbed Distribution function vs. $\beta_i$]{\label{fig:gevpar_betai} At the peak of the electron energization at point $\V{r}_C$ for each $\beta_i$ simulation, we plot the reduced parallel complementary perturbed distribution function for the electrons $g_e(v_\parallel)$, showing that the perturbations are increasingly confined to a more narrow region around $v_\parallel=0$ as $\beta_i$ increases.}
\end{figure}

It is important to emphasize that all of the asymmetric velocity-space energization signatures at point $\V{r}_C$  shown in \figref{fig:exhaust_betas_timestack_panel} are due to the bulk acceleration of the electrons in the out-of-plane direction by the reconnection electric field $E_\parallel$. Although the lower $|v_\parallel|$ boundary of the positive electron energization appears to decrease along with $v_A/v_{te}$ (vertical dashed lines) as $\beta_i$ increases, this does not necessarily indicate a resonant process. The shift in the positive electron energization to lower $|v_\parallel|$ with increasing $\beta_i$ is governed by a narrowing of the complementary perturbed distribution function $g_e(v_\parallel)$ to smaller values of  $|v_\parallel|$ with increasing $\beta_i$. In \figref{fig:gevpar_betai}, we plot the reduced parallel complementary perturbed distribution function $g_e(v_\parallel)$ at the peak of the electron energization at point $\V{r}_C$ for each $\beta_i$ simulation. This plot shows clearly that the perturbations are increasingly confined to a more narrow region around $v_\parallel=0$ as $\beta_i$ increases.

A resonant energization process, such as electron Landau damping, typically generates a velocity-space signature of energization that is more localized in $v_\parallel$ around the resonant parallel phase velocity $v_A/v_{te}$, as seen in previous studies \cite{Klein:2016a,Klein:2017b,Howes:2017c}.  The velocity-space signatures shown in \figref{fig:exhaust_betas_timestack_panel} are significantly more broad in $v_\parallel$ than expected for a resonant energization mechanism, and appear to indicate a bulk acceleration of the electrons, as our modeling demonstrates in \figref{fig:model_xpoint} and \figref{fig:model_exhaust}.

\section{Discussion}
To understand the kinetic physics governing the energization of electrons in collisionless magnetic reconnection in the strong-guide-field limit, it is critical to recognize that the conversion of the initial magnetic energy into electron heat occurs through a two-step process \cite{Howes:2017c,Howes:2018}: (i) first, collisionless interactions transfer energy from electromagnetic fluctuations to microscopic kinetic energy of the electrons, a reversible process; and (ii) subsequently, the energy transferred to the electrons, which exists as free energy in the non-thermal fluctuations of the electron velocity distribution function (VDF), undergoes a linear \cite{Loureiro:2013b, Howes:2017a} or nonlinear \cite{Tatsuno:2009} phase mixing process to sufficiently small scales in velocity space that arbitrarily weak collisions can thermalize those fluctuations, irreversibly converting the energy to electron heat.  Using nonlinear gyrokinetic simulations of collisionless magnetic reconnection, we focus in this investigation on the first step of this process, and we find that work done on the electrons by the (out-of-plane) reconnection electric field dominates the electron energization through  $j_{\parallel,e} E_\parallel$.

The electron current $j_{\parallel,e}$, necessary to support the change in the in-plane magnetic field across the midplane of the simulation, peaks through the x-point and along the separatrices in the reconnection magnetic field geometry.   The (out-of-plane) reconnection electric field, the $E_\parallel$ component in the strong-guide-field limit, is fairly uniform throughout the ion diffusion region (approximately spanning the range $17< x/\rho_i<23$ and $7< y/\rho_i<25$).  When these fields are combined to determine the work done by $E_\parallel$  on the electrons through  $j_{\parallel,e} E_\parallel$, we find that the electron energization during the main phase of magnetic reconnection (from $10 \lesssim t/\tau_A \lesssim 30$ in the $\beta_i=0.01$ simulation) occurs dominantly at the x-point and along the separatrices within the exhaust, as shown clearly in \figref{fig:coordinate2d_panel}.  Thus, we focus specifically on exploring the energization of the electrons by $E_\parallel$ at the x-point and in the exhaust.

We use the field-particle correlation technique to determine the characteristic \emph{velocity-space signature} $C_{E_{\parallel,e}}(v_{\perp}, v_{\parallel})$ of the electron energization at the x-point and at three positions across the midplane in the exhaust. At the x-point, the velocity-space signature is well modeled by energization of the bulk out-of-plane electron flow $U_{\parallel,e}$ (which provides the current required by Maxwell's equations to support the change in the in-plane magnetic field $B_y$ across the mid-plane) by the parallel electric field $E_\parallel$ which drives the reconnection flow in the $(x,y)$ plane.  In the exhaust, the symmetric (about the midplane) spatial pattern of positive electron energization  $j_{\parallel,e} E_\parallel>0 $ arises from a more complicated kinetic picture of the energization.  The combination of the bulk out-of-plane electron flow $U_{\parallel,e}$  with the well-known quadrupolar electron density variation $\delta n_e$ in guide-field magnetic reconnection \cite{Pritchett:2004, Birn:2007, Loureiro:2013b, Fox:2017, Oieroset:2017} leads to a velocity-space signature that is unexpectedly asymmetric across the midplane: in regions of a negative density perturbation (point A in \figref{fig:exhaust_composite}), electrons with $v_\parallel>0$ experience a net gain in energy; in regions of positive density perturbation (point C in \figref{fig:exhaust_composite}), electrons with $v_\parallel<0$ experience a net gain in energy.  Note that, since a density perturbation leads to zero net energization when integrated over $v_\parallel$ (see the lower panel of \figref{fig:model_exhaust}(b)), the net electron energization at all positions through the exhaust is simply due to the bulk acceleration of the out-of-plane electron flow by $E_\parallel$ through the $j_{\parallel,e} E_\parallel>0 $ work.

The velocity-space signatures of electron energization at the x-point, shown in \figref{fig:xpoint}(b), and within the exhaust, shown in \figref{fig:exhaust_composite}(e), (f), and (g), are key results of this study.  In particular, the asymmetric electron velocity-space signature within the exhaust region is potentially a valuable new way of identifying that one is probing along a trajectory through the exhaust of collisionless magnetic reconnection \emph{using only single-point measurements}.  This technique can be applied to either spacecraft observations from missions such as the Magnetospheric Multiscale (MMS) mission \cite{Burch:2016} or laboratory measurements \citep{Nirwan:2020}, and the possibility to probe the physics of particle energization in magnetic reconnection using single-point measurements is a key implication of this work.

A major conclusion of our modeling of the velocity-space signatures is that the electron energization is due to bulk acceleration of the electron flow by the parallel electric field, rather than some resonant acceleration mechanism, in agreement with previous findings on electron energization by $E_\parallel$ in the strong-guide-field limit of collisionless magnetic reconnection \cite{Egedal:2012,Dahlin:2014,Dahlin:2015,Dahlin:2016}.  This finding differs from the interpretation of the electron energization in strong-guide-field magnetic reconnection by Numata and Loureiro \cite{Numata:2015} (hereafter NL15), where it was suggested that the location in $v_\parallel$ of the fluctuations in the electron velocity distribution function implied a Landau resonant mechanism of energization.  Below we discuss these contrasting interpretations in more detail.

First, it is crucial to emphasize that while our study directly analyzes the work done on the electrons by the electric field---the first step in the two-step process of particle energization in weakly collisional plasmas \cite{Schekochihin:2009,Howes:2017c,Howes:2018}---the NL15 analysis focuses in the second step of the process, the collisional thermalization of energy in the electron velocity distribution.  Note that the energy of the electrons changes in the first step when the electric field does reversible work on the electrons collisionlessly, whereas the second step is the irreversible conversion (through collisions) of the energy gained in the first step, from non-thermal free energy in the electron velocity distribution to thermal energy of the electrons.  These two processes occur at different times and different spatial locations during the process of magnetic reconnection.  Phase mixing is the bridge between these two steps, taking the energy transferred to the electrons in the first step, which is represented by fluctuations in the electron velocity distribution, and transporting these fluctuations to sufficiently small scales in velocity-space that arbitrarily weak collisions can smooth out those fluctuations, irreversibly converting the electron energy into heat of the plasma species.  To be specific, below we use the term ``energization'' to refer to the collisionless work done on electrons that changes their energy, and ``heating'' to refer to the collisional thermalization of that energy.

NL15 report that little electron heating occurs during the main phase of reconnection at the x-point and in the reconnection exhaust.  This is consistent with the weakly collisional conditions of the plasma, whereby Ohmic heating, via resistivity acting on the out-of-plane current, is small compared to the subsequent collisional thermalization of phase-mixed fluctuations in the velocity distribution that contain the energy transferred to the electrons by the parallel electric field.  NL15 suggest that a resonant transfer of energy to the electrons occurs due to the projection of the electron motion (along the total magnetic field, which is dominantly out-of-plane) in the $(x,y)$ plane of the simulation, with a resonant condition on the parallel motion given by $v_\parallel/v_{te} \sim (v_A/v_{te})(B_{z0}/B_\perp) \sim 1$ for $\beta_i=0.01$ and mass ratio $m_i/m_e=100$.  The electron heating is found to peak in the island after the dynamical reconnection phase has ended, and they suggest that the localization in $v_\parallel$ of the linearly phase-mixed fluctuations at $v_\parallel/v_{te} \sim 1$ supports their interpretation of a resonant electron energization.  For $\beta_i =1$, the phase-mixed fluctuations are confined to within $v_\parallel/v_{te} < 1$, qualitatively consistent with the resonant parallel phase velocity decreasing relative to $v_{te}$ as $\beta_i$ increases, which they argue is further evidence of a Landau resonant interaction with the electrons.

Several lines of argument support our interpretation that the electron energization instead is non-resonant in nature, and is simply a bulk acceleration of the electrons by $E_\parallel$. First, the velocity-space signatures of electron energization produced by applying the field-particle correlation technique, presented in \figref{fig:xpoint} and \figref{fig:exhaust_composite}, are well modeled by a simple non-resonant bulk acceleration of the electrons by the reconnection electric field $E_\parallel$, as shown in \figref{fig:model_xpoint} and \figref{fig:model_exhaust}.  Second, Landau resonant acceleration of particles typically generates a velocity-space signature of energization that is more localized in $v_\parallel$ around the resonant parallel phase velocity, with $\Delta v_\parallel \lesssim 0.5 v_{ts}$, as seen in previous studies \cite{Klein:2017b,Howes:2017c}.  The electron energization signatures shown in \figref{fig:xpoint} and \figref{fig:exhaust_composite} are quite broad by comparison, with  $\Delta v_\parallel \gtrsim 1.5 v_{te}$. Third, resonant acceleration of particles implies that the propagation of the accelerating electric field remains in phase with particles over a typically confined range of velocities, as typically found in the collisionless damping of electromagnetic waves, such as kinetic \Alfven waves \cite{Howes:2017c,Klein:2017b} or Langmuir waves \cite{Klein:2016a,Howes:2017a}. In the case of collisionless magnetic reconnection, the parallel electric field is relatively constant over the entire ion diffusion region during the main phase of reconnection, as shown in \figref{fig:coordinate2d_panel}(b), and therefore such a resonant interaction, matching the field's parallel phase velocity with particle velocities, seems unlikely in this case.

A final argument against a resonant interpretation is an alternative explanation for the parallel velocity range of the phase-mixed fluctuations in the electron velocity distributions that are presented in NL15.  The energization of electrons in the exhaust peaks on the magnetic field lines just inside the separatrix, which are swept downstream and ultimately constitute the closed field lines of the magnetic islands where NL15 find that the electron heating peaks.  Since the energization is spatially non-uniform along these closed field lines, occurring primarily in the near exhaust within the ion diffusion region ($17< x/\rho_i<23$ and $7< y/\rho_i<25$), the fluctuations in the electron velocity distribution will subsequently phase mix linearly due to the advective term in the Vlasov equation.  The resulting phase-mixed fluctuations will have the largest amplitudes in the range of parallel velocities where the perturbed electron distribution $g_e(v_\parallel)$ is the largest.  In \figref{fig:gevpar_betai}, we plot $g_e(v_\parallel)$ at the peak of the electron energization at point $\V{r}_C$ for each $\beta_i$ simulation, showing that the perturbed distribution is more narrowly confined to an increasingly  small range of  $|v_\parallel|$  about  $v_\parallel=0$ as $\beta_i$ increases.  This smaller range in $v_\parallel$ is consistent with the lower normalized value of $U_{\parallel,e}/v_{te}$ needed to generate the current required by Maxwell's equations to support the in-plane magnetic field change across the mid-plane as  $\beta_i$ is increased.  Thus, the more narrow localization of the phase-mixed fluctuations in $v_\parallel$ may simply be a consequence of the variation in the initial velocity distributions with  $\beta_i$ that feeds the linear phase mixing process.

A future extension of this work is to explore the dynamics of the phase-mixing process that transports the fluctuations in the electron velocity distribution to small velocity scales in the specific context of the reconnection exhaust and downstream island regions. Such a study would connect our analysis of the collisionless energization of electrons in magnetic reconnection to the collisional dissipation leading to electron heating studied by NL15, and should definitively answer the question of whether the electron energization is resonant or non-resonant.

\section{Conclusion}
Here, we present an analysis of the electron energization in collisionless magnetic reconnection in the limit of strong guide field.  Using 2D gyrokinetic simulations of a tearing unstable current sheet, we apply the field-particle correlation technique to investigate the kinetic physics of the electron energization at the x-point and in the exhaust along the separatrices, where the electrons are dominantly energized during reconnection through work done by the parallel (out-of-plane) component of the electric field, $j_{\parallel,e} E_\parallel$. A key result of this study is the  identification of the velocity-space signatures of the electron energization at the x-point in \figref{fig:xpoint}(b) and at three positions on a trajectory though the exhaust in \figref{fig:exhaust_composite}(e), (f), and (g).  Modeling of these velocity-space signatures suggests that the electron energization is dominated by bulk acceleration of the parallel electron flow by the reconnection (parallel) electric field, a non-resonant mechanism.  This interpretation differs from a previous study \cite{Numata:2015}, which suggested a Landau resonant energization of the electrons.  

Although the energization of the electrons in the exhaust by  $j_{\parallel,e} E_\parallel$ has a symmetric spatial pattern across the mid-plane of the reconnection geometry, the underlying kinetic physics shows an unexpected asymmetric signature. This surprising result raises the possibility that this asymmetry in the velocity-space signatures could be a unique test to identify that one is probing along a trajectory through the exhaust of collisionless magnetic reconnection in the strong-guide-field limit using \emph{only} single-point measurements.  Although multi-spacecraft missions, such as the Magnetospheric Multiscale (MMS) mission \cite{Burch:2016a} have been used to identify the location and probe the dynamics of collisionless magnetic reconnection in space \cite{Burch:2016}, single-point methods such as the field-particle correlation technique have the potential to be applied even on single spacecraft missions with appropriate plasma and field instrumentation,  such as Parker Solar Probe \cite{Fox:2016} and Solar Orbiter \cite{Muller:2013}. 

\appendix
\section{Partition of Energization by Species} \label{sec:rxn_energy}
\begin{figure*}
  \setlength{\unitlength}{1in}
  \begin{picture}(2.75,2.75)
    \put(0,0){\includegraphics[height=2.65in]{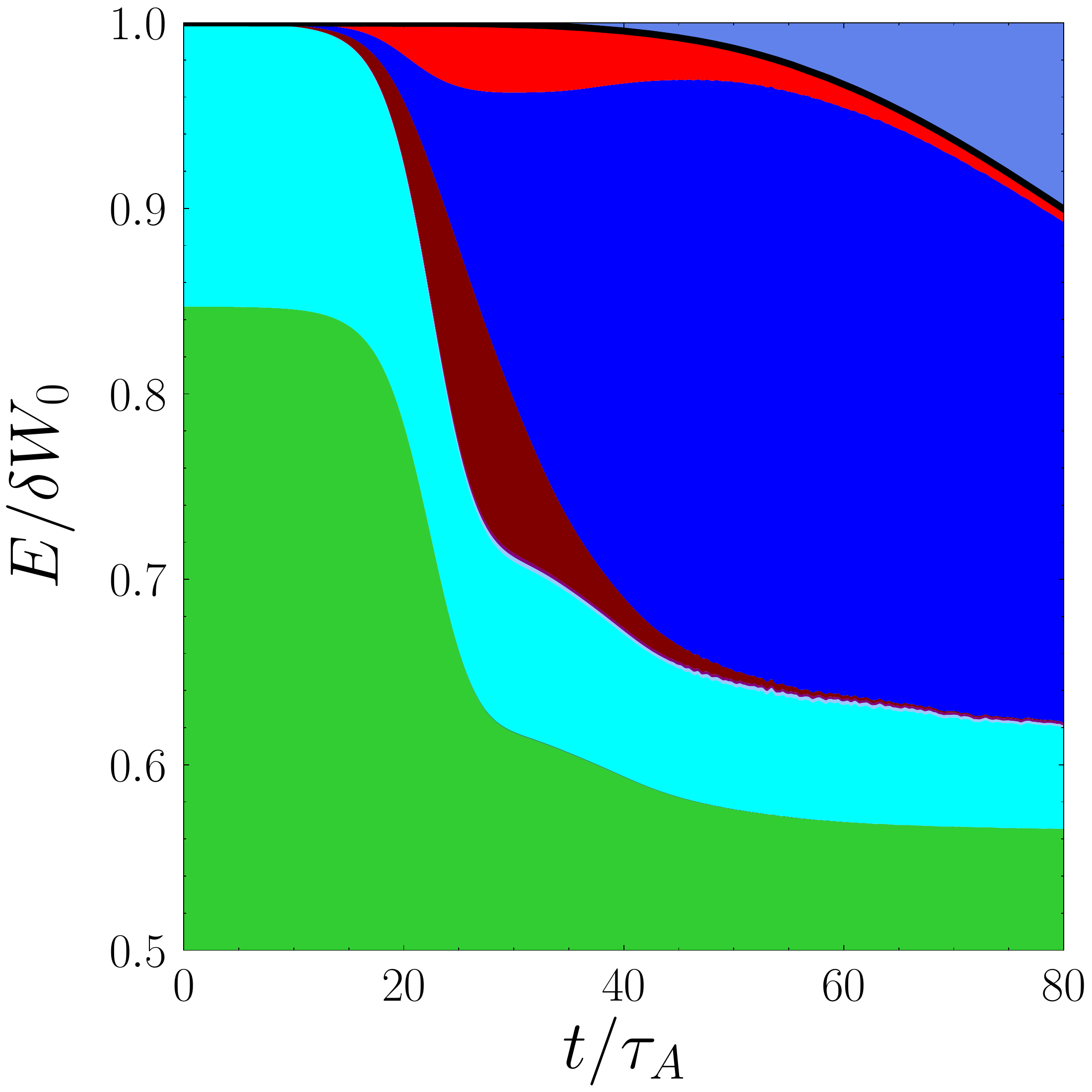}}
    \put(0.3,2.75){(a)\,$\beta_i=0.01$}
  \end{picture}
  \begin{picture}(2.75,2.75)
    \put(0,0){\includegraphics[height=2.65in]{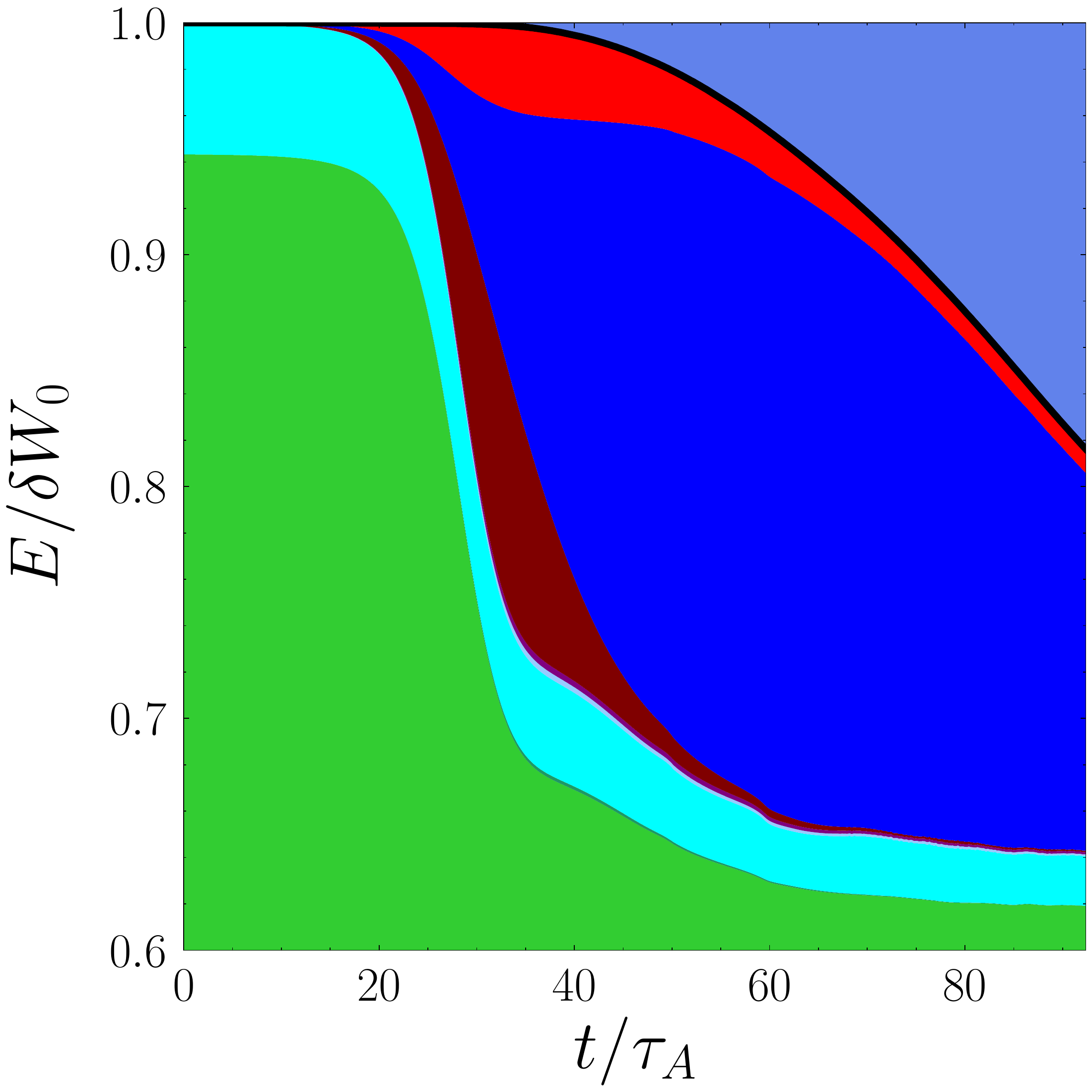}}
     \put(0.3,2.75){(b)\,$\beta_i=0.03$}
  \end{picture}
  \\
  \begin{picture}(2.75,2.75)
    \put(0,0){\includegraphics[height=2.65in]{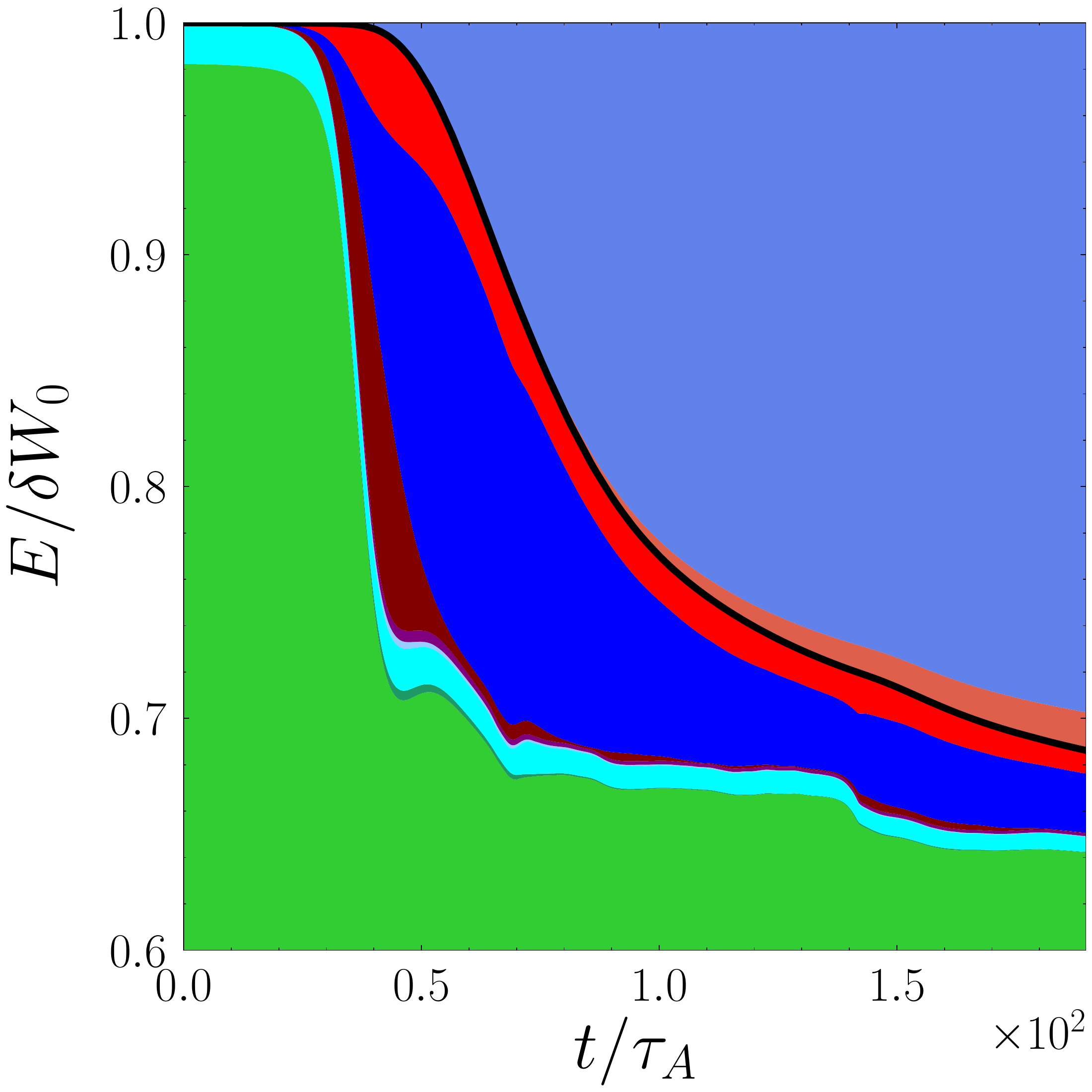}}
     \put(0.3,2.75){(c)\,$\beta_i=0.1$}
  \end{picture}
   \begin{picture}(2.75,2.75)
    \put(0,0){\includegraphics[height=2.65in]{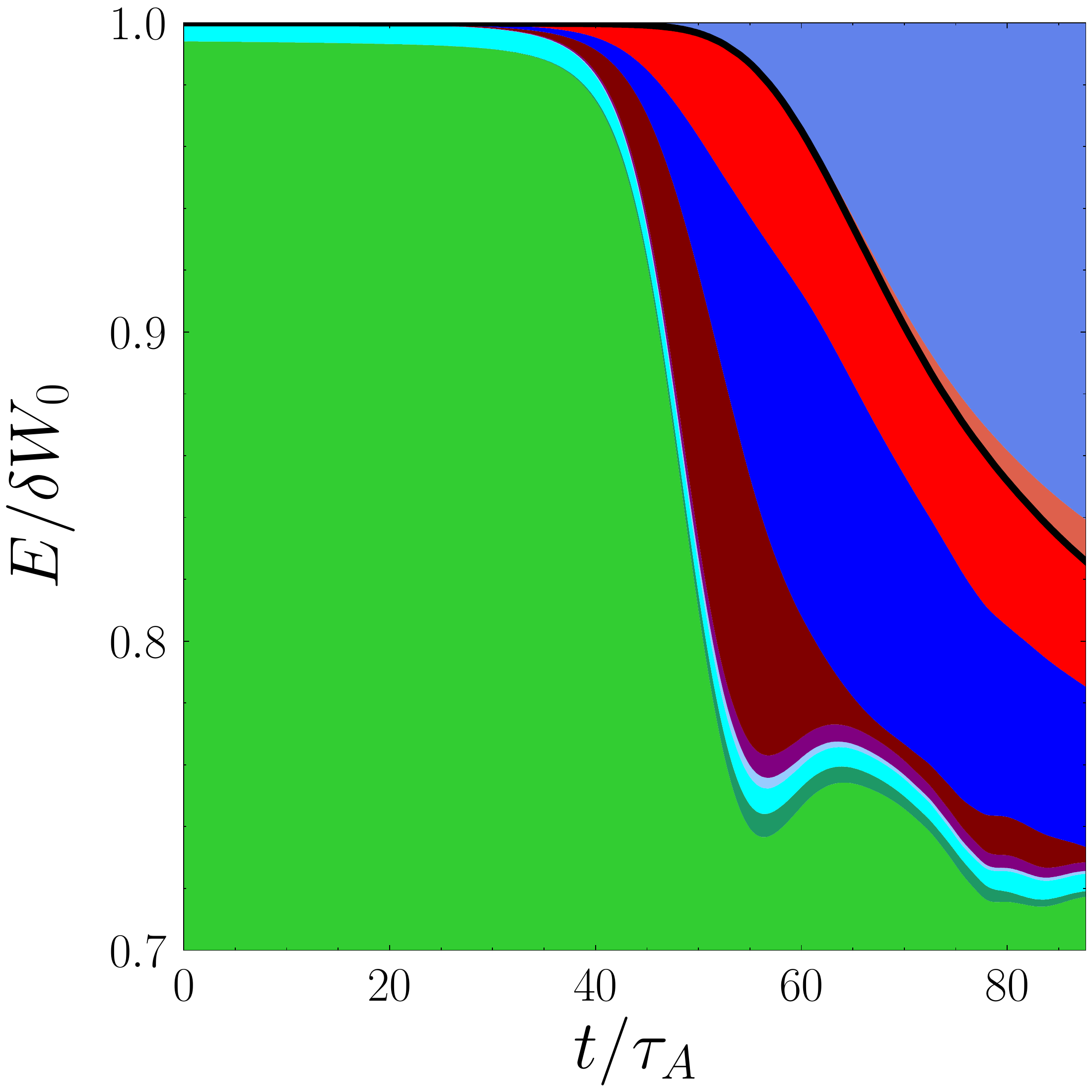}}
    \put(0.35,2.75){(d)\,$\beta_i=0.3$}
  \end{picture}
  \\
  \begin{picture}(2.75,2.75)
    \put(0,0){\includegraphics[height=2.65in]{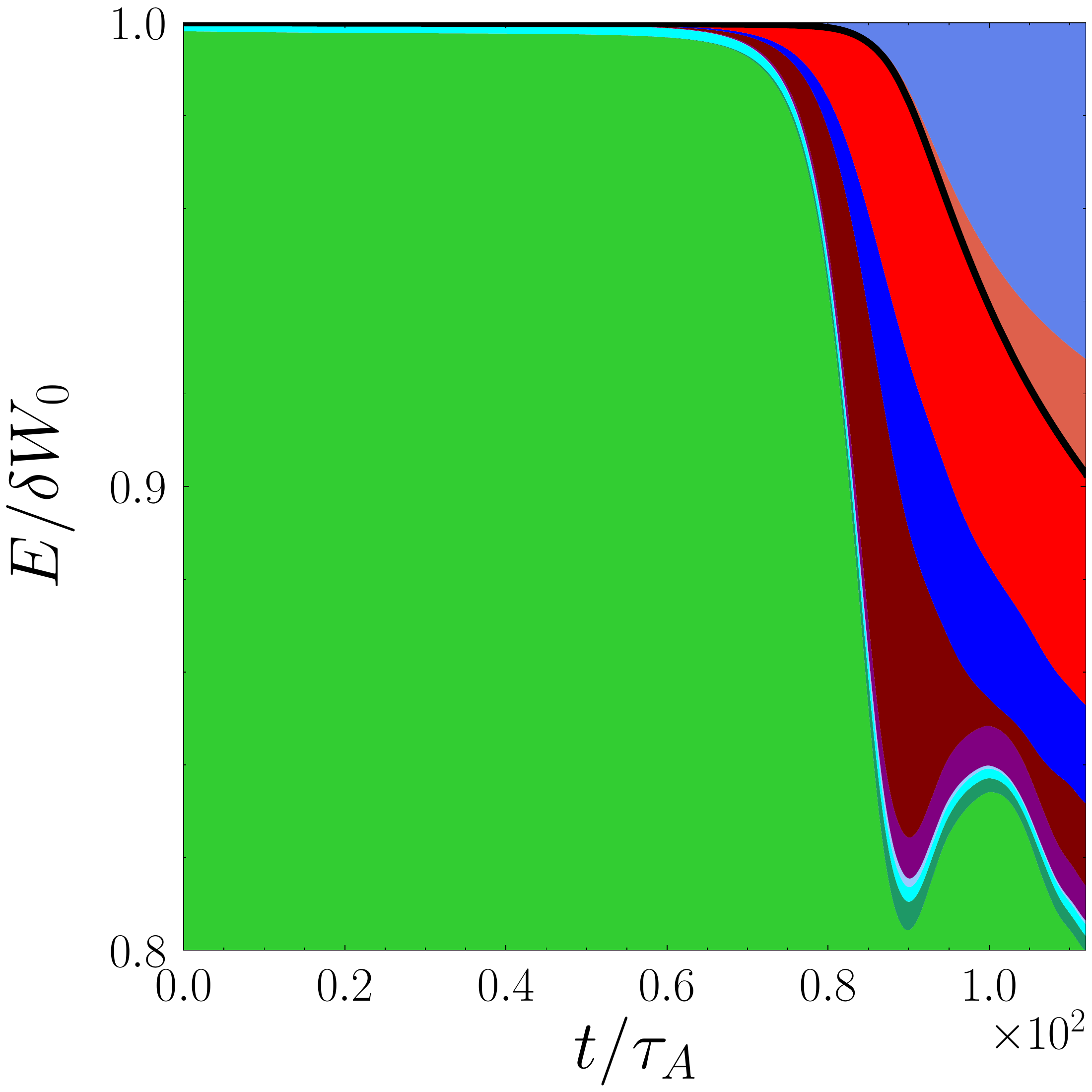}}
    \put(1.2,2.75){(e)\,$\beta_i=1$ }
  \end{picture}
  \caption{Area plots of the energy flow for 2D collisionless magnetic reconnection simulations for (a) $\beta_i=0.01$, (b) $\beta_i=0.03$, (c) $\beta_i=0.1$, (d) $\beta_i=0.3$, and (e)  $\beta_i=1$.  The color coding for the primary contributions to the energy  is $E_{B_\perp}$ (green), $E_{u_\parallel,e}$ (cyan), $E_{u_\perp,i}$ (maroon), $E^{(nt)}_{e}$ (blue), $E^{(nt)}_{i}$ (red), $E_{coll,i}$  (light red), and   $E_{coll,e}$  (light blue).  The other components are  $E_{B_\parallel}$ (dark green),  $E_{u_\perp,e}$ (light purple),  and $E_{u_\parallel,i}$ (medium purple).  The total perturbed energy in the plasma that has not been collisionally thermalized is $\delta W$ (thick solid black line).}
    \label{fig:rxn_energy}
\end{figure*}
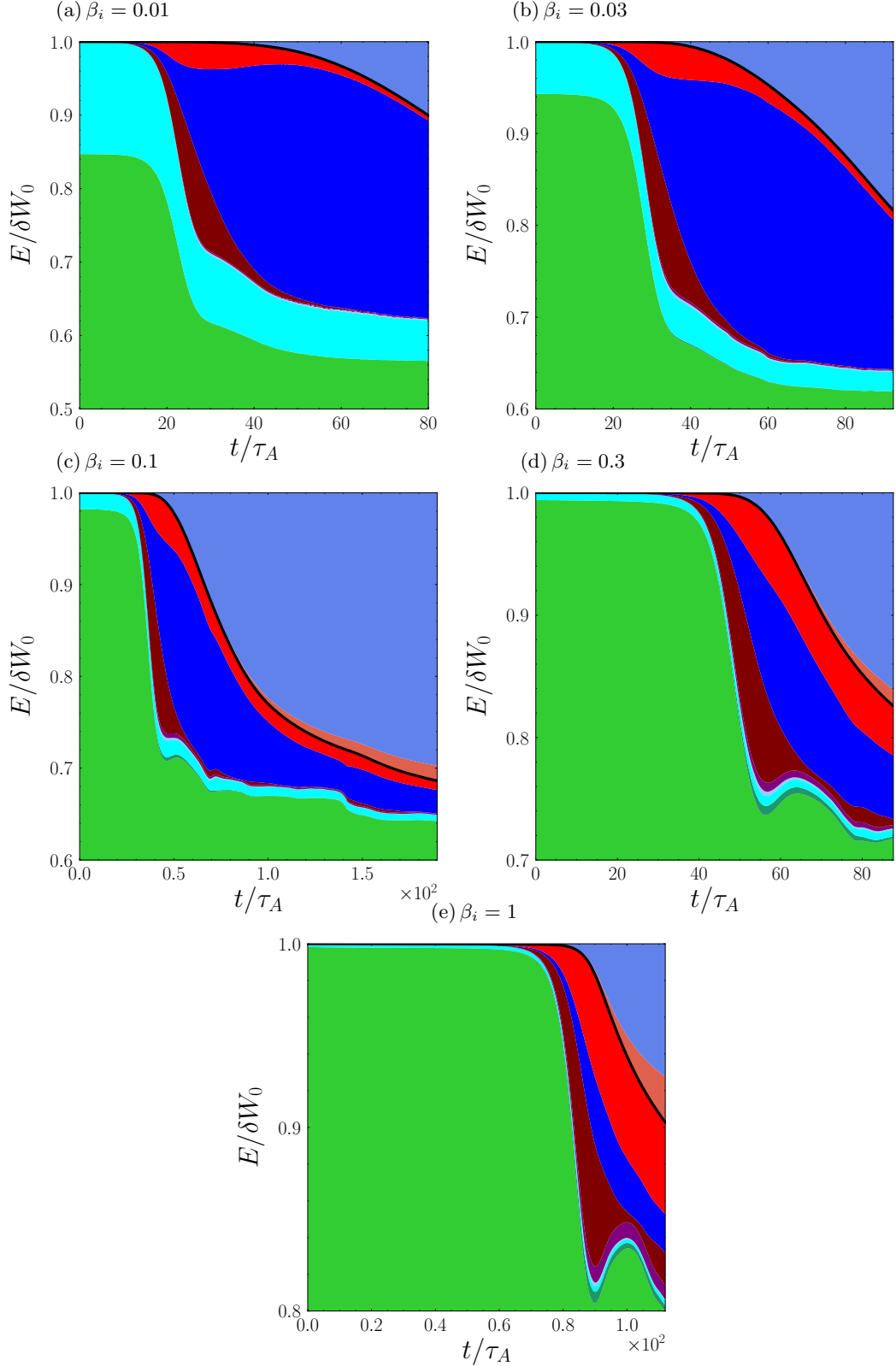

In \T{AstroGK}, there is a full accounting of the particle and field energy partition throughout the simulation. The full energy partition for each simulation with $\beta_i=0.01, 0.03, 0.1, 0.3, 1$ is shown graphically through area plots vs.~time in \figref{fig:rxn_energy}. The energy budget is divided into the different components of the magnetic field energy, non-thermal particle energy, and collisionally thermalized particle energy. In the gyrokinetic limit, the electric field energy is negligible compared to the magnetic energy. \cite{Howes:2006} The following primary partition of energies are: magnetic perpendicular energy $E_{B_\perp}$ (green), parallel electron kinetic energy $E_{u_\parallel,e}$ (cyan), perpendicular ion kinetic energy $E_{u_\perp,i}$ (maroon), non-thermal electron energy $E^{(nt)}_{e}$ (blue), non-thermal ion energy $E^{(nt)}_{i}$ (red), collisional ion energy $E_{coll,i}$  (light red), and  collisional electron energy $E_{coll,e}$  (light blue).  The other components are: parallel magnetic field energy $E_{B_\parallel}$ (dark green), perpendicular electron kinetic energy $E_{u_\perp,e}$ (light purple), parallel ion kinetic energy and $E_{u_\parallel,i}$ (medium purple). We show the fraction of the energy content at the end of each simulation (values $\gtrsim 0.01$) for both species and the magnetic field for each simulation in Table \ref{tab:beta_energy}. In  Table \ref{tab:beta_energy}, $\Sigma E_i$ is the sum of the non-thermal and collisional ion energies, and $\Sigma E_e$ is the sum of the non-thermal and collisional electron energies.
\begin{table*}
    \centering
    \begin{tabular}{ c|c c c| c c c| c c c }
        \hline
        \hline
        $\beta_i$ & $E^{(nt)}_{i}$ &  $E_{coll,i}$ & $\sum E_i$ & $E^{(nt)}_{e}$ & $E_{coll,e}$ & $\sum E_e$ & $E_{B_\perp}$ & $E_{u_\parallel,e}$ & $E_{u_\perp,i}$\\
        \hline
        0.01 & -- & -- & -- & 0.27 & 0.10 & 0.37 & 0.56 & 0.06 & -- \\ 
        \hline
        0.03 & 0.01 & -- & 0.01 & 0.16 & 0.18 & 0.34 & 0.62 & 0.02 & -- \\
        \hline
        0.1 & 0.01 & 0.02 & 0.03 & 0.03 & 0.30 & 0.33 & 0.64 & 0.01 & -- \\
        \hline
        0.3 & 0.04 & 0.01 & 0.05 & 0.05 & 0.16 & 0.21 & 0.72 & -- & -- \\
        \hline
        1.0 & 0.05 & 0.02 & 0.07 & 0.02 & 0.07 & 0.09 & 0.80 & -- & 0.02\\
        \hline
        \hline
    \end{tabular}
    \caption{Final energy partition for each $\beta_i$ simulation normalized by total energy. Note values do not add up to 1 as only the largest values are included (\emph{i.e.} the value is $\gtrsim 0.01$).}
    \label{tab:beta_energy}
\end{table*}

Initially, the majority of the energy in the simulation is contained in the perpendicular (in-plane) component of the magnetic field. The reconnection dynamics then releases some fraction of this initial magnetic energy, leading rapidly to non-thermal energization of the electrons and ions, and some perpendicular bulk acceleration of the ions. Once reconnection begins, the magnetic field energy is quickly transferred to the particles (almost exponential growth of energization), consistent with the fast ramp-up and decline of phase-space energy density rate shown by the field-particle correlation analysis. The parallel bulk kinetic energy of the electrons stays fairly constant in time, even after the primary reconnection phase in each of the simulations. 

It is not until well after the primary reconnection phase commences that thermalization processes begin and the collisionally thermalized energy of the particles (light blue and light red, above the solid black line) begins to increase. At this time, the reconnection has essentially ceased, except for $\beta_e \ge 0.1$, where the formation of a plasmoid at the x-point allows for secondary reconnection. However, the only partition affected during the secondary reconnection is the perpendicular ion bulk energy, which decreases as the perpendicular magnetic energy increases. Once all reconnection has ceased, there is little energization due to the fields. At late times, thermalization is ongoing, as evidenced by the black line, which indicates the total amount of energy in the simulations that has not been collisionally thermalized, retaining a non-zero slope in each plot of \figref{fig:rxn_energy}. 

The thermalization of ion energy is significantly slower than for electrons due to two factors. First, the linear phase mixing that drives non-thermal energy in the particle velocity distributions functions is proportional to the species thermal velocity, and is therefore a factor of $(m_e/m_i)^{1/2}$ slower for the ions than for the electrons. Second, like-species collisions that dominate the thermalization of each species scale as $\nu_{ii}/\nu_{ee} \propto (m_e/m_i)^{1/2}$.  Thus, at the end of each simulation, the ions have collisionally thermalized a significantly smaller fraction of their non-thermal energy than the electrons.

At low  $\beta_i\ll 1$, the electrons receive nearly all of the released magnetic energy.  As $\beta_i$ increases, the ions receive an increasing share of the released magnetic energy, reaching nearly equipartition with the ions at $\beta_i=1$.
\section{Energization Following a Fluid Element} \label{sec:exhaust_trajectory}
\begin{figure*}
    \setlength{\unitlength}{1in}
    \begin{picture}(4.5,3.6)
        \put(0.1,0){\includegraphics[width=4.0in]{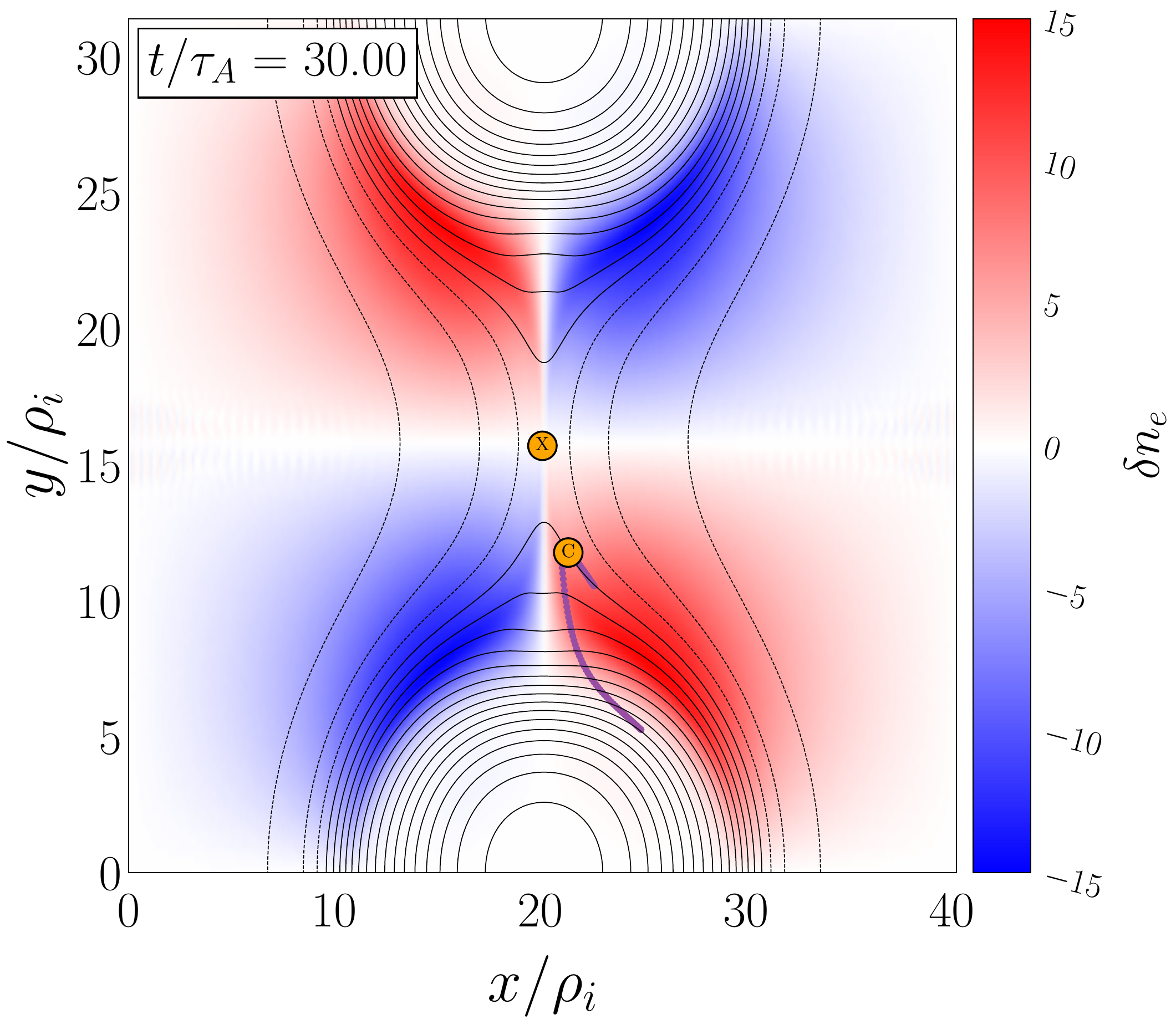}}
        \put(0.1,3.5){(a)}
    \end{picture}
    \\
    \begin{picture}(4.5,4.0)
        \put(0.1,0){ \includegraphics[width=4.0in]{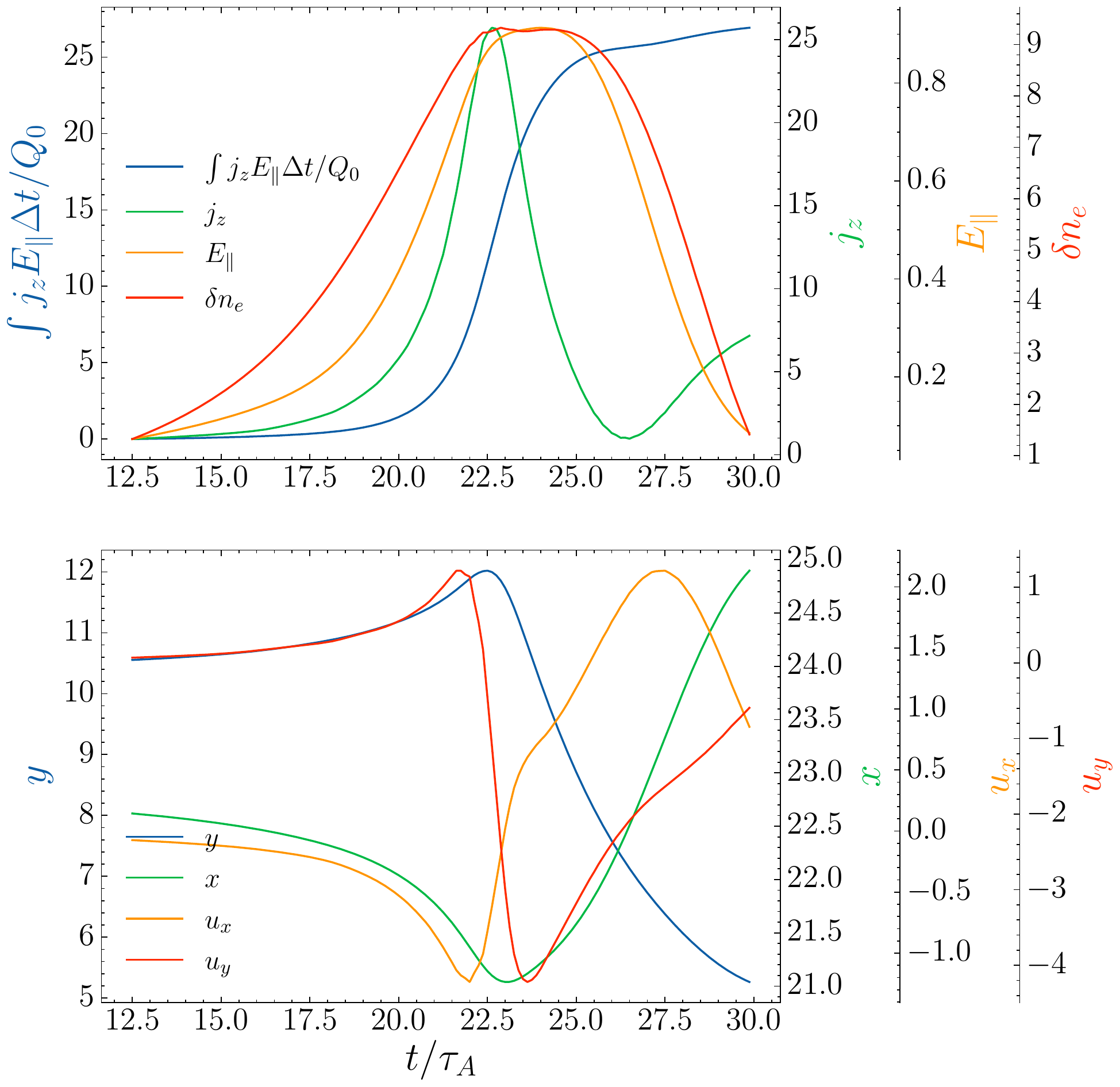}}
        \put(0.1,3.9){(b)}
        \put(0.1,2.0){(c)}
    \end{picture}
    \caption[Electron fluid element trajectory]{\label{fig:exhaust_trajectory}(a) Coordinate trajectory of an electron fluid element beginning on the high density side of the mid-plane. (b) The plasma parameters and field values along the electron fluid element trajectory. (c) The position and electron fluid velocity along the electron fluid element trajectory.}
\end{figure*}

If we follow a fluid element of the electrons along a characteristic trajectory, shown in \figref{fig:exhaust_trajectory}(a), we can identify the incremental cumulative sum of the energization $\sum j_z E_{\parallel} \Delta t / Q_0$ in \figref{fig:exhaust_trajectory}(b). The fluid element initially travels along the in-plane field until it traverses through $\mathbf{r_C}$, where it experiences an increase in parallel acceleration. Once the tearing instability growth rate becomes large, the energization grows with it exponentially in time to its peak. The maximum acceleration occurs at $t/\tau_A = 22.5$, consistent with the overall maximum net energization in $j_{\parallel,e} E_{\parallel}$. Once the magnetic field configuration energy is exhausted after reconnection, the fluid element energization plateaus as the current and parallel electric field drop. 


%
%

%

\begin{acknowledgments}
This work was supported by NSF Grant AGS-1842561. Some figures were generated using modified Matplotlib \cite{Hunter:2007} styles from \citet{SciencePlots}.
\end{acknowledgments}

\section*{Data Availability}
The data that support the findings of this study are available from
the corresponding author upon reasonable request.

\section*{Author Declarations}
The authors have no conflicts to disclose.

\bibliography{abbrev2,reconnection}

\end{document}